\documentclass[%
 aip,
 jmp,%
 amsmath,amssymb,
 reprint,%
]{revtex4-2}

\usepackage{graphicx}
\usepackage{dcolumn}
\usepackage{bm}
\usepackage{xcolor}
\def\[#1\]{\begin{align}#1\end{align}}

\def \nn {\nonumber}
\def \a {\alpha}
\def \b {\beta}

\def \dd{\mathrm{d}}
\def \D {\Delta}
\def \e {\epsilon}

\def \pd{\partial}

\def \ra{\rangle}
\def \la{\langle}

\def \mo{\mathcal{O}}

\def \tr{\text{tr}}
\def \T{\mathcal{T}}

\begin{document}

\title{Pseudo-entropy for descendant operators in two-dimensional conformal field theories}

\author{Song He}
 \email{hesong@jlu.edu.cn}
 \affiliation{%
 Center for Theoretical Physics and College of Physics, Jilin University, Changchun 130012, People's Republic of China
}%
\affiliation{%
 Max Planck Institute for Gravitational Physics (Albert Einstein Institute), Am M\"uhlenberg 1, 14476 Golm, Germany
}%
\author{Jie Yang}
 \email{yangjie@cnu.edu.cn}

\affiliation{%
School of Mathematical Sciences, Capital Normal University, Beijing 100048, People's Republic of China
}%
\author{Yu-Xuan Zhang}
 \email{yuxuanz20@mails.jlu.edu.cn}
\author{Zi-Xuan Zhao}
 \email{zzx23@mails.jlu.edu.cn}
\affiliation{%
 Center for Theoretical Physics and College of Physics, Jilin University, Changchun 130012, People's Republic of China
}%

\date{\today}

\begin{abstract}
We study the late-time behaviors of pseudo-(R\'enyi) entropy of locally excited states in rational conformal field theories (RCFTs).
To construct the transition matrix, we utilize two non-orthogonal locally excited states that are created by the application of different descendant operators to the vacuum. We show that when two descendant operators are generated by a single Virasoro generator acting on the same primary operator, the late-time excess of pseudo-entropy and pseudo-R\'enyi entropy  corresponds to the logarithmic of the quantum dimension of the associated primary operator, in agreement with the case of entanglement entropy. However, for linear combination operators generated by the generic summation of Virasoro generators, we obtain a distinct late-time excess formula for the pseudo-(R\'enyi) entropy compared to that for (R\'enyi) entanglement entropy. As the mixing of holomorphic and antiholomorphic generators enhances the entanglement, in this case, the pseudo-(R\'enyi) entropy can receive an additional contribution. The additional contribution can be expressed as the pseudo-(R\'enyi) entropy of an effective transition matrix in a finite-dimensional Hilbert space.
\end{abstract}

\maketitle

\section{Introduction}
The discovery of the AdS/CFT correspondence \cite{Maldacena:1997re, Gubser:1998bc, Witten:1998qj} has motivated much research related to quantum information theory in the high-energy physics community in recent years. Among them, quantum entanglement, as a carrier of quantum information, plays an increasingly significant role in probing the structure of quantum field theories (QFTs) \cite{Casini:2004bw,Calabrese:2004eu,Kitaev:2005dm,Casini:2016fgb,Nishioka:2018khk,Witten:2018zxz,Casini:2022rlv}, the emergence of geometry \cite{VanRaamsdonk:2010pw,Maldacena:2013xja,Rangamani:2016dms}, the black hole information paradox \cite{Hawking:1976ra,Mathur:2009hf,Almheiri:2012rt,Penington:2019npb,Almheiri:2019psf}.

{Recently, a new entanglement measure, called \textit{pseudo-entropy}, was proposed in \cite{Nakata:2020luh} as a generalization of entanglement entropy. Specifically, pseudo-entropy is a two-state vector version of entanglement entropy, defined as follows. Given two non-orthogonal states $|\psi\rangle$ and $|\varphi\rangle$ in the Hilbert space $\mathcal{H}_{\mathcal{S}}$ of a composed quantum system $\mathcal{S}=A\cup B$, we first construct an operator called the \textit{ transition matrix} acting on $\mathcal{H}_{\mathcal{S}}$\cite{Nakata:2020luh,Guo:2022jzs},
\[
\T^{\psi|\varphi}\equiv\frac{|\psi\ra\la\varphi|}{\la\varphi|\psi\ra}=\frac{\rho_\psi\rho_\varphi}{\tr[\rho_\psi\rho_\varphi]}.\label{transitionmatrix}
\]
The pseudo-entropy of subsystem $A$, then, is obtained by calculating the von Neumann entropy of the reduced transition matrix $\mathcal{T}_A^{\psi|\varphi}\equiv\tr_{B}[\T^{\psi|\varphi}]$,
\[
S(\T^{\psi|\varphi}_A)=-\tr[\T^{\psi|\varphi}_A\log \T^{\psi|\varphi}_A].\label{pseudoentropy}
\]
Generally, the reduced transition matrix is non-Hermitian, requiring careful consideration when discussing pseudo-entropy in systems with infinite-dimensional Hilbert spaces (such as in QFTs) since taking the logarithm of a generic operator requires choosing a radial line in the complex plane that does not intersect the spectrum. \footnote{We thank the anonymous referee for bringing this to our attention.} 
To avoid dealing directly with the logarithm of the non-Hermitian matrix, in practice,  one usually computes a quantity called \textit{pseudo-R\'enyi entropy},
\[
S_A^{(n)}\equiv S^{(n)}(\mathcal{T}_A^{\psi|\varphi}):=\frac{1}{1-n}\log\tr\big[\big(\mathcal{T}_A^{\psi|\varphi}\big)^n\big],\label{pseudorenyi}
\]
instead of pseudo-entropy and the branch of the logarithm function is chosen to be $-\pi<\text{Im}[\log(z)]<\pi$. 
For $n\in N^+$, $n\geq2$, \eqref{pseudorenyi} admits an alternative expression:
\[
S^{(n)}(\mathcal{T}_A^{\psi|\varphi}):=\frac{1}{1-n}\log\big[\sum\limits_j\lambda_j\big(\mathcal{T}_A^{\psi|\varphi}\big)^n\big],\quad \big(\sum_j\lambda_j\big(\mathcal{T}_A^{\psi|\varphi}\big)=1\big),\label{eigenvalue}
\]
where $\lambda_j\big(\mathcal{T}_A^{\psi|\varphi}\big)$ are the eigenvalues of $\mathcal{T}_A^{\psi|\varphi}$, directly following from a Jordan decomposition of $\mathcal{T}_A^{\psi|\varphi}$. In this paper, we mainly focus on the pseudo-R\'enyi entropy for general $n~(n\geq2)$.  When discussing pseudo-entropy, we refer to the Shannon entropy defined by the eigenvalues of $\mathcal{T}_A^{\psi|\varphi}$,
\[
-\sum\limits_j \lambda_j \big(\mathcal{T}_A^{\psi|\varphi}\big)\log[\lambda_j\big(\mathcal{T}_A^{\psi|\varphi}\big)].\label{pseudoentropy2}
\]
For finite-dimensional systems, it is clear that \eqref{pseudoentropy} equals \eqref{pseudoentropy2}, and the latter can be obtained from taking an analytic continuation of $n\to1$ on \eqref{eigenvalue}. However, as previously mentioned, for infinite-dimensional systems, the definition of pseudo-entropy in \eqref{pseudoentropy} may not be well-defined in general.
}

Pseudo-entropy was originally proposed from the study of the generalization of holography entanglement entropy \cite{Nakata:2020luh}. In the AdS/CFT context, the pseudo-entropy of a boundary subsystem is proposed to be dual to the area of a  minimal surface in a Euclidean time-dependent AdS space  \cite{Nakata:2020luh}. In addition, it is found that pseudo-entropy is closely related to postselection experiments in quantum
information \cite{Nakata:2020luh,Akal:2021dqt} (i.e., in addition to the initial state, the system's final state is also specified \cite{aharonov2008two}). {There are also many research interests and prospects driving the study of pseudo-entropy in QFTs \cite{Mollabashi:2020yie,Camilo:2021dtt,Mollabashi:2021xsd,Nishioka:2021cxe,Goto:2021kln,Mukherjee:2022jac,Guo:2022sfl}. See \cite{Miyaji:2021lcq,Ishiyama:2022odv,Bhattacharya:2022wlp,Guo:2022jzs,Doi:2022iyj,Li:2022tsv,Jiang:2023loq,Jiang:2023ffu,Wang:2018jva} for other related developments of pseudo-entropy.}

{ Nonequilibrium dynamics in quantum many-body systems is a subject of intensive research \cite{kamenev2005manybody,stefanucci2013nonequilibrium}. One of the recurring themes is how quantum entanglement arises and propagates in non-equilibrium processes, known as entanglement dynamics. Research shows that chaotic quantum many-body systems can non-locally disrupt quantum information. The scrambling of quantum information will at least lead to the loss of local initial state information and lead to thermalization \cite{deutsch1991quantum,srednicki1994chaos,rigol2008thermalization}. A typical non-equilibrium process in quantum many-body systems is quantum quench \cite{Calabrese:2005in,calabrese2007quantum}. The process usually involves two steps: first prepare an initial state $|\psi\ra$, which can be the ground state of a certain Hamiltonian $H$, and then evolve it with a different Hamiltonian $H'$. One can also quench the system by a local perturbation (generally called a local quench \cite{Calabrese:2007mtj,eisler2007evolution}),
for instance acting on a local operator (generally called a local operator quench \cite{alcaraz2011entanglement,Nozaki:2014hna}). Then the entanglement dynamics are diagnostic about the nature
of this excitation.} 

{ The present paper aims to study the properties of pseudo-(R\'enyi) entropy 
of states obtained by acting on the vacuum with a descendant of a local primary operator (also referred to  as descendant states in this paper) in two-dimensional conformal field theories (2D CFTs). Our study can be traced back to the research on entanglement entropy in local operator quantum quenches in 2D CFTs \cite{Alcaraz:2011tn,Nozaki:2014hna,He:2014mwa,Nozaki:2014uaa,Caputa:2014vaa,Caputa:2014eta,Guo:2015uwa,Caputa:2015tua,Chen:2015usa,Caputa:2016tgt,Numasawa:2016kmo,He:2017lrg,Guo:2018lqq,Apolo:2018oqv,Caputa:2019avh,Bianchi:2022ulu}.\footnote{See \cite{Miyaji:2015woj, Miyaji:2016fse,Zhang:2019kwu,Wen:2015qwa,Kudler-Flam:2018qjo,Kudler-Flam:2019oru,Kudler-Flam:2020url,Kudler-Flam:2020yml,Kudler-Flam:2020xqu} for studies on other information quantities (such as information metric, negativity, reflected entropy, etc) in local or global quantum quenches in CFTs.}
The local operator quench exhibits broad applicability in measuring scrambling and thermalization effects in  CFTs with large central charge \cite{Caputa:2014eta,Caputa:2014vaa,Asplund:2014coa,Asplund:2015eha,Caputa:2016tgt},  which can be regarded as a manifestation of quantum chaos,  as well as in probing the bulk geometry \cite{Suzuki:2019xdq} and characterizing bulk dynamics \cite{Nozaki:2013wia,Caputa:2019avh,Kusuki:2019avm,Kawamoto:2022etl} in the context of AdS/CFT correspondence—an essential avenue for comprehending quantum gravity. }
It is found that the excess of R\'enyi entropy of the local primary or descendant excited states in rational conformal field theories (RCFTs) saturates
to a constant equal to the logarithm of the quantum dimension \cite{Moore:1988ss} of the local operator’s conformal family \cite{He:2014mwa,Caputa:2015tua,Chen:2015usa}. Such saturation is well explained by the picture of quasiparticle pair propagation \cite{Nozaki:2014hna}.  The  related research  has been extended to the pseudo-entropy in parallel \cite{Guo:2022sfl}. Specifically, when studying the real-time evolution of the pseudo-R\'enyi entropy, such as the 2nd pseudo-R\'enyi entropy, for locally primary excited states in RCFTs, the conformal block at early times relies on  the spatial positions of two identical primary operators, leading to a model-dependent pseudo-R\'enyi entropy. Nevertheless, the pseudo-R\'enyi entropy shows a universal behavior at late times, which only depends on the quantum dimension of the primary operator, just like the entanglement entropy. The result suggests that the picture of quasiparticle pair propagation is preserved in the pseudo-entropy. {We generalize the previous study \cite{Guo:2022sfl} on the pseudo-(R\'enyi) entropy to descendant operators in this paper to understand the intricate connections between fusion rules and entanglement properties\cite{He:2023syy}, where fusion rules play a fundamental role in characterizing algebraic and structural properties of a CFT \cite{Shi:2019mlt,Verlinde:1988sn}. The algebraic and structural properties would be encoded in the dynamics of entanglement.} Specifically, we would like to explore the late-time behavior of pseudo-R\'enyi entropy of two descendant operators in RCFTs. We construct the transition matrix using two locally excited states created by the operator
\[
V_\alpha(x)=\sum_{\{n_i\},\{\bar{n}_j\}  }\alpha_{\{n_i\}\{\bar{n}_j\} }\cdot\prod_{i,j}L_{-n_i}\bar{L}_{-\bar{n}_j}\mathcal{O}(x)\label{equ4}
\]
and evaluate the pseudo-R\'enyi entropy using the replica method \cite{Calabrese:2004eu} and conformal mapping. In $\eqref{equ4}$, $\mathcal{O}(x)$ is a primary operator in Schr$\ddot{\text{o}}$dinger picture with chiral and anti-chiral conformal dimension $\Delta$, $L_{-n}$ ($\bar{L}_{-n}$) are holomorphic (anti-holomorphic) Virasoro generators, and $\alpha_{ \{n_i\},\{\bar{n}_j\} }\in\mathbb{C}$ are superposition coefficients. Since the two-point function between descendant operators of different levels does not vanish, the transition matrices we are permitted to construct have more degrees of freedom than the cases of the primary operator.\footnote{Due to the orthogonality of primary operators with different conformal dimensions in the sense of two-point function, we usually have to choose the same primary operators for constructing the transition matrix in specific models \cite{Guo:2022sfl}.} It is interesting to see whether the late-time behavior of the pseudo-(R\'enyi) entropy of subsystems corresponding to these transition matrices has contributions other than the quantum dimension.

The rest of this paper is organized as follows. In section \ref{section2}, we briefly review the replica method for locally excited states in 2D CFTs and provide our convention and some useful formulae for the later calculations. In section \ref{section3:2ndPEE}, we mainly focus on the late-time behavior of the 2nd pseudo-R\'enyi entropy of locally descendant excited states. For simplicity, we study the cases in that a single holomorphic Virasoro generator generates the descendants. More general and complicated situations are discussed in section \ref{section4:kthPEE}, where we derive the late-time behavior of the $k$-th pseudo-R\'enyi entropy for the generic descendant states. We end with conclusions and prospects in section \ref{section5:conclusion}. Some calculation details are presented in the appendices.

\section{Setup in 2D CFTs}\label{section2}
\subsection{Replica method with local operators}
Our focus is on the pseudo-R\'enyi entropy  of locally excited  states created by acting with the operator $V_\a$ \eqref{equ4} on the ground state in RCFTs, {and the subsystem $A$ under consideration in this paper is always taken to be the interval $[0,\infty)$. In this scenario, the pseudo-R\'enyi entropy can be formulated in the path integral formalism using the replica method.} Given an RCFT that lives on a plane and has a vacuum state $|\Omega\ra$, we first prepare two locally excited states using $V_{\alpha}$ to construct a real-time evolved transition matrix $\mathcal{T}^{1|2}(t)$,
\begin{align}
|\psi_1\rangle\equiv e^{-\epsilon H}V_\alpha(x_1)|\Omega\rangle,\quad
|\psi_2\rangle\equiv e^{-\epsilon H}V_{\beta}(x_2)|\Omega\rangle,\quad \mathcal{T}^{1|2}(t)\equiv e^{-iH t}\frac{|\psi_1\rangle\langle\psi_2|}{\langle\psi_2|\psi_1\rangle}e^{iHt}.
\end{align}
Notice that an infinitesimally small parameter  $\epsilon$ has been introduced to suppress the high energy modes \cite{Calabrese:2005in}. We can obtain the reduced transition matrix of subsystem $A$ at time $t$ by tracing out the degrees of freedom of $A^c$ (the complement of $A$), $\mathcal{T}^{1|2}_A(t)=\tr_{A^c}[\T^{1|2}(t)]$.  It turns out that the excess of the $n$-th pseudo-R\'enyi entropy of $A$ with respect to the ground state, defined as $\Delta S^{(n)}(T_{A}^{1|2}(t)):=S^{(n)}(T_A^{1|2}(t))-S^{(n)}\big(\tr_{A^c}\big[|\Omega\rangle\langle\Omega|\big]\big)$, is of the form \cite{Guo:2022sfl}
 \begin{align}
\Delta S^{(n)}(\mathcal{T}_A^{1|2}(t))=\frac{1}{1-n}\big[&\log\langle \prod_{k=1}^{n} V_{\alpha}(w_{2k-1},\bar{w}_{2k-1})V_{\beta}^{\dagger}(w_{2k},\bar{w}_{2k})\rangle_{\Sigma_n}\nn\\
-&n\log \langle V_{\alpha}(w_1,\bar{w}_1)V_\beta^{\dagger}(w_2,\bar{w}_2) \rangle_{\Sigma_1}\big]\label{e4}
\end{align}
using the replica method. In $\eqref{e4}$, $\Sigma_n$ denotes a  $n$-sheeted Riemann surface with cuts on each copy corresponding to $A$, and  $(w_{2k-1},\bar{w}_{2k-1})$and $(w_{2k},\bar{w}_{2k})$  are coordinates on the $k$th-sheet surface. The term in the first line is given by a $2n$-point correlation function on $\Sigma_n$, while a two-point function gives the one in the second line on $\Sigma_1$. We have
\[
w_{2k-1}&=x_1+t-i\epsilon,\quad w_{2k}=x_2+t+i\epsilon,\nn\\
\bar{w}_{2k-1}&=x_1-t+i\epsilon,\quad\bar{w}_{2k}=x_2-t-i\epsilon.\quad(k=1,2,...,n)\label{equ11}
\]
\subsection{Convention and useful formulae}

The $2n$-point correlation function on $\Sigma_n$ in Eq.\eqref{e4} can be evaluated with the help of a conformal mapping
of $\Sigma_n$ to the complex plane $\Sigma_1$. We can then map $\Sigma_n$ to $\Sigma_1$ using the simple conformal mapping
\[
w=z^{n}\label{equ12}.
\]
Let us first focus on the case of $n=2$. The calculation of $\Delta S^{(2)}(\mathcal{T}_A^{1|2}(t))$ is related to the four-point function known pretty well for exactly solvable CFTs. In our convention, using Eq.$\eqref{equ12}$, the four points $z_1,z_2,z_3,z_4$ in the complex plane are given by
\[
&z_1=-z_3=i\sqrt{-x_1-t+i\epsilon},\quad \bar z_1=-\bar z_3=-i\sqrt{-x_1+t-i\epsilon},\nn\\
&z_2=-z_4=i\sqrt{-x_2-t-i\epsilon},\quad \bar{z}_2=-\bar z_4=-i\sqrt{-x_2+t+i\epsilon}.
\label{equ6}
\]
The key point is that one should treat $t\pm i\epsilon$ as a pure imaginary number in all algebraic calculations and take $t$ to be real only in the final expression of the pseudo-R\'enyi entropy. To evaluate the four-point correlation function, it is useful to focus on the cross ratios \cite{Guo:2022sfl}
\[
&\eta:=\frac{z_{12}z_{34}}{z_{13}z_{24}}=\frac{(x_1+x_2+2t)+2\sqrt{(x_1+t)(x_2+t)+\epsilon^2+i\epsilon(x_1-x_2)}}{4\sqrt{(x_1+t)(x_2+t)+\epsilon^2+i\epsilon(x_1-x_2)}},\nn\\
&\bar{\eta}:=\frac{\bar{z}_{12}\bar{z}_{34}}{\bar{z}_{13}\bar{z}_{24}}=\frac{(x_1+x_2-2t)+2\sqrt{(x_1-t)(x_2-t)+\epsilon^2-i\epsilon(x_1-x_2)}}{4\sqrt{(x_1-t)(x_2-t)+\epsilon^2-i\epsilon(x_1-x_2)}},
\label{equ7}
\]
where $z_{ij}=z_i-z_j$, and a useful relation is
\[
1-\eta=\frac{z_{14}z_{23}}{z_{13}z_{24}}.
\]
Since we are mainly interested in the late-time ($t\to\infty$) behavior of pseudo-R\'enyi entropy, one can find some useful late-time formulae from $\eqref{equ6}$
\[
\lim_{t\to\infty}z_1\sim& \lim_{t\to\infty}z_4\sim-\sqrt{t},\quad ~~\lim_{t\to\infty}z_2\sim \lim_{t\to\infty}z_3\sim\sqrt{t},\nn\\
\lim_{t\to\infty}z_{12}\sim& \lim_{t\to\infty}z_{13}\sim-\sqrt{t},\quad \lim_{t\to\infty}z_{24}\sim \lim_{t\to\infty}z_{34}\sim\sqrt{t},\nn\\
\lim_{t\to\infty}z_{14}\sim& \lim_{t\to\infty}z_{23}\sim\sqrt{\frac{1}{t}}.
\label{equ9}
\]
For the cross ratios $(\eta,\bar{\eta})$, as shown in \cite{Guo:2022sfl}, we  have
\[
&\lim_{t\to\infty}(\eta,\bar{\eta})=(1+\frac{(x_2-x_1+2i\epsilon)^2}{16t^2},-\frac{(x_2-x_1-2i\epsilon)^2}{16t^2})\simeq(1,0),\nn\\
&\pd_i\eta\sim\frac{1}{t^{\frac{3}{2}}},\quad \pd_i\pd_j\eta\sim\frac{1}{t}~,\quad \pd_i\pd_j\pd_k\eta\sim\frac{1}{t^{\frac{5}{2}}},\quad\pd_i\pd_j\pd_k\pd_l\eta\sim\frac{1}{t^2}(i\neq j\neq k\neq l).
\label{equ10}
\]
For general $n$-th pseudo-R\'enyi entropy, the $2n$ points $z_1, z_2,...,z_{2n}$ in the $z$-coordinates are given by
\[
z_{2k+1}=&\text{e}^{2\pi i\frac{k+1/2}{n}}(-x_1-t+i\e)^{\frac{1}{n}},\quad \bar z_{2k+1}=\text{e}^{-2\pi i\frac{k+1/2}{n}}(-x_1+t-i\e)^{\frac{1}{n}},\nn\\
z_{2k+2}=&\text{e}^{2\pi i\frac{k+1/2}{n}}(-x_2-t-i\e)^{\frac{1}{n}},\quad \bar z_{2k+2}=\text{e}^{-2\pi i\frac{k+1/2}{n}}(-x_2+t+i\e)^{\frac{1}{n}},\quad (k=0,...,n-1).\label{A=INFzcoordin}
\]

\section{Second pseudo-R\'enyi entropy $\D S_A^{(2)}$ for descendant operators}\label{section3:2ndPEE}
In RCFTs, it is known that the excess of the R\'enyi entropy for the primary/descendant operator saturates to a constant equal to the logarithm of the quantum dimension of the inserted primary operator\cite{He:2014mwa,Caputa:2015tua,Chen:2015usa}. To study the entanglement entropy of local operators, one needs to use two identical operators with the same spatial coordinates to generate the density matrix. However, as mentioned in the introduction, pseudo entropy provides us with greater flexibility---we can use descendant operators of different levels, and with different spatial coordinates, to construct the transition matrix. This section will explore the 2nd pseudo-R\'enyi entropy for some specific descendant operators.
\subsection{$\D S_A^{(2)}$ for $V_{\alpha}=L_{-1}\mathcal{O},~V_{\beta}=\mathcal{O}$}
Let us initially examine the simplest scenario that deviates from the previous studies \cite{Guo:2022sfl}: $V_{\alpha}(x_1)=L_{-1}\mathcal{O}(x_1),~V_{\beta}(x_2)=\mathcal{O}(x_2)$. The 2nd pseudo-R\'enyi entropy, according to $\eqref{e4}$, is related to a four-point function on $\Sigma_2$,
\begin{align}
e^{-\Delta S^{(2)}(\mathcal{T}_A^{1|2}(t))}=&\frac{\langle L_{-1}\mathcal{O}(w_1,\bar{w}_1)\mathcal{O}^{\dagger}(w_2,\bar{w}_2)L_{-1}\mathcal{O}(w_3,\bar{w}_3)\mathcal{O}^{\dagger}(w_4,\bar{w}_4)\rangle_{\Sigma_2}}{\langle L_{-1}\mathcal{O}(w_1,\bar{w}_1)\mathcal{O}^{\dagger}(w_2,\bar{w}_2)\rangle^2_{\Sigma_1}}.\label{e7}
\end{align}
For the first descendant operators, the transformation law under the conformal mapping $w=z^2$ is given by
\begin{align}
\partial\mathcal{O}(w_i,\bar{w}_i)=&(w_i')^{-\Delta}(\bar{w}_i')^{-\Delta}\left((w'_i)^{-1}\partial \mathcal{O}(z_i,\bar{z}_i)-\Delta\frac{w_i''}{(w_i')^2}\mathcal{O}(z_i,\bar{z}_i)\right),
\end{align}
where the prime denotes the derivative with respect to $z$ or $\bar{z}$.  Then the four-point function in $\eqref{e7}$ can be written in terms of correlators on the plane as
\begin{align}
&\langle L_{-1}\mathcal{O}(w_1,\bar{w}_1)\mathcal{O}^{\dagger}(w_2,\bar{w}_2)L_{-1}\mathcal{O}(w_3,\bar{w}_3)\mathcal{O}^{\dagger}(w_4,\bar{w}_4)\rangle_{\Sigma_2}\nonumber\\
=&\big(\prod_{i=1}^{4}|w'_i|^{-2\Delta}\big)\cdot\Big(\frac{\partial z_1}{\partial w_1}\frac{\partial z_3}{\pd w_3}\langle\partial\mathcal{O}(1)\mathcal{O}^{\dagger}(2)\partial\mathcal{O}(3)\mathcal{O}^{\dagger}(4)\rangle_{\Sigma_1}+\Delta^2\big(\frac{\partial z_1}{\pd w_1}\big)^2\frac{\partial^2 w_1}{\pd z_1^2}\big(\frac{\partial z_3}{\pd w_3}\big)^2\frac{\partial^2 w_3}{\pd z_3^2}\langle\mathcal{O}(1)\mathcal{O}^{\dagger}(2)\mathcal{O}(3)\mathcal{O}^{\dagger}(4)\rangle_{\Sigma_1}\nonumber\\
&-\Delta\frac{\pd z_1}{\pd w_1}\big(\frac{\partial z_3}{\pd w_3}\big)^2\frac{\partial^2 w_3}{\pd z_3^2}\langle\partial\mathcal{O}(1)\mathcal{O}^{\dagger}(2)\mathcal{O}(3)\mathcal{O}^{\dagger}(4)\rangle_{\Sigma_1}-\Delta\frac{\pd z_3}{\pd w_3}\big(\frac{\partial z_1}{\pd w_1}\big)^2\frac{\partial^2 w_1}{\pd z_1^2}\langle\mathcal{O}(1)\mathcal{O}^{\dagger}(2)\partial\mathcal{O}(3)\mathcal{O}^{\dagger}(4)\rangle_{\Sigma_1}\Big),\label{eq9}
\end{align}
where we use the notation $\mathcal{O}(i)\equiv\mathcal{O}(z_i,\bar{z}_i)$. Due to the conformal symmetry, we can express the four-point functions involved in \eqref{eq9}  as follows
\begin{align}
\langle \mo(1)\mo^\dagger(2)\mo(3)\mo^{\dagger}(4)\rangle_{\Sigma_1}=&|z_{13}z_{24}|^{-4\Delta}G(\eta,\bar{\eta}),\nonumber\\
\langle \partial \mo(1)\mo^{\dagger}(2)\mo(3)\mo^{\dagger}(4)\rangle_{\Sigma_1}=&|z_{13}z_{24}|^{-4\Delta}\partial_{z_1}G(\eta,\bar{\eta})-\frac{2\Delta}{z_{13}}|z_{13}z_{24}|^{-4\Delta}G(\eta,\bar{\eta}),\nonumber\\
\langle \mo(1)\mo^{\dagger}(2)\partial \mo(3)\mo^{\dagger}(4)\rangle_{\Sigma_1}=&|z_{13}z_{24}|^{-4\Delta}\partial_{z_3}G(\eta,\bar{\eta})+\frac{2\Delta}{z_{13}}|z_{13}z_{24}|^{-4\Delta}G(\eta,\bar{\eta}),\nonumber\\
\langle \partial \mo(1)\mo^{\dagger}(2)\partial \mo(3)\mo^{\dagger}(4)\rangle_{\Sigma_1}=&|z_{13}z_{24}|^{-4\Delta}\partial_{z_1}\partial_{z_3}G(\eta,\bar{\eta})+\frac{2\Delta}{z_{13}}|z_{13}z_{24}|^{-4\Delta}(\partial_{z_1}-\partial_{z_3})G(\eta,\bar{\eta})\nonumber\\
&+\frac{-2\Delta(2\Delta+1)}{z_{13}^2}|z_{13}z_{24}|^{-4\Delta}G(\eta,\bar{\eta}),\label{equa10}
\end{align}
where
\[
G(\eta,\bar{\eta}):=\lim_{z\to\infty}|z|^{4\Delta}\langle \mo(z,\bar{z})\mo(1,1)\mo(\eta,\bar{\eta})\mo(0,0)\rangle_{\Sigma_1}.
\]
Under the conformal mapping between $\Sigma_2$ and $\Sigma_1$, we  have
\[
&\langle L_{-1}\mathcal{O}(w_1,\bar{w}_1)\mathcal{O}^{\dagger}(w_2,\bar{w}_2)L_{-1}\mathcal{O}(w_3,\bar{w}_3)\mathcal{O}^{\dagger}(w_4,\bar{w}_4)\rangle_{\Sigma_2}\nonumber\\
=&2^{-8\Delta}|z_1z_2z_3z_4|^{-2\Delta}|z_{13}z_{24}|^{-4\Delta}\cdot\bigg\{\frac{1}{4z_1z_3}\Big[\pd_{z_1}\pd_{z_3}+\frac{2\Delta}{z_{13}}(\pd_{z_1}-\pd_{z_3})-\frac{2\Delta(2\Delta+1)}{z_{13}^2}\Big]G(\eta,\bar{\eta})\nonumber\\
&+\frac{\Delta^2}{4z_1^2z_3^2}G(\eta,\bar{\eta})-\frac{\Delta}{4z_1z_3^2}\Big[\pd_{z_1}-\frac{2\Delta}{z_{13}}\Big]G(\eta,\bar{\eta})-\frac{\Delta}{4z_1^2z_3}\Big[\pd_{z_3}+\frac{2\Delta}{z_{13}}\Big]G(\eta,\bar{\eta})\bigg\}.\label{eq12}
\]
At late times ($t\to\infty$), as shown in \cite{Guo:2022sfl}, $\eta$ and $\bar{\eta}$ approach 1 and 0, respectively,
which leads to the following late time behavior of $G(\eta,\bar{\eta})$ for RCFTs
\[
\lim_{t\to\infty}G(\eta,\bar{\eta})\sim d^{-1}_{\mo}(1-\eta)^{-2\Delta}\bar{\eta}^{-2\Delta},
\]
{
where $d_{\mo}$ is so called \textit{quantum dimension}  and by using modular S matrix $\mathcal{S}_{ab}$ this is given by $d_{\mo_a}=\mathcal{S}_{0a}/\mathcal{S}_{00}$} \cite{Verlinde:1988sn,fuchs1991quantum}. Hence we can obtain
\[
\lim_{t\to\infty}\partial_{z_1}G(\eta,\bar{\eta})\sim&\frac{2\Delta\partial_{z_1}\eta}{1-\eta}d_{\mathcal{O}}^{-1}(1-\eta)^{-2\Delta}\bar{\eta}^{-2\Delta},\quad \lim_{t\to\infty}\partial_{z_3}G(\eta,\bar{\eta})\sim\frac{2\Delta\partial_{z_3}\eta}{1-\eta}d_{\mathcal{O}}^{-1}(1-\eta)^{-2\Delta}\bar{\eta}^{-2\Delta},\nonumber\\
\lim_{t\to\infty}\partial_{z_1}\partial_{z_3}G(\eta,\bar{\eta})\sim&\frac{2\Delta\partial_{z_{1}}\partial_{z_3}\eta}{1-\eta}d_{\mathcal{O}}^{-1}(1-\eta)^{-2\Delta}\bar{\eta}^{-2\Delta}+\frac{2\Delta(2\Delta+1)\partial_{z_1}\eta\partial_{z_3}\eta}{(1-\eta)^2}d_{\mathcal{O}}^{-1}(1-\eta)^{-2\Delta}\bar{\eta}^{-2\Delta}.\label{e15}
\]
On the other hand, the two-point function in $\eqref{e7}$ is
\[
\langle L_{-1}\mo(w_1,\bar{w}_1)\mo^{\dagger}(w_2,\bar{w}_2)\rangle_{\Sigma_1}=\partial_{w_1}\frac{1}{|w_{12}|^{4\Delta}}=\frac{-2\Delta}{w_{12}}\cdot\frac{1}{|w_{12}|^{4\Delta}}.\label{e16}
\]
Substituting \eqref{eq12}, \eqref{e15} and \eqref{e16} into \eqref{e7} and setting $z_3=-z_1$, $z_4=-z_2$, we obtain, at late times,
\[
e^{-\Delta S^{(2)}(\mathcal{T}_A^{1|2}(t))}\sim&\frac{w^2_{12}}{4\Delta^2}\eta^{2\Delta}(1-\bar{\eta})^{2\Delta}\bigg\{\frac{-1}{4z_1^2}\Big[\frac{2\Delta\frac{z_1^2+z_2^2}{8z_1^3z_2}}{(1-\eta)d_{\mo}}-\frac{2\Delta(2\Delta+1)\frac{(z_1^2-z_2^2)^2}{64z_1^4z_2^2}}{(1-\eta)^2d_\mo}+\frac{2\Delta^2\frac{z_2^2-z_1^2}{4z_1^2z_2}}{z_{1}(1-\eta)d_{\mo}}-\frac{\Delta(2\Delta+1)}{2z_{1}^2d_{\mo}}\Big]\nonumber\\
&+\frac{\Delta^2}{4z_1^4d_{\mo}}-\frac{\Delta}{4z_1^3}\Big[\frac{2\Delta\frac{z_2^2-z_1^2}{8z_1^2z_2}}{(1-\eta)d_{\mo}}-\frac{\Delta}{z_{1}d_{\mo}}\Big]+\frac{\Delta}{4z_1^3}\Big[\frac{2\Delta\frac{z_1^2-z_2^2}{8z_1^2z_2}}{(1-\eta)d_{\mo}}+\frac{\Delta}{z_{1}d_{\mo}}\Big]\bigg\}\nonumber\\
\sim& d_\mo^{-1}.
\]
In going from the second to the third line, we use Eq.\eqref{equ6} and perform the Laurent expansion at infinity.
The late-time limit of the 2nd pseudo-R\'enyi entropy is thus given by
\[
\lim_{t\to\infty}\Delta S^{(2)}\big(\mathcal{T}^{1|2}(t)\big)=\log d_\mo.
\]
In this simplest case, the late-time behavior of the 2nd pseudo-R\'enyi entropy of $L_{-1}\mo$ with $\mathcal{O}$ is the same as that of the primary operator $\mo$. Note that the four-point functions in the plane in Eq.$\eqref{eq9}$ are also encountered when studying the entanglement entropy of $L_{-1}\mathcal{O}$ \cite{Chen:2015usa}. However, they are discarded as sub-leading terms. Our finding shows that  these sub-leading correlators  can also  reproduce the result of $\log d_{\mathcal{O}}$,  as long as we consider the pseudo-R\'enyi entropy instead of the R\'enyi  entropy.
\subsection{$\D S_A^{(2)}$ for $V_\a=L_{-n}\mo,~V_\b=\mo$}
We next consider a more complicated case that $V_\a$ is a general $n$-level descendant associated with the Virasoro generator $L_{-n}$, and $V_{\b}$ is still a primary. The two-point function of $V_\a$ and $V_\b$ reads \cite{DiFrancesco:1997nk}
\[
\langle L_{-n}\mo(w_1,\bar{w}_1)\mo(w_2,\bar{w}_2)\rangle_{\Sigma_1}=\frac{(n+1)\D}{w_{21}^n}|w_{12}|^{-4\D}.\label{eq39}
\]
We then compute the four-point function on $\Sigma_2$. Under the conformal transformation, the level $n$ descendant transforms as
\[
L_{-n}O(w_i,\bar{w}_i)=(w_i')^{-(\D+n)}(\bar{w}_i')^{-\D}L_{-n}\mo(z_i,\bar{z}_i)+...
\label{equ26}
\]
The ellipsis stands for operators with lower conformal dimensions, contributing to lower-order singularities in the correlation functions, that is, we have
\[
&\la\mo^{(-n)}(w_1,\bar{w}_1)\mo^{\dagger}(w_2,\bar{w}_2)\mo^{(-n)}(w_3,\bar{w}_3)\mo^{\dagger}(w_4,\bar{w}_4)\ra_{\Sigma_2}\nn\\
\sim&\big(\prod_{i=1}^{4}|w_i'|^{-2\D}\big)(w'_1)^{-n}(w'_3)^{-n}\la\mo^{(-n)}(1)\mo^\dagger(2)\mo^{(-n)}(3)\mo^{\dagger}(4)\ra_{\Sigma_1}\label{eq41}
\]
at late times. We next pick out the most singular terms of the four-point function on the $z$-plane in \eqref{eq41}. According to $\eqref{equ9}$ and $\eqref{app-eq1}$ in appendix \ref{sec:app1}, the leading contribution at late times in  $\la\mo^{(-n)}(1)\mo^\dagger(2)\mo^{(-n)}(3)\mo^{\dagger}(4)\ra_{\Sigma_1}$ should be
\[
&\la\mo^{(-n)}(1)\mo^\dagger(2)\mo^{(-n)}(3)\mo^{\dagger}(4)\ra_{\Sigma_1}\nonumber\\
=&\frac{(n-1)\D}{z_{41}^n}\la\mo(1)\mo^\dagger(2)\mo^{(-n)}(3)\mo^{\dagger}(4)\ra_{\Sigma_1}+\frac{-\pd_{z_4}}{z^{n-1}_{41}}\la\mo(1)\mo^\dagger(2)\mo^{(-n)}(3)\mo^{\dagger}(4)\ra_{\Sigma_1}+...\nn\\
=&\Big(\frac{(n-1)\D}{z_{41}^n}-\frac{\pd_{z_4}}{z_{41}^{n-1}}\Big)\Big(\frac{(n-1)\D}{z_{23}^n}-\frac{\pd_{z_2}}{z_{23}^{n-1}}\Big)\la\mo(1)\mo^\dagger(2)\mo(3)\mo^{\dagger}(4)\ra_{\Sigma_1}+...\nn\\
=&|z_{13}z_{24}|^{-4\D}d_{\mo}^{-1}(1-\eta)^{-2\D}\bar{\eta}^{-2\D}\nn\\
&\times\bigg(\frac{(n-1)^2\D^2}{z_{41}^nz_{23}^n}-\frac{(n-1)\D}{z_{41}^nz_{23}^{n-1}}\cdot\frac{2\D\pd_{z_2}\eta}{1-\eta}-\frac{(n-1)\D}{z_{41}^{n-1}z_{23}^{n}}\cdot\frac{2\D\pd_{z_4}\eta}{1-\eta}\nn\\
&+\frac{1}{z_{41}^{n-1}z_{23}^{n-1}}\cdot\Big(\frac{2\D(2\D+1)\pd_{z_2}\eta\cdot\pd_{z_4}\eta}{(1-\eta)^2}+\frac{2\D\pd_{z_2}\pd_{z_4}\eta}{1-\eta}\Big)\bigg)+...\label{equ42}
\]
Again, the ellipsis represents the terms that give rise to lower-order singularities in the correlation functions.

Combining  \eqref{eq39}-\eqref{equ42} and taking the limit $t\to\infty$, the leading-order behavior of $\exp\{-\D S^{(2)}(\T_A^{1|2}(t))\}$ is given by
\[
&\lim_{t\to\infty}e^{-\D S^{(2)}\big(\T_A^{1|2}(t)\big)}\nn\\
=& \frac{w_{12}^{2n}}{(n+1)^2\D^2} \times\frac{1}{4^nz_1^nz_3^n}d_{\mo}^{-1}\bigg(\frac{(n-1)^2\D^2}{z_{41}^nz_{23}^n}-\frac{(n-1)\D}{z_{41}^nz_{23}^{n-1}}\cdot\frac{2\D\pd_{z_2}\eta}{1-\eta}-\frac{(n-1)\D}{z_{41}^{n-1}z_{23}^{n}}\cdot\frac{2\D\pd_{z_4}\eta}{1-\eta}\nn\\
&+\frac{1}{z_{41}^{n-1}z_{23}^{n-1}}\cdot\Big(\frac{2\D(2\D+1)\pd_{z_2}\eta\cdot\pd_{z_4}\eta}{(1-\eta)^2}+\frac{2\D\pd_{z_2}\pd_{z_4}\eta}{1-\eta}\Big)\bigg)+...\nn\\
=&\frac{1}{d_\mo}+...
\]
The ellipsis here denotes the sub-leading terms that vanish as $t$ goes to infinity. Hence the late-time limit of the 2nd pseudo-R\'enyi entropy of the transition matrix constructed by a primary $\mo$ and its $n$-level descendant $L_{-n}\mo$ is still $\log d_\mo$.
\subsection{$\D S_A^{(2)}$ for $V_\a=L_{-n}\mo$, $V_\beta=L_{-m}\mo$}
In this subsection, we use the conformal block and operator product expansion (OPE) to show the phenomenon discovered in previous subsections is true for a general case: $V_\a=L_{-n}\mo$, $V_\beta=L_{-m}\mo$.

In terms of \cite{DiFrancesco:1997nk}, the two-point function of $V_\a$ and $V_\b$ reads\footnote{Here the following equation to simplify the result has been used
 
 $\sum \limits_{k=1}^{m-1} \frac{(k+m) (-k+m+1) (k+n-1)!}{(k+1)! (n-2)!}=\frac{2 \left(m^2 n+m \left(2 n^2-1\right)-n (n+1)\right) \Gamma (m+n)}{\Gamma (m+1) \Gamma (n+2)}-m (m+1) n+2$
 }
\[
&\la L_{-n}\mo(w_1,\bar{w}_1)L_{-m}\mo(w_2,\bar{w}_2)\ra_{\Sigma_1}\nn\\
=&\frac{1}{12} (-1)^n (w_1-w_2)^{-m-n}\frac{1}{|w_{12}|^{4\D}} \nn\\
&\left(\frac{\Gamma (m+n) \left(c m \left(m^2-1\right) n \left(n^2-1\right)+24 \Delta  (m+n) (m+n+1) (m n-1)\right)}{\Gamma (m+2) \Gamma (n+2)}+12 \Delta  (\Delta  (m+1) (n+1)+2)\right).
\label{equ44}
\]
The late-time behavior of the four-point function on $\Sigma_2$ of $\eqref{e4}$ can be derived according to $\eqref{equ26}$
\[
&\la\mo^{(-n)}(1)\mo^{(-m)\dagger}(2)\mo^{(-n)}(3)\mo^{(-m)\dagger}(4)\ra_{\Sigma_2}\nn\\
\sim&\big(\prod_{i=1}^{4}|w_i'|^{-2\D}\big)(w'_1)^{-n}(w'_2)^{-m}(w'_3)^{-n}(w'_4)^{-m}\la\mo^{(-n)}(1)\mo^{(-m)\dagger}(2)\mo^{(-n)}(3)\mo^{(-m)\dagger}(4)\ra_{\Sigma_1}.
\label{equ32}
\]
We can next pick out the most singular terms of the four-point function on the $z$-plane in $\eqref{equ32}$. According to $\eqref{equ9}$ , the leading contribution at late times in  $\la\mo^{(-n)}(1)\mo^{(-m)\dagger}(2)\mo^{(-n)}(3)\mo^{(-m)\dagger}(4)\ra_{\Sigma_1}$ comes from the OPE of $\mo^{(-n)}(1)\mo^{(-m)\dagger}(4)$ and $\mo^{(-m)\dagger}(2)\mo^{(-n)}(3)$ and its complete result is given by \eqref{equ45} in appendix \ref{sec:app2}.


Combining  $\eqref{equ44}$ and $\eqref{equ45}$ and taking the limit $t\to\infty$, the leading-order behavior of $e^{-\D S^{(2)}(\T_A^{1|2}(t))}$ is
\[
&\lim_{t\to\infty}e^{-\D S^{(2)}\big(\T_A^{1|2}(t)\big)}
=\frac{1}{d_\mo}+...\label{equ46}
\]
Again, the ellipsis denotes the sub-leading terms that vanish as $t\to\infty$. The late-time limit of the 2nd pseudo-R\'enyi entropy of the transition matrix constructed by a $m$-level descendant operator $L_{-m}\mo$ and a $n$-level descendant operator $L_{-n}\mo$ is $\log d_\mo$, being consistent with the studies in previous sections.

\section{$k$-th pseudo-R\'enyi entropy for generic descendant states}\label{section4:kthPEE}
In the previous section, we find that the 2nd pseudo-R\'enyi entropy corresponding to  $L_{-n}\mo$ and $L_{-m}\mo$ is the same as the 2nd pseudo-R\'enyi entropy of the corresponding primary operator $\mathcal{O}$ at late times, i.e., the logarithm of the quantum dimension of the primary operator $\mo$.  In this section, we shall investigate the $k$-th pseudo-R\'enyi entropy for general descendant states and take $k\to 1$ to obtain the corresponding pseudo-entropy.

\subsection{$\D S_A^{(k)}$ for $V_\a=L_{-n}\mo$, $V_\beta=L_{-m}\mo$}
We begin with studying the case discussed above: $V_\a=L_{-n}\mo$, $V_\beta=L_{-m}\mo$.
According to $\eqref{equ26}$, the $2k$-point function on $\Sigma_k$ at late times can be reformulated as the  $2k$-point function on $\Sigma_1$  as follows, 
\[
&\la\mo^{(-n)}(w_1,\bar w_1)\mo^{(-m)\dagger}(w_2,\bar w_2)\dots\mo^{(-n)}(w_{2k-1},\bar w_{2k-1})\mo^{(-m)\dagger}(w_{2k},\bar w_{2k})\ra_{\Sigma_k},\nn\\
\sim&\mathcal{F}(w_1,w_2,\dots,w_{2k},m,n,\Delta)\la\mo^{(-n)}(1)\mo^{(-m)\dagger}(2)\dots\mo^{(-n)}(2k-1)\mo^{(-m)\dagger}(2k)\ra_{\Sigma_1}+\dots,
\label{equ47}
\]
where
\[
\mathcal{F}(w_1,w_2,\dots,w_{2k},m,n,\Delta)=\big(\prod_{i=1}^{2k}|w_i'|^{-2\D}\big)(w'_1)^{-n}(w'_2)^{-m}\dots(w'_{2k-1})^{-n}(w'_{2k})^{-m}
\label{equ48}
\]
is the leading factor coming from the conformal transformation between correlation functions on $\Sigma_k$ and correlation functions on $\Sigma_1$ and the ellipsis denotes terms contributing to lower-order singularity in the correlation functions. 

\begin{figure}[h]
\centering
\includegraphics[width=0.7\linewidth]{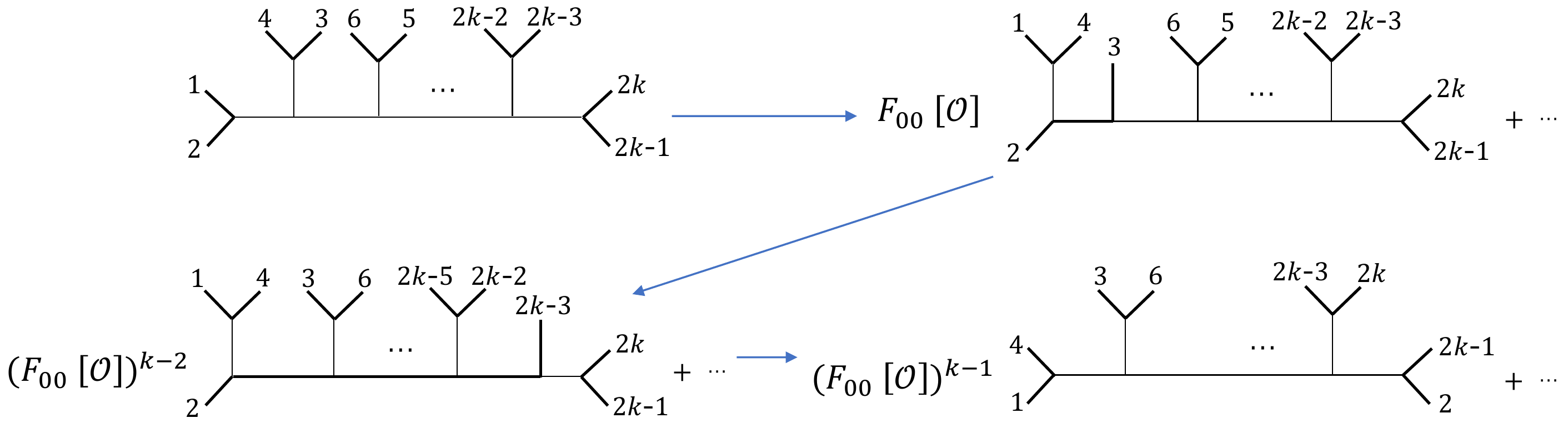}
\caption{$k-1$~fusion transformations to obtain $\Delta S_A^{(k)}$}\label{fusion}
\end{figure}

Based on $\eqref{A=INFzcoordin}$, it can be found that during the late time, $2k$ holomorphic coordinates and $2k$ anti-holomorphic coordinates approach each other in distinct pairings \cite{Guo:2022sfl}.  
\[
\lim\limits_{t\to\infty}(z_{2j+1}-z_{2j+4})&\sim
e^{2\pi i\frac{j+1}{k}}\frac{w_1-w_2}{kt^{1-\frac{1}{k}}}\sim 0, \nn\\
\lim\limits_{t\to\infty}(\bar{z}_{2j+1}-\bar{z}_{2j+2})&\sim e^{-2\pi i\frac{j+\frac{1}{2}}{k}}\frac{\bar w_2-\bar w_1}{kt^{1-\frac{1}{k}}}\sim 0,\quad(j=0,1,...,k-1;z_{2k+2}\equiv z_{2}).
\label{equ49}
\]
Hence, at late times, the most divergent part of the $2k$-point correlation function on the plane in \eqref{equ47} arises from the OPE of $\mo^{(-n)}(2j+1)\mo^{(-m)\dagger}(2j +4)$, i.e.,
\[
&\la\mo^{(-n)}(1)\mo^{(-m)\dagger}(2)\dots\mo^{(-n)}(2k-1)\mo^{(-m)\dagger}(2k)\ra_{\Sigma_1} \nn\\
\sim&\mathcal{D}_{1,4}\mathcal{D}_{3,6}\dots\mathcal{D}_{2k-3,2k}\mathcal{D}_{2k-1,2}\la\mo(1)\mo^{\dagger}(2)\dots\mo(2k-1)\mo^{\dagger}(2k)\ra_{\Sigma_1},
\label{equ37}
\]
where $\mathcal{D}_{2i+1,2i+4}$ is a derivative operator that only contains constants related to the information of two descendant operators and derivatives coming from the most singular part of the OPE of $\mo^{(-n)}(2i+1)\mo^{(-m)\dagger}(2i+4)$, i.e., $\mathcal{D}_{2i+1,2i+4}=\mathcal{D}(\partial_{2i+1},\partial_{2i+4};m,n,c,\Delta)$. 
See appendix \ref{sec:app2} for a concrete example of the $\mathcal{D}$-operator. We need to pick up the proper channel to expand the 2$k$-point function into the holomorphic and the anti-holomorphic part, as graphically shown in figure \ref{fusion}.
In each channel, only the identity operator contributes to the final result. Hence, the 2$k$-point function breaks up into $k$ two-point functions for the holomorphic part (and $k$ for the anti-holomorphic part).
\[
&\la\mo^{(-n)}(1)\mo^{(-m)\dagger}(2)\dots\mo^{(-n)}(2k-1)\mo^{(-m)\dagger}(2k)\ra_{\Sigma_1} \nn\\
\sim&(F_{00}[\mo])^{k-1}\mathcal{D}_{1,4}\dots\mathcal{D}_{2k-3,2k}\mathcal{D}_{2k-1,2}\la\mo(z_1)\mo^{\dagger}(z_4)\ra_{\Sigma_1}\dots\la\mo(z_{2k-3})\mo^{\dagger}(z_{2k})\ra_{\Sigma_1}\la\mo(z_{2k-1})\mo^{\dagger}(z_2)\ra_{\Sigma_1}\nn\\
&\times\la\mo(\bar z_1)\mo^{\dagger}(\bar z_2)\ra_{\Sigma_1}\dots\la\mo(\bar z_{2k-3})\mo^{\dagger}(\bar z_{2k-2})\ra_{\Sigma_1}\la\mo(\bar z_{2k-1})\mo^{\dagger}(\bar z_{2k})\ra_{\Sigma_1}\nonumber\\
\sim&(F_{00}[\mo])^{k-1}\la\mo^{(-n)}(z_1)\mo^{(-m)\dagger}(z_4)\ra_{\Sigma_1}\dots\la\mo^{(-n)}(z_{2k-3})\mo^{(-m)\dagger}(z_{2k})\ra_{\Sigma_1}\la\mo^{(-n)}(z_{2k-1})\mo^{(-m)\dagger}(z_2)\ra_{\Sigma_1}\nn\\
&\times\la\mo(\bar z_1)\mo^{\dagger}(\bar z_2)\ra_{\Sigma_1}\dots\la\mo(\bar z_{2k-3})\mo^{\dagger}(\bar z_{2k-2})\ra_{\Sigma_1}\la\mo(\bar z_{2k-1})\mo^{\dagger}(\bar z_{2k})\ra_{\Sigma_1},
\label{equ50}
\]
where we fomally decompose the operator $\mo(z,\bar z)$ into a product of a holomorphic operator $\mo(z)$ and an anti-holomorphic operator $\mo(\bar z)$, in the sense of the two-point function $\la\mo(1)\mo(2)\ra=z_{12}^{-2\Delta}\bar z_{12}^{-2\Delta}=\la\mo(z_1)\mo(z_2)\ra\la\mo(\bar z_1)\mo(\bar z_2)\ra$.
In the last line, the fact that $\mathcal{D}_{2i+1,2i+4}$ is a linear operator, and coordinates $z_i$ and $z_j$ are independent for $i\neq j$ has been applied.

Changing back into the $w$-coordinate, with the leading divergent term being transformed homogeneously and keeping the most divergent term, we find that
\[
&\la\mo^{(-n)}(w_1,\bar w_1)\mo^{(-m)\dagger}(w_2,\bar w_2)\dots\mo^{(-n)}(w_{2k-1,\bar w_{2k-1}})\mo^{(-m)\dagger}(w_{2k},\bar w_{2k})\ra_{\Sigma_k}\nn\\
\sim&(F_{00}[\mo])^{k-1}\mathcal{F}(w_1,w_2,\dots,w_{2k},m,n,\Delta)\nn\\
&\la\mo^{(-n)}(z_1)\mo^{(-m)\dagger}(z_4)\ra_{\Sigma_1}\dots\la\mo^{(-n)}(z_{2k-3})\mo^{(-m)\dagger}(z_{2k})\ra_{\Sigma_1}\la\mo^{(-n)}(z_{2k-1})\mo^{(-m)\dagger}(z_2)\ra_{\Sigma_1}\nn\\
&\times\la\mo(\bar z_1)\mo^{\dagger}(\bar z_2)\ra_{\Sigma_1}\dots\la\mo(\bar z_{2k-3})\mo^{\dagger}(\bar z_{2k-2})\ra_{\Sigma_1}\la\mo(\bar z_{2k-1})\mo^{\dagger}(\bar z_{2k})\ra_{\Sigma_1}\nn\\
\sim&(F_{00}[\mo])^{k-1}
\la\mo^{(-n)}(w_1)\mo^{(-m)\dagger}(w_4)\ra_{\Sigma_k}\dots\la\mo^{(-n)}(w_{2k-3})\mo^{(-m)\dagger}(w_{2k})\ra_{\Sigma_k}\la\mo^{(-n)}(w_{2k-1})\mo^{(-m)\dagger}(w_2)\ra_{\Sigma_k}\nn\\
&\times\la\mo(\bar w_1)\mo^{\dagger}(\bar w_2)\ra_{\Sigma_k}\dots\la\mo(\bar w_{2k-3})\mo^{\dagger}(\bar w_{2k-2})\ra_{\Sigma_k}\la\mo(\bar w_{2k-1})\mo^{\dagger}(\bar w_{2k})\ra_{\Sigma_k}.
\label{equ52}
\]
By utilizing Eq.\eqref{A=INFzcoordin}, \eqref{equ26} and \eqref{equ49}, we find that in the late-time limit, the correlation functions of both holomorphic and anti-holomorphic two-point functions on $\Sigma_k$ are equal to those on $\Sigma_1$, up to a unitary factor.
\[
&\la\mo^{(-n)}(w_{2j+1})\mo^{(-m)\dagger}(w_{2j+4})\ra_{\Sigma_k}\sim e^{-2\pi i(1+j)(2\Delta+m+n)}\la\mo^{(-n)}(w_{1})\mo^{(-m)\dagger}(w_{2})\ra_{\Sigma_1},\nn\\
&\la\mo(\bar w_{2j+1})\mo^{\dagger}(\bar w_{2j+2})\ra_{\Sigma_k}\sim e^{2\pi i(1+j)2\Delta}\la\mo(\bar w_{1})\mo^{\dagger}(\bar w_{2})\ra_{\Sigma_1},\quad(j=1,2,...,k-1;w_{2k+2}\equiv w_2).
\label{equ53}
\]
Substituting \eqref{equ53} into \eqref{equ52}, the $2k$-point function on $\Sigma_k$ is reduced to
\[
&\la\mo^{(-n)}(w_1,\bar w_1)\mo^{(-m)\dagger}(w_2,\bar w_2)\dots\mo^{(-n)}(w_{2k-1,\bar w_{2k-1}})\mo^{(-m)\dagger}(w_{2k},\bar w_{2k})\ra_{\Sigma_k}\nn\\
\sim& d_{\mathcal{O}}^{1-k}\la\mo^{(-n)}(w_1,\bar w_1)\mo^{(-m)\dagger}(w_2,\bar w_2)\ra^k_{\Sigma_1},\label{KTO2}
\]
where we use the relation between the quantum dimension and the fusion matrix: $d_{\mo}=1/F_{00}[\mo]$.

Finally, in accordance with Eq.\eqref{KTO2}, the excess of the $k$-th pseudo-R\'enyi entropy of $L_{-n}\mo$ and $L_{-m}\mo$ at late times can be deduced as equal to
\[
\lim_{t\to\infty}\Delta S^{(k)}(\mathcal{T}_A^{1|2}(t))=&\lim_{t\to\infty}\frac{1}{1-k}\log\frac{\la\mo^{(-n)}(w_1,\bar w_1)\dots\mo^{(-m)\dagger}(w_{2k},\bar w_{2k})\ra_{\Sigma_k}}{\la\mo^{(-n)}(w_1,\bar w_1)\mo^{(-m)\dagger}(w_2,\bar w_2)\ra^k_{\Sigma_1}}\nonumber\\
=&\log d_{\mo},
\]
which is independent of the level $k$ and consistent with the results of the 2nd pseudo-R\'enyi entropy in the previous sections. Based on the above results, we can conclude that the late-time excess of the pseudo-entropy of $L_{-n}\mo$ and $L_{-m}\mo$ is consistent with the entanglement entropy of $L_{-n}\mo$ and also equals $\log d_{\mo}$.
\subsection{$\D S_A^{(k)}$ for Linear combination of descendant operators }

Let us consider two linear combination operators constructed by operators in $\mo$'s conformal family.
\[
V_\alpha(w,\bar{w})=&\sum_{i=1}^MC_iV_i(w,\bar{w}),~\quad V_i(w,\bar{w})=L_{-\{K_i\}}\bar{L}_{-\{\bar{K}_i\}}\mo(w,\bar{w}),\nn\\
V_\beta(w,\bar{w})=&\sum_{j=1}^{M'}C'_jV'_j(w,\bar{w}),\quad V'_j(w,\bar{w})=L_{-\{K'_j\}}\bar{L}_{-\{\bar{K}'_j\}}\mo(w,\bar{w}),
\]
where $L_{-\{K_i\}}\equiv L_{-k_{i1}}L_{-k_{i2}}...L_{-k_{in_i}}$, $(0\leq k_{i1}\leq k_{i2}\leq...\leq k_{in_i})$, and $L_{-\{\bar K_i\}}\equiv L_{-\bar k_{i1}}L_{-\bar k_{i2}}...L_{-\bar k_{i\bar{n}_i}}$, $(0\leq \bar k_{i1}\leq \bar k_{i2}\leq...\leq \bar k_{i\bar{n}_i})$. Likewise for $L_{-\{K'_j\}}$ and $L_{-\{\bar{K}'_j\}}$. If the combination coefficients $C_i$ $(C'_i)$ are required to be dimensionless, all $V_i(w,\bar{w})$ $\big(V'_i(w,\bar{w})\big)$ should have the same mass dimension, denoted as $N$ $(N')$. This indicates that $\{K_i\}$ and $\{ K'_i\}$ satisfy
\[
|K_i|+|\bar K_i|=N,\quad  |K'_i|+|\bar K'_i|=N',\quad\big(|K_i|\equiv\sum_{j=1}^{n_i}k_{ij},~ |\bar K_i|\equiv\sum_{j=1}^{\bar n_i}\bar k_{ij}\big).
\label{equ56}
\]
Firstly, the two-point function of $V_\alpha$ and $V_\beta^\dagger$ on $\Sigma_1$ is given by
\[
&\la V_\alpha(w_1,\bar{w}_1)V^\dagger_\beta(w_2,\bar{w}_2)\ra_{\Sigma_1}\nn\\
=&\sum\limits_{i=1}^M\sum\limits_{j=1}^{M'}C_i C^{\prime*}_j \la L_{-\{K_i\}}\mo(w_1)L_{-\{K'_j\}}\mo^\dagger(w_2)\ra_{\Sigma_1}\la\bar{L}_{-\{\bar{K}_i\}}\mo(\bar{w}_1) \bar{L}_{-\{\bar{K}'_j\}}\mo^\dagger(\bar{w}_2)\ra_{\Sigma_1}\nn\\
=&\sum\limits_{i=1}^M\sum\limits_{j=1}^{M'}C_i C^{\prime*}_j\frac{c_0(\{K_i\},\{K'_j\})}{(w_1-w_2)^{2\Delta+|K_i|+|K'_j|}}\frac{\bar c_0(\{\bar K_i\},\{\bar{K}'_j\})}{(\bar{w}_1-\bar{w}_2)^{2\Delta+|\bar K_i|+|\bar K'_j|}}.
\label{equ54}
\]
Similar to the previous subsection, in the above, we formally decompose the operator $L_{-\{K_i\}}\bar L_{-\{\bar K_i\}}O(w,\bar w)$ into a holomorphic operator $L_{-\{K_i\}}O(w)$ and an antiholomorphic operator $\bar L_{-\{\bar K_i\}}O(\bar w)$ in the sense of the two-point function.
$c_0$ and $\bar{c}_0$ in Eq.\eqref{equ54} are respectively the coefficients of the holomorphic and antiholomorphic two-point correlation function.

We then cope with the $2k$-point function on $\Sigma_k$. At late times, the $2k$-point function is given by
\[
&\la V_\alpha(w_1,\bar{w}_1)V^\dagger_\beta(w_2,\bar{w}_2)\dots V_\alpha(w_{2k-1},\bar{w}_{{2k-1}})V_\beta^\dagger(w_{{2k}},\bar{w}_{2k})\ra_{\Sigma_k}\nn\\
=&\sum\limits_{i_1,i_3,...,i_{2k-1}}\sum\limits_{j_2,j_4,...j_{2k}}C_{i_1} C^{\prime*}_{j_2}\dots C_{i_{2k-1}} C^{\prime*}_{j_{2k}} \la V_{i_1}(w_1,\bar w_1)V_{j_2}^{'\dagger}(w_2,\bar w_2)... V_{i_{2k-1}}(w_{2k-1},\bar w_{2k-1})V_{j_{2k}}^{'\dagger}(w_{2k},\bar w_{2k})\ra_{\Sigma_k}\nn\\
\sim&d_{\mo}^{1-k}\sum\limits_{i_1,i_3,...,i_{2k-1}}\sum\limits_{j_2,j_4,...j_{2k}}C_{i_1} C^{\prime*}_{j_2}\dots C_{i_{2k-1}} C^{\prime*}_{j_{2k}} \nn\\
&\times\la L_{-\{K_
{i_1}\}}\mo(w_1)L_{-\{K'_{j_4}\}}\mo^\dagger(w_4)\ra_{\Sigma_k}\dots \la L_{-\{K_
{i_{2k-1}}\}}\mo(w_{2k-1})L_{-\{K'_{j_{2k+2}}\}}\mo^\dagger(w_{2k+2})\ra_{\Sigma_k}\quad(2k+2\equiv2)\nn\\
&\times\la\bar{L}_{-\{\bar{K}_{i_1}\}}\mo(\bar{w}_1) \bar{L}_{-\{\bar{K}'_{j_2}\}}\mo^\dagger(\bar{w}_2)\ra_{\Sigma_k}\dots\la\bar{L}_{-\{\bar{K}_{i_{2k-1}}\}}\mo(\bar{w}_{2k-1}) \bar{L}_{-\{\bar{K}'_{j_{2k}}\}}\mo^\dagger(\bar{w}_{2k})\ra_{\Sigma_k}\nn\\
\sim&d_{\mo}^{1-k}
\sum\limits_{i_1,i_3,...,i_{2k-1}}\sum\limits_{j_2,j_4,...j_{2k}}
C_{i_1} C^{\prime*}_{j_2}\dots C_{i_{2k-1}} C^{\prime*}_{j_{2k}}\nn\\
&\times\frac{c_0(\{K_{i_1}\},\{K'_{j_4}\})}{(w_1-w_2)^{2\Delta+|K_{i_1}|+|K'_{j_4}|}}\dots \frac{c_0(\{K_{i_{2k-3}}\},\{K'_{j_{2k}}\})}{(w_{1}-w_{2})^{2\Delta+|K_{i_{2k-3}}|+|K'_{j_{2k}}|}}\frac{c_0(\{K_{i_{2k-1}}\},\{K'_{j_{2}}\})}{(w_{1}-w_{2})^{2\Delta+|K_{i_{2k-1}}|+|K'_{j_{2}}|}}\nn\\
&\times\frac{\bar c_0(\{\bar K_{i_1}\},\{\bar{K}'_{j_2}\})}{(\bar{w}_1-\bar{w}_2)^{2\Delta+|\bar K_{i_1}|+|\bar K'_{j_2}|}}\dots\frac{\bar c_0(\{\bar K_{i_{2k-1}}\},\{\bar{K}'_{j_{2k}}\})}{(\bar{w}_{1}-\bar{w}_{2})^{2\Delta+|\bar K_{i_{2k-1}}|+|\bar K'_{j_{2k}}|}}.
\label{equ55}
\]
In the above derivation, from the first equation to the first tilde, we follow the approach outlined in the preceding subsection: first, we map the $2k$-point function on $\Sigma_k$ to the plane through conformal transformation $w=z^k$ and extract its leading behavior; subsequently, using Eq.\eqref{equ49} and fusion transformation $k-1$ times, we decompose the leading $2k$-point function on the plane into $k$ holomorphic two-point functions and $k$ antiholomorphic two-point functions, and finally, map the $2k$ two-point functions back to $\Sigma_k$. From the first tilde to the second tilde, we utilize a late-time relation similar to \eqref{equ53} as follows:
\[
&\la L_{-\{K\}}\mo(w_{2j+1})L_{-\{K'\}}\mo^{\dagger}(w_{2j+4})\ra_{\Sigma_k}\sim e^{-2\pi i(1+j)(2\Delta+|K|+|K'|)}\la L_{-\{K\}}\mo(w_{1})  L_{-\{K'\}}\mo^{\dagger}(w_{2})\ra_{\Sigma_1},\nn\\
&\la L_{-\{\bar K\}}\mo(\bar w_{2j+1}) L_{-\{\bar K'\}}\mo^{\dagger}(\bar w_{2j+2})\ra_{\Sigma_k}\sim e^{2\pi i(1+j)(2\Delta+|\bar K|+|\bar K'|)}\la L_{-\{\bar K\}}\mo(\bar w_{1}) L_{-\{\bar K'\}}\mo^{\dagger}(\bar w_{2})\ra_{\Sigma_1},\label{xxxx}\\
&(j=1,2,...,k-1;w_{2k+2}\equiv w_2)\nn
\]
which can be readily derived using Eq.\eqref{A=INFzcoordin}, \eqref{equ26} and \eqref{equ49}.

\begin{figure}[htbp]
  \centering
  \includegraphics[width=0.6\linewidth]{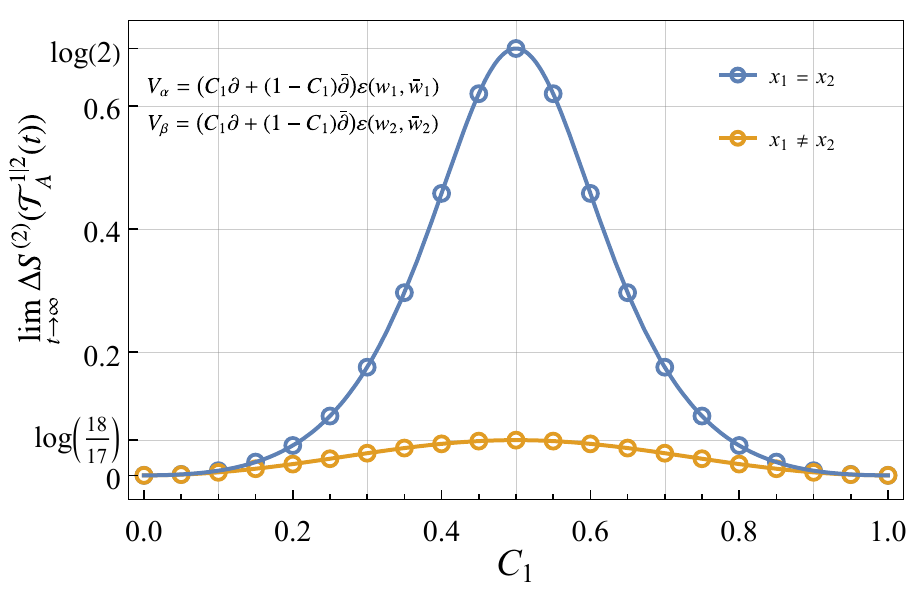}
  \caption{The late-time excess of the 2nd R\'enyi entropy (in blue) or the 2nd pseudo-R\'enyi entropy (in orange) of the linear combination operator $(C_1\partial+(1-C_1)\bar\partial)\mathcal{\varepsilon}$, where $\varepsilon$ is the energy density operator in the critical Ising model. we have $d_{\varepsilon}=1$. The hollow circles represent the numerical data obtained by using the known four-point function of $\varepsilon$, while the solid lines represent the theoretical result obtained by using Eq.\eqref{kthpee2}(or \eqref{kthpee3}) (corresponding to the blue line) and Eq.\eqref{kthpee1}  (corresponding to the orange line). It should be noted that when the linear combination operator is the equally-weighted sum of $L_{-1}\varepsilon$ and $\bar L_{-1}\varepsilon$, i.e., $C_1=1/2$, the late-time excess of the R\'enyi entropy ($x_1=x_2$) is $\log2$, while the late-time excess of the pseudo-R\'enyi entropy ($x_1\neq x_2$) is $\log\frac{18}{17}\approx0.057$.}\label{figsigma}
\end{figure}

Upon substituting Eq.\eqref{equ54} and \eqref{equ55} into the $k$-th pseudo-R\'enyi entropy expression \eqref{e4} and attempting to eliminate $w_{1,2}$, we encounter some subtleties. Specifically, after analytic continuation, the expressions for $w_{1,2}$ and $\bar w_{1,2}$ \eqref{equ11} imply that when $x_1 \neq x_2$, we have $w_{1}-w_{2}=\bar{w}_{1}-\bar{w}_2=x_1-x_2$ in the limit $\epsilon\to0$. Consequently, based on the initial constraint \eqref{equ56}, we can extract the power of $x_1-x_2$ in \eqref{equ55} from the summation, which is equal to $(x_{1}-x_2)^{-k(4\Delta+N+N')}$ (For Eq.\eqref{equ54} it is $(x_{1}-x_2)^{-(4\Delta+N+N')}$ ). The late-time excess formula of the $k$-th pseudo-R\'enyi entropy is thus given by 
\[
&\lim_{t\to\infty}\Delta S^{(k)}(\mathcal{T}_A^{1|2}(t))=\log d_{\mo}+\nn\\
&\frac{1}{1-k}\log\left(\frac{\sum^{M}\limits_{i_1,i_3,...,i_{2k-1}=1} \sum^{M'}\limits_{j_2,j_4,...,j_{2k}=1}\prod\limits_{u=1}^{k}C_{i_{2u-1}}C^{'*}_{j_{2u}}c_0(\{K_{i_{2u-1}}\},\{K'_{j_{2u+2}}\})\bar c_0(\{\bar K_{i_{2u-1}}\},\{\bar{K}'_{j_{2u}}\})}{\Big(\sum\limits_{i}\sum\limits_{j}C_i C^{\prime^*}_j c_0(\{K_i\},\{K'_j\})\bar c_0(\{\bar K_i\},\{\bar{K}'_j\})\Big)^k}\right),
\nn\\
&(\text{for}~x_1\neq x_2;~2k+2\equiv2).\label{kthpee1}
\]
However, things become slightly different when $x_1=x_2$. This is because in this case, $w_1-w_2=-(\bar w_1-\bar w_2)=-2i\epsilon$. When attempting to eliminate the normalization parameter $\epsilon$ by dividing Eq.\eqref{equ55} by the $k$-th power of Eq.\eqref{equ54}, we will be left with a negative power in both the numerator and denominator summations, leading to another late-time excess formula for the $k$-th pseudo-R\'enyi entropy.
\[
&\lim_{t\to\infty}\Delta S^{(k)}(\mathcal{T}_A^{1|2}(t))=\log d_{\mathcal{O}}+\nn\\
&\frac{1}{1-k}\log\left(\frac{\sum^M\limits_{i_1,i_3,...,i_{2k-1}=1}\sum\limits_{j_2,j_4,...j_{2k}=1}^{M'}\prod\limits_{u=1}^{k}(-1)^{|K_{i_{2u-1}}|+|K'_{j_{2u}}|}C_{i_{2u-1}}C_{j_{2u}}^{'*} c_0(\{K_{i_{2u-1}}\},\{K'_{j_{2u+2}}\})\bar c_0(\{\bar K_{i_{2u-1}}\},\{\bar K'_{j_{2u}}\})}{\left(\sum_{i,j}(-1)^{|K_i|+|K'_j|}C_iC^{'*}_jc_0\big([K_i],[K'_j]\big)\bar{c}_0\big([\bar K_i],[\bar K'_j]\big)\right)^k}\right),\nn\\
&(\text{for}~x_1=x_2;~2k+2\equiv2).\label{kthpee2}
\]
Indeed, one can absorb such negative power in Eq.\eqref{kthpee2} into the coefficient of the holomorphic two-point function, to obtain a formula similar to Eq.\eqref{kthpee1}.
\[
&\lim_{t\to\infty}\Delta S^{(k)}(\mathcal{T}_A^{1|2}(t))=\log d_{\mathcal{O}}+\nn\\
&\frac{1}{1-k}\log\left(\frac{\sum^M\limits_{i_1,i_3,...,i_{2k-1}=1}\sum\limits_{j_2,j_4,...j_{2k}=1}^{M'}\prod\limits_{u=1}^{k}C_{i_{2u-1}}C_{j_{2u}}^{'*} c_0(\{K'_{j_{2u+2}}\},\{K_{i_{2u-1}}\})\bar c_0(\{\bar K_{i_{2u-1}}\},\{\bar K'_{j_{2u}}\})}{\left(\sum_{i,j}C_iC^{'*}_jc_0\big(\{K'_j\},\{K_i\}\big)\bar{c}_0\big(\{\bar K_i\},\{\bar K'_j\}\big)\right)^k}\right),\nn\\
&(\text{for}~x_1=x_2;~2k+2\equiv2),\label{kthpee3}
\]
where $c_0(\{K'_j\},\{K_i\}):=(-1)^{2\Delta+|K_i|+|K'_j|}c_0(\{K_i\},\{K'_j\})$.
When we delve into a detailed analysis of these two formulas, we may find that  Eq.\eqref{kthpee2} (or Eq.\eqref{kthpee3}) is compatible with the late-time excess formula for entanglement entropy given in \cite{Chen:2015usa}. This can be verified by simply removing the prime from Eq.\eqref{kthpee2} (or Eq.\eqref{kthpee3}). However, generally speaking, since Eq.\eqref{kthpee1} is not equal to Eq.\eqref{kthpee2}, Eq.\eqref{kthpee1} cannot be reduced to the formula for entanglement entropy. We verify the discontinuity of the pseudo-R\'enyi entropy in these two cases (i.e., $x_1=x_2$ and $x_1\neq x_2$) through numerical calculations in the critical Ising model, see figure \ref{figsigma}. Mathematically, we can attribute this discontinuity of pseudo-R\'enyi entropy to the non-commutativity of the limits as $\epsilon\to0$ and $x_1\to x_2$. It would be interesting to comprehend this point from a physical perspective.

More importantly, regardless of the case (whether $x_1=x_2$ or $x_1\neq x_2$), the late-time excess of the pseudo-R\'enyi entropy of two linear combination operators is composed of two parts. The first part is the logarithm of the quantum dimension of the corresponding primary operator, which reflects the entanglement properties of the primary/descendant operators used to construct the linear combination operators. The second part involves the coefficients of the superposition $C_i$, the coefficients of the holomorphic and antiholomorphic two-point functions, which reflect the additional entanglement generated by the process of linear combination. We can express this part of the additional entanglement contribution as the entanglement entropy (pseudo-entropy) of an effective  density (transition) matrix in a finite-dimensional Hilbert space. Taking Eq.\eqref{kthpee1} as an example (Eq.\eqref{kthpee2} shares a similar treatment), we use the superposition coefficients, the coefficients of holomorphic and antiholomorphic two-point functions to define the following $M\times M$  matrix $\mathcal{T}_{\text{eff}}$,
\[
X_{ij}:=\sqrt{C_{i}C^{'*}_j}\cdot c_0(\{K_i\},\{K'_j\}),\quad\bar X_{ij}:=\sqrt{C_{i}C^{'*}_j}\cdot  \bar c_0(\{\bar K_i\},\{\bar K'_j\}),\nn\\
\mathcal{T}_{\text{eff}}:=\frac{X\bar X^{\mathrm{T}}}{\tr[X\bar X^{\mathrm{T}}]},
\quad(i=1,2,...,M;~j=1,2,...,M').\label{efftm}
\]
We refer to $\mathcal{T}_{\text{eff}}$ as an effective transition matrix, because $\mathcal{T}_{\text{eff}}$ is usually non-Hermitian, and we will see that it characterizes the additional pseudo-R\'enyi entropy at the late time. With the help of $\mathcal{T}_{\text{eff}}$, Eq.\eqref{kthpee1} can be equivalently written as 
 \[
\lim_{t\to\infty}\Delta S^{(k)}(\mathcal{T}_A^{1|2}(t))=\log d_{\mo}+\frac{1}{1-k}\log{\tr[(\mathcal{T}_{\text{eff}})^k]}.
\]
From the above equation, it is clear that the additional pseudo-R\'enyi entropy generated by the linear combination process equals the pseudo-R\'enyi entropy of an effective transition matrix in a $M$-dimentional Hilbert space, and the additional pseudo-entropy thus is equal to $-\tr[\mathcal{T}_{\text{eff}}\log \mathcal{T}_{\text{eff}}]$.

It is evident that the late-time additional contributions for all levels of pseudo-R\'enyi entropy resulting from a linear combination are zero only when $\mathcal{T}_{\text{eff}}$ possesses a single non-zero eigenvalue of 1. We show that the physical origin of these additional corrections is attributed to the mixing of holomorphic Virasoro generators and antiholomorphic Virasoro generators. Considering 
\[
V_{\alpha}=\sum_{i}C_iL_{-\{K_i\}}\bar L_{-\{\bar K\}}\mo,\quad V_{\beta}=\sum_{j}C'_jL_{-\{K'_j\}}\bar L_{-\{\bar K'\}}\mo,
\] 
the holomorphic and antiholomorphic Virasoro generators in $V_{\alpha(\beta)}$ appear in the form of product (not mixed). Based on Eq.\eqref{efftm}, we can write down the matrix element of $\mathcal{T}_{\text{eff}}$
\[
(\mathcal{T}_{\text{eff}})_{ij}=\frac{(\vec{a})_i(\vec{b})_j}{\vec{a}\cdot \vec{b}}&,\nn\\
(\vec{a})_i:=\sqrt{C_i}\sum\limits_{l}C^{'*}_lc_0(\{K_i\},\{K'_l\}),\quad &(\vec{b})_j:=\bar c_0(\{\bar K\},\{\bar K'\})\sqrt{C_j}.
\]
Evidently, $\mathcal{T}_{\text{eff}}$ under this scenario takes the form of the pure state transition matrix \eqref{transitionmatrix}, resulting in a single non-zero eigenvalue of 1 for $\mathcal{T}_{\text{eff}}$. Therefore, we prove that the linear combination process does not generate any additional correction in this case. To provide a heuristic understanding of this results, we may draw an analogy to \cite{Nozaki:2014hna}: when $V_{\alpha(\beta)}(w,\bar{w})$ takes the form of $\sum_{i}C_iL_{-\{K_i\}}\bar L_{-\{\bar K\}}\mo(w,\bar w)$, $V_{\alpha(\beta)}$ can be decomposed into a holomorphic operator $\sum_{i}C_iL_{-\{K_i\}}\mo(w)$ and an antiholomorphic operator $\bar L_{-\{\bar K\}}\mo(\bar {w})$ in the sense of the two-point function $\la V_{\alpha}V^\dagger_{\beta}\ra$, producing a product state $\sum_{i}C_i\big|L_{-\{K_i\}}\mo\big\rangle_{\mathcal{H}}\otimes\big|\bar L_{-\{\bar K\}}\mo\big\rangle_{\bar{\mathcal{H}}}$ in the Verma module $\mathcal{H}\otimes\bar{\mathcal{H}}$ when acting on the vacuum state. However, when $V_{\alpha(\beta)}(w,\bar{w})=\sum_{i}C_iL_{-\{K_i\}}\bar L_{-\{\bar K_i\}}\mo(w,\bar w)$, where $L_{-\{K_i\}}\bar L_{-\{\bar K_i\}}\mo(w,\bar w)$ can each be decomposed into a holomorphic operator $L_{-\{K_i\}}\mo(w)$ and an antiholomorphic operator  $\bar L_{-\{\bar K_i\}}\mo(\bar w)$ in the sense of two-point functions, $V_\alpha$ acting on the vacuum state produces an entangled state $\sum_iC_i\big|L_{-\{K_i\}}\mo\big\rangle_{\mathcal{H}}\otimes\big|\bar L_{-\{\bar K_i\}}\mo\big\rangle_{\bar{\mathcal{H}}}$, enhancing the entanglement.

Finally, we  consider the
late-time excess formula \eqref{kthpee1} for the 2nd pseudo-R\'enyi entropy
\[
&\lim_{t\to\infty}\Delta S^{(2)}(\mathcal{T}^{1|2}_A(t))= \log d_{\mo}\nn\\
&-\log\left(\frac{\sum\limits_{i_1,i_3} \sum\limits_{j_{2},j_4}C_{i_1}C_{i_3}  C^{\prime*}_{j_{2}}C^{'*}_{j_4}c_0(\{K_{i_1}\},\{K'_{j_4}\})c_0(\{K_{i_{3}}\},\{K'_{j_{2}}\})\bar c_0(\{\bar K_{i_1}\},\{\bar{K}'_{j_2}\}) \bar c_0(\{\bar K_{i_{3}}\},\{\bar{K}'_{j_{4}}\})}{(\sum\limits_{i}\sum\limits_{j}C_i C^{\prime*}_j  c_0(\{K_i\},\{K'_j\})\bar c_0(\{\bar K_i\},\{\bar{K}'_j\}))^2}\right)
\label{equ64}
\]
to show the phenomenon of ``mixing enhances the entanglement".
\begin{itemize}
\item Example 1 with $V_\alpha(w_1,\bar{w}_1)=(L_{-1}+\bar{L}_{-1})\mo(w_1,\bar{w}_1)$ , $V_\beta(w_2,\bar{w}_2)=(L_{-1}+\bar{L}_{-1})\mo(w_2,\bar{w}_2)$\\
The two-point function is
\[
&\la V_\alpha(w_1,\bar{w}_1)V^\dagger_\beta(w_2,\bar{w}_2)\ra_{\Sigma_1}=\frac{-4\Delta(4\Delta+1)}{(x_1-x_2)^{4\Delta+2}}.
\label{equ58}
\]
Formula $\eqref{equ58}$ is easy to check. Here, we replace $x_1+t$ and $x_2+t$ into $w_1$ and $w_2$ in the final result. The four-point function at the late time is
\[
&\la V_\alpha(w_1,\bar{w}_1)V^\dagger_\beta(w_2,\bar{w}_2)V_\alpha(w_3,\bar{w}_3)V^\dagger_\beta(w_4,\bar{w}_4)\ra_{\Sigma_2}\sim\frac{1}{d_{\mo}}\frac{ 8\Delta^2(1+16\Delta(1+2\Delta)}{(x_1-x_2)^{8\Delta+4}}.
\label{equ59}
\]
From $\eqref{equ58}$ and $\eqref{equ59}$, we have
\[
\lim\limits_{t\to\infty}\Delta S^{(2)}_A(t)= \log d_{\mo}-\log\left(1-\frac{1}{2(1+4\Delta)^2}\right) .
\]
\begin{figure}[htbp]
  \centering
  \includegraphics[width=0.6\linewidth]{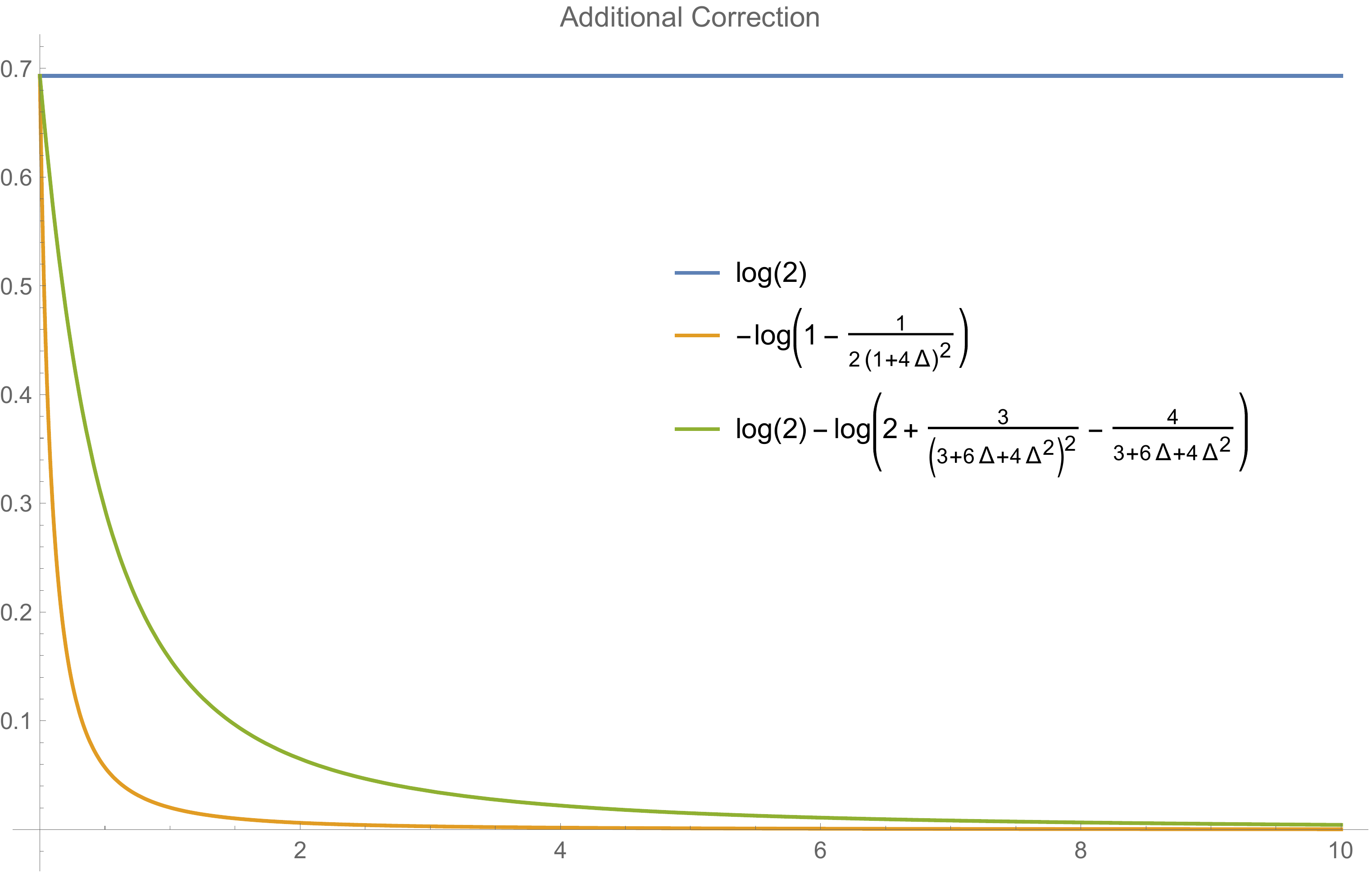}
  \caption{Additional correction of the late-time $\Delta S^{(2)}_A$ due to the mixing of holomorphic and anti holomorphic Virasoro generators. The horizontal axis is the conformal dimension of the primary operator $\mo$.}\label{figex1}
\end{figure}
In this case, the correlation function of $V_\alpha$ and $V_\beta$ can not be divided into the product of the holomorphic part and antiholomorphic part, and $\Delta S^{(2)}$ contains an extra correction $\log\left(1-\frac{1}{2(1+4\Delta)^2}\right)$ besides $\log d_{\mo}$. The relation between extra correction and the conformal weight has shown in figure \ref{figex1} (the orange curve). Note that  the extra correction will be $\log 2$ when we consider another late-time excess formula \eqref{kthpee2}, reproducing the result of entanglement entropy in \cite{Chen:2015usa}.
\item Example 2 with $V_\alpha(w_1,\bar{w}_1)=(L_{-1}L_{-1}+\bar{L}_{-1}\bar{L}_{-1})\mo(w_1,\bar{w}_1)$ , $V_\beta(w_2,\bar{w}_2)=(L_{-1}L_{-1}+\bar{L}_{-1}\bar{L}_{-1})\mo(w_2,\bar{w}_2)$\\
The two-point function is 
\[
\la V_\alpha(w_1,\bar{w}_1)V_\beta(w_2,\bar{w}_2)\ra_{\Sigma_1}=\frac{8\Delta(1+2\Delta)(3+6\Delta+4\Delta^2)}{(x_1-x_2)^{-4(1+\Delta)}}.\label{example2.1}
\]
The four-point function at the late time is
\[
\la V_\alpha(w_1,\bar{w}_1)V_\beta(w_2,\bar{w}_2)V_\alpha(w_3,\bar{w}_3)V_\beta(w_4,\bar{w}_4)\ra_{\Sigma_2}\sim\frac{32\Delta^2(1+2\Delta)^2(9+8\Delta(3+2\Delta)(2+\Delta(3+2\Delta)))}{(x_1-x_2)^{-8(1+\Delta)}}.\label{example2.2}
\]
Combine \eqref{example2.1} and \eqref{example2.2}, the 2nd pseudo-R\'enyi entropy is given by
\[
\lim\limits_{t\to\infty}\Delta S_A^{(2)}(t)=\log d_{\mo}+\log2-\log\left[2+\frac{3}{(3+6\Delta+4\Delta^2)^2}-\frac{4}{3+6\Delta+4\Delta^2}\right].
\]
Notice that there is an additional correction depending on the conformal weight of the corresponding primary operator and its relation with the conformal weight can be seen in figure \ref{figex1} (the green curve). For two general linear combination operators, its pseudo-R\'enyi entropy may also acquire extra correction depending on the central charge and conformal weight of the theory at the late time, and one can calculate the extra correction in general cases. 
\end{itemize}

\section{Conclusion and prospect}\label{section5:conclusion}
In this paper, we investigate the pseudo-R\'enyi entropy of local descendant operators in RCFTs, extending the previous studies in \cite{Guo:2022sfl}\cite{He:2014mwa}\cite{Chen:2015usa}. In \cite{Guo:2022sfl}\cite{Chen:2015usa}, it has been found that the late-time excess of the pseudo-R\'enyi entropy of two primary states and the R\'enyi entropy of a descendant state equal to the logarithmic quantum dimension of the primary operator in RCFTs. It is a natural question to consider the pseudo-R\'enyi entropy of the descendant states.

Firstly, we show that in some special cases: $V_\alpha=L_{-1}\mo$, $V_\beta=\mo$ and $V_\alpha=L_{-n}\mo$, $V_\beta=\mo$ with $\mo$ being primary, the late-time excess of the 2nd pseudo-R\'enyi entropy \eqref{e4} is still logarithmic of the quantum dimension of the primary operator. Using the conformal block and operator product expansion, we compute the 2nd pseudo-R\'enyi entropy constructed by two descendant operators with different Virasoro generators. We show that their 2nd pseudo-R\'enyi entropy is the same as their primaries for such states. Although the calculation looks quite complicated, the leading divergent terms in the late time limit are simple, behaving as the one for primary operators.

Further, we compute $k$-th pseudo-R\'enyi entropy with two descendant operators $L_{-n}\mo$ and $L_{-m}\mo$. We extract the most divergent term of the 2$k$-point function on $\Sigma_k$ with an overall factor $\mathcal{F}$ $\eqref{equ47}$, and then associate the 2$k$-point function of descendant operators with the 2$k$-point function of primary operators $\eqref{equ37}$ with some derivative operators of the form
\[
\mathcal{D}_{2i+1,2i+4}=\mathcal{D}(\partial_{2i+1},\partial_{2i+4};m,n,c,\Delta).
\]
We find the $2k$-point function breaks up into $k$ two-point functions for the holomorphic part(and $k$ for the anti-holomorphic part). The two-point function only depends on the conformal weight and some constant $\eqref{equ53}$. As a result, in this case, the pseudo-entropy of the descendant operators is the same as primaries.

Finally, we discuss the most generic descendant operators, which are two linear combination operators constructed by operators in $\mo$'s conformal family
\[
V_\alpha(w_1,\bar{w}_1)=\sum_{i}C_iV_i(w_1,\bar{w}_1),\quad
V_\beta(w_2,\bar{w}_2)=\sum_{j}C'_jV_j(w_2,\bar{w}_2).
\]
We derive the formula for $k$-th pseudo-R\'enyi entropy of linear combination operators at the late time
\[
&\lim_{t\to\infty}\Delta S^{(k)}(\mathcal{T}^{1|2}(t))=\log d_{\mo}+\nn\\
&\frac{1}{1-k}\log\left(\frac{\sum^{M}\limits_{i_1,i_3,...,i_{2k-1}=1} \sum^{M'}\limits_{j_2,j_4,...,j_{2k}=1}\prod\limits_{u=1}^{k}C_{i_{2u-1}}C^{'*}_{j_{2u}}c_0(\{K_{i_{2u-1}}\},\{K'_{j_{2u+2}}\})\bar c_0(\{\bar K_{i_{2u-1}}\},\{\bar{K}'_{j_{2u}}\})}{\Big(\sum\limits_{i}\sum\limits_{j}C_i C^{\prime^*}_j c_0(\{K_i\},\{K'_j\})\bar c_0(\{\bar K_i\},\{\bar{K}'_j\})\Big)^k}\right),\nn\\
&(\text{for}~x_1\neq x_2;~2k+2\equiv2),
\label{c111}\]
which is pretty different from the formula derived when $x_1\neq x_2$ 
\[
&\lim_{t\to\infty}\Delta S^{(k)}(\mathcal{T}^{1|2}(t))=\log d_{\mathcal{O}}+\nn\\
&\frac{1}{1-k}\log\left(\frac{\sum^M\limits_{i_1,i_3,...,i_{2k-1}=1}\sum\limits_{j_2,j_4,...j_{2k}=1}^{M'}\prod\limits_{u=1}^{k}(-1)^{|K_{i_{2u-1}}|+|K'_{j_{2u}}|}C_{i_{2u-1}}C_{j_{2u}}^{'*} c_0(\{K_{i_{2u-1}}\},\{K'_{j_{2u+2}}\})\bar c_0(\{\bar K_{i_{2u-1}}\},\{\bar K'_{j_{2u}}\})}{\left(\sum_{i,j}(-1)^{|K_i|+|K'_j|}C_iC^{'*}_jc_0\big([K_i],[K'_j]\big)\bar{c}_0\big([\bar K_i],[\bar K'_j]\big)\right)^k}\right),\nn\\
&(\text{for}~x_1= x_2;~2k+2\equiv2).\label{c222}
\]
For convenience, we also introduce an effective transition  matrix $\mathcal{T}_{\text{eff}}$
\[
X_{ij}:=\sqrt{C_{i}C^{'*}_j}c_0(\{K_i\},\{K'_j\}),\quad\bar X_{ij}:=\sqrt{C_{i}C^{'*}_j}\bar c_0(\{\bar K_i\},\{\bar K'_j\}),\nn\\
\mathcal{T}_{\text{eff}}:=\frac{X\bar X^{\mathrm{T}}}{\tr[X\bar X^{\mathrm{T}}]},
\quad(i=1,2,...,M;~j=1,2,...,M')
\]
to simplify the formula \eqref{c111} (Eq.\eqref{c222} shares a similar treatment).
Using the formula \eqref{kthpee1}, we find that the pseudo-R\'enyi entropy of linear combination operators is generally different from that of the primary operator $\mo$. The pseudo-R\'enyi entropies are the same as the ones of the primary when the correlation function of $V_\alpha$ and $V_\beta$ can be divided into the product of the holomorphic part and the anti-holomorphic part. A typical example is 
\[
V_\alpha(w_1,\bar{w}_1)=\sum\limits_i C_i L_{-\{K_i\}}\bar{L}_{-\{\bar{K}_1\}}\mo(w_1,\bar{w}_1), \quad V_\beta(w_2,\bar{w}_2)=\sum\limits_j C'_j L_{-\{K'_j\}}\bar{L}_{-\{\bar{K}'_2\}}\mo(w_2,\bar{w}_2).
\]
Otherwise, there is an extra contribution due to the mixing of the holomorphic and antiholomorphic Virasoro generators. A typical example of extra contribution is
\[
V_\alpha(w_1,\bar{w}_1)=(L_{-1}+\bar{L}_{-1})\mo(w_1,\bar{w}_1), \quad V_\beta(w_2,\bar{w}_2)=(L_{-1}+\bar{L}_{-1})\mo(w_2,\bar{w}_2).
\]
In general, the $k$-th pseudo-R\'enyi entropy for two linear combination operators at the late time only depends on the quantum dimension and the contribution from a finite-dimensional Hilbert space,
\[
\lim_{t\to\infty}\Delta S^{(k)}(\mathcal{T}^{1|2}(t))=\log d_{\mo}+\frac{1}{1-k}\log\tr [\left(\mathcal{T}_{\text{eff}}\right)^k].
\]
Noticing the current results in RCFTs, one can directly calculate the pseudo-entropy of generic local operators in Liouville CFT, holographic CFTs, non-diagonal CFTs, etc. Since the spectra in such theories have different structures, the associated pseudo-entropy will be highly different from those in RCFTs. In particular, since holomorphic and antiholomorphic conformal blocks have different structures in non-diagonal CFTs, the late-time behavior of the entanglement entropy and pseudo-entropy associated with locally excited states will not be the same as the ones demonstrated in the current paper. We would like to leave them to future work.

\begin{acknowledgments}
We thank Wu-zhong Guo, Linlin Huang, Pak Hang Chris Lau, Yang Liu, Yuan Sun, and Long Zhao for the valuable discussions. S.H. would appreciate the financial support from Jilin University, Max Planck Partner group, and Natural Science Foundation of China Grants (No.12075101, No.12047569).
\end{acknowledgments}

\appendix

\section{Reduction of $\la\mo^{(-n)}(1)\mo^{\dagger}(2)\mo^{(-n)}(3)\mo^{\dagger}(4)\ra_{\Sigma_1}$}\label{sec:app1}
Following the standard way \cite{DiFrancesco:1997nk}, we can compute the four-point function $\la\mo^{(-n)}(1)\mo^{\dagger}(2)\mo^{(-n)}(3)\mo^{\dagger}(4)\ra_{\Sigma_1}$ in this section.
\[
&\la\mo^{(-n)}(1)\mo^\dagger(2)\mo^{(-n)}(3)\mo^\dagger(4)\ra_{\Sigma_1}\nn\\
=&-\frac{1}{2\pi i}\sum_{i=2}^{4}\oint_{\mathcal{C}(z_i)}\dd z(z-z_1)^{-n+1}\la T(z)\mo(1)\mo^{\dagger}(2)\mo^{(-n)}(3)\mo^{\dagger}(4)\ra_{\Sigma_1}\nn\\
=&\frac{-1}{2\pi i}\oint_{\mathcal{C}(z_2)}\frac{\dd z}{(z-z_1)^{n-1}}\bigg\{\frac{\D\la\mo(1)\mo^\dagger(2)\mo^{(-n)}(3)\mo^\dagger(4)\ra_{\Sigma_1}}{(z-z_2)^2}+\frac{\pd_{z_2}\la\mo(1)\mo^\dagger(2)\mo^{(-n)}(3)\mo^\dagger(4)\ra_{\Sigma_1}}{z-z_2}+\text{reg}(z-z_2)\bigg\}\nn\\
&+\frac{-1}{2\pi i}\oint_{\mathcal{C}(z_3)}\frac{\dd z}{(z-z_1)^{n-1}}\bigg\{\frac{n(n^2-1)c/12+2n\D}{(z-z_3)^{n+2}}\la\mo(1)\mo^\dagger(2)\mo(3)\mo^\dagger(4)\ra_{\Sigma_1}\nn\\
&~+\sum_{k=1}^{n-1}\frac{(n+k)\la\mo(1)\mo^\dagger(2)\mo^{(-(n-k))}(3)\mo^\dagger(4)\ra_{\Sigma_1}}{(z-z_3)^{k+2}}+\frac{(\D+n)\la\mo(1)\mo^\dagger(2)\mo^{(-n)}(3)\mo^\dagger(4)\ra_{\Sigma_1}}{(z-z_3)^2}\nn\\
&~+\frac{\pd_{z_3}\la\mo(1)\mo^\dagger(2)\mo^{(-n)}(3)\mo^\dagger(4)\ra_{\Sigma_1}}{z-z_3}+\text{reg}(z-z_3)\bigg\}\nn\\
&+\frac{-1}{2\pi i}\oint_{\mathcal{C}(z_4)}\frac{\dd z}{(z-z_1)^{n-1}}\bigg\{\frac{\D\la\mo(1)\mo^\dagger(2)\mo^{(-n)}(3)\mo^\dagger(4)\ra_{\Sigma_1}}{(z-z_4)^2}+\frac{\pd_{z_4}\la\mo(1)\mo^\dagger(2)\mo^{(-n)}(3)\mo^\dagger(4)\ra_{\Sigma_1}}{z-z_4}+\text{reg}(z-z_4)\bigg\}\nn\\
=&\frac{(n-1)\D}{z_{21}^n}\la\mo(1)\mo^\dagger(2)\mo^{(-n)}(3)\mo^\dagger(4)\ra_{\Sigma_1}+\frac{-\pd_{z_2}}{z_{21}^{n-1}}\la\mo(1)\mo^\dagger(2)\mo^{(-n)}(3)\mo^\dagger(4)\ra_{\Sigma_1}\nn\\
&+\frac{(n-1)\D}{z_{41}^n}\la\mo(1)\mo^\dagger(2)\mo^{(-n)}(3)\mo^\dagger(4)\ra_{\Sigma_1}+\frac{-\pd_{z_4}}{z_{41}^{n-1}}\la\mo(1)\mo^\dagger(2)\mo^{(-n)}(3)\mo^\dagger(4)\ra_{\Sigma_1}\nn\\
&+(-1)^n\frac{\big(n(n^2-1)c/12+2n\D\big)(2n-1)!}{(n+1)!(n-2)!}\frac{\la\mo(1)\mo^\dagger(2)\mo(3)\mo^\dagger(4)\ra_{\Sigma_1}}{z_{13}^{2n}}\nn\\
&+(-1)^n\sum_{k=1}^{n-1}\frac{(n+k)!}{(k+1)!(n-2)!}\frac{\la\mo(1)\mo^\dagger(2)\mo^{(-(n-k))}(3)\mo^\dagger(4)\ra_{\Sigma_1}}{z_{13}^{n+k}}\nn\\
&+\frac{(n-1)(\D+n)}{z_{31}^n}\la\mo(1)\mo^\dagger(2)\mo^{(-n)}(3)\mo^\dagger(4)\ra_{\Sigma_1}+\frac{-\pd_{z_3}}{z_{31}^{n-1}}\la\mo(1)\mo^\dagger(2)\mo^{(-n)}(3)\mo^\dagger(4)\ra_{\Sigma_1}\label{app-eq1}
\]
\section{Reduction of$\la\mo^{(-n)}(1)\mo^{(-m)\dagger}(2)\mo^{(-n)}(3)\mo^{(-m)\dagger}(4)\ra_{\Sigma_2}$}\label{sec:app2}
In terms of $\eqref{equ9}$, the most divergent term of $\la\mo^{(-n)}(1)\mo^{(-m)\dagger}(2)\mo^{(-n)}(3)\mo^{(-m)\dagger}(4)\ra_{\Sigma_1}$ should only contain $z_{14}$ and $z_{23}$, as any terms containing $z_{13}$,$z_{24}$,$z_{12}$ and $z_{34}$ are subleading. So we can firstly expand $\mo(1)$'s Virasoro generator,
\[
&\la\mo^{(-n)}(1)\mo^{(-m)\dagger}(2)\mo^{(-n)}(3)\mo^{(-m)\dagger}(4)\ra_{\Sigma_1}\nn\\
\sim&-\frac{1}{2\pi i}\oint_{\mathcal{C}(z_4)}\frac{\dd z}{(z-z_1)^{n-1}}\la T(z)\mo(1)\mo^{(-m)\dagger}(2)\mo^{(-n)}(3)\mo^{(-m)\dagger}(4)\ra_{\Sigma_1}\nn\\
\sim& -\frac{1}{2\pi i}\oint_{\mathcal{C}(z_4)}\frac{\dd z}{(z-z_1)^{n-1}}\la \mo(1)\mo^{(-m)\dagger}(2)\mo^{(-n)}(3)(\frac{m(m^{2}-1)c/12+2m\Delta}{(z-z_{4})^{m+2}}\mo^{\dagger}(4)\nn\\ &+\sum\limits_{k=1}^{m-1}\frac{(m+k)}{(z-z_4)^{k+2}}\mo^{-(m-k)\dagger}(4)+\frac{(\Delta+m)}{(z-z_4)^2}\mo^{(-m)\dagger}(4)+\frac{\partial_4\mo^{(-m)\dagger}(4)}{z-z_4})\ra_{\Sigma_1}\nn\\
\sim&(-1)^m\frac{(n+m-1)!}{(m+1)!(n-2)!}\frac{m(m^2-1)c/12+2m\Delta}{z_{41}^{n+m}}\la\mo(1)\mo^{(-m)\dagger}(2)\mo^{(-n)}(3)\mo^{\dagger}(4)\ra_{\Sigma_1}\nn\\
&+(-1)^n\sum\limits_{k=1}^{m-1}\frac{(n+k-1)!}{(k+1)!(n-2)!}\frac{(m+k)}{z_{14}^{n+k}}\la\mo(1)\mo^{(-m)\dagger}(2)\mo^{(-n)}(3)\mo^{(-(m-k))\dagger}(4)\ra_{\Sigma_1}\nn\\
&+\frac{(n-1)(\Delta+m)}{z_{41}^n}\la\mo(1)\mo^{(-m)\dagger}(2)\mo^{(-n)}(3)\mo^{(-m)\dagger}(4)\ra_{\Sigma_1}\nn\\
&-\frac{\partial_4}{z_{41}^{n-1}}\la\mo(1)\mo^{(-m)\dagger}(2)\mo^{(-n)}(3)\mo^{(-m)\dagger}(4)\ra_{\Sigma_1}\label{app-eq2}.
\]
The correlation function with four Virasoro generators is deformed into correlation functions containing no more than 3 Virasoro generators. We can then expand $\mo(4)$'s Virasoro generator,
\[
&\la\mo(1)\mo^{(-m)\dagger}(2)\mo^{(-n)}(3)\mo^{(-m)\dagger}(4)\ra_{\Sigma_1}\nn\\
\sim&-\frac{1}{2\pi i}\oint_{c(z_1)}\frac{\dd z}{(z-z_4)^{m-1}}\la T(z)\mo(1)\mo^{(-m)\dagger}(2)\mo^{(-n)}(3)\mo^{\dagger}(4)\ra_{\Sigma_1}\nn\\
\sim&-\oint_{c(z_1)}\frac{\dd z}{(z-z_4)^{m-1}}\la(\frac{\Delta}{(z-z_1)^2}\mo(1)+\frac{\partial_1 O(1)}{Z-Z_1})\mo^{(-m)\dagger}(2)\mo^{(-n)}(3)\mo^{\dagger}(4)\ra_{\Sigma_1}\nn\\
\sim&\frac{(m-1)\Delta}{z_{14}^{m}}\la(\mo(1)\mo^{(-m)\dagger}(2)\mo^{(-n)}(3)\mo^{\dagger}(4)\ra_{\Sigma_1}-\frac{\partial_1}{z_{14}^{m-1}}\la(\mo(1)\mo^{(-m)\dagger}(2)\mo^{(-n)}(3)\mo^{\dagger}(4)\ra_{\Sigma_1}\label{app-eq3}.
\]
Form $\eqref{app-eq2}$ and $\eqref{app-eq3}$, we can read the exact form of $\mathcal{D}_{1,4}$ introduced in \eqref{equ37}
\[
\mathcal{D}_{1,4}=&(-1)^m\frac{(n+m-1)!}{(m+1)!(n-2)!}\frac{m(m^2-1)c/12+2m\Delta}{z_{41}^{n+m}}\nn\\
&+(-1)^n\sum\limits_{k=1}^{m-1}\frac{(n+k-1)!}{(k+1)!(n-2)!}\frac{(m+k)}{z_{14}^{n+k}}(\frac{(m-k-1)\Delta}{z_{14}^{m-k}}-\frac{\partial_1}{z_{14}^{m-k-1}})\nn\\
&+\frac{(n-1)(\Delta+m)}{z_{41}^n}(\frac{(m-1)\Delta}{z_{14}^{m}}-\frac{\partial_1}{z_{14}^{m-1}})\nn\\
&-\frac{\partial_4}{z_{41}^{n-1}}(\frac{(m-1)\Delta}{z_{14}^{m}}-\frac{\partial_1}{z_{14}^{m-1}}).
\]
We can expand $\mo(2)$'s Virasoro generator and $\mo(3)$'s Virasoro generator in a similar way,
\[
&\la\mo(1)\mo^{(-m)\dagger}(2)\mo^{(-n)}(3)\mo^{\dagger}(4)\ra_{\Sigma_1}\nn\\
\sim &-\oint_{c(z_3)}\frac{\dd z}{(z-z_2)^{m-1}}\la\mo(1)T(z)\mo^{\dagger}(2)\mo^{(-n)}(3)\mo^{\dagger}(4)\ra_{\Sigma_1}\nn\\
\sim &-\oint_{c(z_3)}\frac{\dd z}{(z-z_2)^{m-1}}\la\mo(1)\mo^{\dagger}(2)(\frac{n(n^2-1)c/12+2n\Delta}{(z-z_3)^{n+2}}\mo(3)\nn\\
&+\sum\limits_{l=1}^{n-1}\frac{(n+l)}{(z-z_3)^{l+2}}\mo^{-(n-l)}(3)+\frac{(\Delta+n)}{(z-z_3)^2}\mo^{-n}(3)+\frac{\partial_3}{z-z_3}\mo^{(-n)}(3))\mo^{\dagger}(4)\ra_{\Sigma_1}\nn\\
\sim&(-1)^n\frac{(n+m-1)!}{(n+1)!(m-2)!}\frac{n(n^2-1)c/12+2n\Delta}{z_{32}^{m+n}}\la\mo(1)\mo^{\dagger}(2)\mo(3)\mo^{\dagger}(4)\ra_{\Sigma_1}\nn\\ &+(-1)^m\sum\limits_{l=1}^{n-1}\frac{(m+l-1)!}{(l+1)!(m-2)!}\frac{(n+l)}{z_{23}^{m+l}}\la\mo(1)\mo^{\dagger}(2)\mo^{(-(n-l)}(3)\mo^{\dagger}(4)\ra_{\Sigma_1}\nn\\
&+\frac{(m-1)(\Delta+n)}{z_{32}^m}\la\mo(1)\mo^{\dagger}(2)\mo^{(-n)}(3)\mo^{\dagger}(4)\ra_{\Sigma_1}-\frac{\partial_3}{z_{32}^{m-1}}\la\mo(1)\mo^{\dagger}(2)\mo^{(-n)}(3)\mo^{\dagger}(4)\ra_{\Sigma_1}\label{app-eq4}.
\]
Finally, we can have
\[
&\la\mo(1)\mo^{\dagger}(2)\mo^{(-n)}(3)\mo^{\dagger}(4)\ra_{\Sigma_1}\sim \frac{(n-1)\Delta}{z_{23}^n}G-\frac{\partial_2}{z_{23}^{n-1}}G\label{app-eq5},
\]
where $G$ is $\la\mo(1)\mo^{\dagger}(2)\mo(3)\mo^{\dagger}(4)\ra_{\Sigma_1}$,\ and in late-time limit, it's\ $d_{\mo}^{-1}(1-\eta)^{-2\Delta}\bar{\eta}^{-2\Delta}$.

Combining $\eqref{app-eq2}$,$\eqref{app-eq3}$,$\eqref{app-eq4}$ and $\eqref{app-eq5}$, we have
\[
&\la\mo^{(-n)}(1)\mo^{(-m)\dagger}(2)\mo^{(-n)}(3)\mo^{(-m)\dagger}(4)\ra_{\Sigma_1}\nn\\
    	\sim&(-1)^m\frac{(n+m-1)!}{(m+1)!(n-2)!}\frac{m(m^2-1)c/12+2m\Delta}{z_{41}^{n+m}}\{(-1)^n\frac{(n+m-1)!}{(n+1)!(m-2)!}\frac{n(n^2-1)c/12+2n\Delta}{z_{32}^{m+n}}G\nn\\
    	&+(-1)^m\sum\limits_{l=1}^{n-1}\frac{(m+l-1)!}{(l+1)!(m-2)!}\frac{(n+l)}{z_{23}^{m+l}}(\frac{(n-l-1)\Delta}{z_{23}^{n-l}}G-\frac{\partial_2}{z_{23}^{n-l-1}}G)\nn\\ &+\frac{(m-1)(\Delta+n)}{z_{32}^m}(\frac{(n-1)\Delta}{z_{23}^n}G-\frac{\partial_2}{z_{23}^{n-1}}G)-\frac{1}{z_{32}^{m-1}}(\frac{n(n-1)\Delta}{z_{23}^{n+1}}G+\frac{(n-1)\Delta\partial_3}{z_{23}^n}G-\frac{(n-1)\partial_2}{z_{23}^n}G-\frac{\partial_3\partial_2}{z_{23}^{n-1}}G)\} \nn\\
    	&+(-1)^n\sum\limits_{k=1}^{m-1}\frac{(n+k-1)!}{(k+1)!(n-2)!}\frac{(m+k)}{z_{14}^{n+k}}\{\frac{(m-k-1)\Delta}{z_{14}^{m-k}}[
    	(-1)^n\frac{(n+m-1)!}{(n+1)!(m-2)!}\frac{n(n^2-1)c/12+2n\Delta}{z_{32}^{m+n}}G\nn\\
    	&+(-1)^m\sum\limits_{l=1}^{n-1}\frac{(m+l-1)!}{(l+1)!(m-2)!}\frac{(n+l)}{z_{23}^{m+l}}(\frac{(n-l-1)\Delta}{z_{23}^{n-l}}G-\frac{\partial_2}{z_{23}^{n-l-1}}G)\nn\\ &+\frac{(m-1)(\Delta+n)}{z_{32}^m}(\frac{(n-1)\Delta}{z_{23}^n}G-\frac{\partial_2}{z_{23}^{n-1}}G)-\frac{1}{z_{32}^{m-1}}(\frac{n(n-1)\Delta}{z_{23}^{n+1}}G+\frac{(n-1)\Delta\partial_3}{z_{23}^n}G-\frac{(n-1)\partial_2}{z_{23}^n}G-\frac{\partial_3\partial_2}{z_{23}^{n-1}}G)
    	]\nn\\
    	&-\frac{1}{z_{14}^{m-k-1}}[
    	(-1)^n\frac{(n+m-1)!}{(n+1)!(m-2)!}\frac{n(n^2-1)c/12+2n\Delta}{z_{32}^{m+n}}\partial_1G+\nn\\
    	&(-1)^m\sum\limits_{l=1}^{n-1}\frac{(m+l-1)!}{(l+1)!(m-2)!}\frac{(n+l)}{z_{23}^{m+l}}(\frac{(n-l-1)\Delta}{z_{23}^{n-l}}\partial_1G-\frac{\partial_1\partial_2}{z_{23}^{n-l-1}}G)+\frac{(m-1)(\Delta+n)}{z_{32}^m}(\frac{(n-1)\Delta}{z_{23}^n}\partial_1G-\frac{\partial_1\partial_2}{z_{23}^{n-1}}G)\nn\\ &-\frac{1}{z_{32}^{m-1}}(\frac{n(n-1)\Delta}{z_{23}^{n+1}}\partial_1G+\frac{(n-1)\Delta\partial_1\partial_3}{z_{23}^n}G-\frac{(n-1)\partial_1\partial_2}{z_{23}^n}G-\frac{\partial_1\partial_3\partial_2}{z_{23}^{n-1}}G)
    	]\} \nn\\
    	&+\frac{(n-1)(\Delta+m)}{z_{41}^n}\{ \frac{(m-1)\Delta}{z_{14}^m}[(-1)^n\frac{(n+m-1)!}{(n+1)!(m-2)!}\frac{n(n^2-1)c/12+2n\Delta}{z_{32}^{m+n}}G\nn\\
    	&+(-1)^m\sum\limits_{l=1}^{n-1}\frac{(m+l-1)!}{(l+1)!(m-2)!}\frac{(n+l)}{z_{23}^{m+l}}(\frac{(n-l-1)\Delta}{z_{23}^{n-l}}G-\frac{\partial_2}{z_{23}^{n-l-1}}G)\nn\\ &+\frac{(m-1)(\Delta+n)}{z_{32}^m}(\frac{(n-1)\Delta}{z_{23}^n}G-\frac{\partial_2}{z_{23}^{n-1}}G)-\frac{1}{z_{32}^{m-1}}(\frac{n(n-1)\Delta}{z_{23}^{n+1}}G+\frac{(n-1)\Delta\partial_3}{z_{23}^n}G-\frac{(n-1)\partial_2}{z_{23}^n}G-\frac{\partial_3\partial_2}{z_{23}^{n-1}}G)]\nn\\
    	&-\frac{1}{z_{14}^{m-1}}[(-1)^n\frac{(n+m-1)!}{(n+1)!(m-2)!}\frac{n(n^2-1)c/12+2n\Delta}{z_{32}^{m+n}}\partial_1G\nn\\
    	&+(-1)^m\sum\limits_{l=1}^{n-1}\frac{(m+l-1)!}{(l+1)!(m-2)!}\frac{(n+l)}{z_{23}^{m+l}}(\frac{(n-l-1)\Delta}{z_{23}^{n-l}}\partial_1G-\frac{\partial_1\partial_2}{z_{23}^{n-l-1}}G)+\frac{(m-1)(\Delta+n)}{z_{32}^m}(\frac{(n-1)\Delta}{z_{23}^n}\partial_1G-\frac{\partial_1\partial_2}{z_{23}^{n-1}}G)\nn\\ &-\frac{1}{z_{32}^{m-1}}(\frac{n(n-1)\Delta}{z_{23}^{n+1}}\partial_1G+\frac{(n-1)\Delta\partial_1\partial_3}{z_{23}^n}G-\frac{(n-1)\partial_1\partial_2}{z_{23}^n}G-\frac{\partial_1\partial_3\partial_2}{z_{23}^{n-1}}G)]\}\nn\\
    	&-\frac{1}{z_{41}^{n-1}}\{\frac{m(m-1)\Delta}{z_{14}^{m+1}}[(-1)^n\frac{(n+m-1)!}{(n+1)!(m-2)!}\frac{n(n^2-1)c/12+2n\Delta}{z_{32}^{m+n}}G\nn\\
    	&+(-1)^m\sum\limits_{l=1}^{n-1}\frac{(m+l-1)!}{(l+1)!(m-2)!}\frac{(n+l)}{z_{23}^{m+l}}(\frac{(n-l-1)\Delta}{z_{23}^{n-l}}G-\frac{\partial_2}{z_{23}^{n-l-1}}G)+\frac{(m-1)(\Delta+n)}{z_{32}^m}(\frac{(n-1)\Delta}{z_{23}^n}G-\frac{\partial_2}{z_{23}^{n-1}}G)\nn\\
    	&-\frac{1}{z_{32}^{m-1}}(\frac{n(n-1)\Delta}{z_{23}^{n+1}}G+\frac{(n-1)\Delta\partial_3}{z_{23}^n}G-\frac{(n-1)\partial_2}{z_{23}^n}G-\frac{\partial_3\partial_2}{z_{23}^{n-1}}G)]\nn
    	\]
    	\[
    	&+\frac{(m-1)\Delta}{z_{14}^m}[(-1)^n\frac{(n+m-1)!}{(n+1)!(m-2)!}\frac{n(n^2-1)c/12+2n\Delta}{z_{32}^{m+n}}\partial_4G\nn\\
    	&+(-1)^m\sum\limits_{l=1}^{n-1}\frac{(m+l-1)!}{(l+1)!(m-2)!}\frac{(n+l)}{z_{23}^{m+l}}(\frac{(n-l-1)\Delta}{z_{23}^{n-l}}\partial_4G-\frac{\partial_4\partial_2}{z_{23}^{n-l-1}}G)\nn\\ &+\frac{(m-1)(\Delta+n)}{z_{32}^m}(\frac{(n-1)\Delta}{z_{23}^n}\partial_4G-\frac{\partial_4\partial_2}{z_{23}^{n-1}}G)\nn\\
    	&-\frac{1}{z_{32}^{m-1}}(\frac{n(n-1)\Delta}{z_{23}^{n+1}}\partial_4G+\frac{(n-1)\Delta\partial_4\partial_3}{z_{23}^n}G-\frac{(n-1)\partial_4\partial_2}{z_{23}^n}G-\frac{\partial_4\partial_3\partial_2}{z_{23}^{n-1}}G)]\nn\\
    	&-\frac{(m-1)}{z_{14}^m}[(-1)^n\frac{(n+m-1)!}{(n+1)!(m-2)!}\frac{n(n^2-1)c/12+2n\Delta}{z_{32}^{m+n}}\partial_1G\nn\\
    	&+(-1)^m\sum\limits_{l=1}^{n-1}\frac{(m+l-1)!}{(l+1)!(m-2)!}\frac{(n+l)}{z_{23}^{m+l}}(\frac{(n-l-1)\Delta}{z_{23}^{n-l}}\partial_1G-\frac{\partial_1\partial_2}{z_{23}^{n-l-1}}G)+\frac{(m-1)(\Delta+n)}{z_{32}^m}(\frac{(n-1)\Delta}{z_{23}^n}\partial_1G-\frac{\partial_1\partial_2}{z_{23}^{n-1}}G)\nn\\ &-\frac{1}{z_{32}^{m-1}}(\frac{n(n-1)\Delta}{z_{23}^{n+1}}\partial_1G+\frac{(n-1)\Delta\partial_1\partial_3}{z_{23}^n}G-\frac{(n-1)\partial_1\partial_2}{z_{23}^n}G-\frac{\partial_1\partial_3\partial_2}{z_{23}^{n-1}}G)]\nn\\
    	&-\frac{1}{z_{14}^{m-1}}[(-1)^n\frac{(n+m-1)!}{(n+1)!(m-2)!}\frac{n(n^2-1)c/12+2n\Delta}{z_{32}^{m+n}}\partial_4\partial_1G\nn\\
    	&+(-1)^m\sum\limits_{l=1}^{n-1}\frac{(m+l-1)!}{(l+1)!(m-2)!}\frac{(n+l)}{z_{23}^{m+l}}(\frac{(n-l-1)\Delta}{z_{23}^{n-l}}\partial_4\partial_1G-\frac{\partial_4\partial_1\partial_2}{z_{23}^{n-l-1}}G)\nn\\ &+\frac{(m-1)(\Delta+n)}{z_{32}^m}(\frac{(n-1)\Delta}{z_{23}^n}\partial_4\partial_1G-\frac{\partial_4\partial_1\partial_2}{z_{23}^{n-1}}G)\nn\\
    	&-\frac{1}{z_{32}^{m-1}}(\frac{n(n-1)\Delta}{z_{23}^{n+1}}\partial_4\partial_1G+\frac{(n-1)\Delta\partial_4\partial_1\partial_3}{z_{23}^n}G-\frac{(n-1)\partial_4\partial_1\partial_2}{z_{23}^n}G-\frac{\partial_4\partial_1\partial_3\partial_2}{z_{23}^{n-1}}G)]\}.
    	\label{app-eq6}   	
\]
The correlation function of four descendant operators becomes the correlation functions of their corresponding primary operators with some constants and derivatives.\\
For $i\neq j\neq k \neq l$, we can have
\[
&\partial_i G=\frac{2\Delta\partial_i \eta}{1-\eta}G,\nn\\
&\partial_j \partial_i G=\frac{2\Delta\partial_j \partial_i \eta}{1-\eta}G+\frac{2\Delta(2\Delta+1)\partial_j \eta \partial_i \eta}{(1-\eta)^2}G,\nn\\
&\partial_k \partial_j \partial_i G=\frac{2\Delta \partial_k \partial_j\partial_i \eta}{1-\eta}G+2\Delta(2\Delta+1)[\frac{\partial_j\partial_i\eta\partial_k\eta+\partial_j\partial_k\eta\partial_i\eta+\partial_k\partial_i\eta\partial_j\eta}{(1-\eta)^2}G+\frac{(2\Delta+2)\partial_j\eta\partial_i\eta\partial_k\eta}{(1-\eta)^3}G]\nn\\
&\sim2\Delta(2\Delta+1)[\frac{\partial_j\partial_i\eta\partial_k\eta+\partial_j\partial_k\eta\partial_i\eta+\partial_k\partial_i\eta\partial_j\eta}{(1-\eta)^2}G+\frac{(2\Delta+2)\partial_j\eta\partial_i\eta\partial_k\eta}{(1-\eta)^3}G],\nn\\
&\partial_l\partial_k\partial_j\partial_i G=\frac{2\Delta\partial_l\partial_k\partial_j\partial_i\eta}{1-\eta}G+\frac{(2\Delta(2\Delta+1))}{(1-\eta)^2}(\partial_k\partial_j\partial_i\eta\partial_l\eta+\partial_k\partial_j\partial_l\eta\partial_i\eta+\partial_l\partial_j\partial_i\eta\partial_k\eta+\partial_k\partial_l\partial_i\eta\partial_j\eta\nn\\
&+\partial_i\partial_j\eta\partial_l\partial_k\eta+\partial_i\partial_k\eta\partial_j\partial_l\eta+\partial_i\partial_l\eta\partial_k\partial_j\eta)G+\frac{2\Delta(2\Delta+1)(2\Delta+2)}{(1-\eta)^3}(\partial_j\partial_i\eta\partial_k\eta\partial_l\eta+\partial_j\partial_k\eta\partial_i\eta\partial_l\eta\nn\\
&+\partial_k\partial_i\eta\partial_j\eta\partial_l\eta+\partial_l\partial_i\eta\partial_k\eta\partial_j\eta+\partial_j\partial_l\eta\partial_k\eta\partial_i\eta+\partial_l\partial_k\eta\partial_i\eta\partial_j\eta)G\nn\\
&+\frac{2\Delta(2\Delta+1)(2\Delta+1)(2\Delta+3)}{(1-\eta)^4}\partial_i\eta\partial_j\eta\partial_k\eta\partial_l\eta G\nn\\
&\sim \frac{(2\Delta(2\Delta+1))}{(1-\eta)^2}(\partial_i\partial_j\eta\partial_l\partial_k\eta+\partial_i\partial_k\eta\partial_j\partial_l\eta+\partial_i\partial_l\eta\partial_k\partial_j\eta)G+\frac{2\Delta(2\Delta+1)(2\Delta+1)}{(1-\eta)^3}(\partial_j\partial_i\eta\partial_k\eta\partial_l\eta+\partial_j\partial_k\eta\partial_i\eta\partial_l\eta\nn\\
&+\partial_k\partial_i\eta\partial_j\eta\partial_l\eta+\partial_l\partial_i\eta\partial_k\eta\partial_j\eta+\partial_j\partial_l\eta\partial_k\eta\partial_i\eta+\partial_l\partial_k\eta\partial_i\eta\partial_j\eta)G+\frac{2\Delta(2\Delta+1)(2\Delta+2)(2\Delta+3)}{(1-\eta)^4}\partial_i\eta\partial_j\eta\partial_k\eta\partial_l\eta G.
\label{app-eq7}
\]
{From \eqref{app-eq6}, \eqref{app-eq7}, \eqref{eq9}, \eqref{equ10} and \eqref{equ32} we derive the leading behavior of $\la\mo^{(-n)}(1)\mo^{(-m)\dagger}(2)\mo^{(-n)}(3)\mo^{(-m)\dagger}(4)\ra_{\Sigma_2}$ at the late time limit,
\[
&\la\mo^{(-n)}(1)\mo^{(-m)\dagger}(2)\mo^{(-n)}(3)\mo^{(-m)\dagger}(4)\ra_{\Sigma_2}\nn\\
\sim&\big(\prod_{i=1}^{4}|w_i'|^{-2\D}\big)(w_1-w_2)^{-m-n}(w_3-w_4)^{-m-n}G(\eta,\bar{\eta})(-1)^{m+n}\nn\\
&\Big(-\frac{1}{12 \Gamma (m+2)^2 \Gamma (n+2)^2}\Delta   (-1)^{-m} ((m (m+1) n-2) \Gamma (m+2) \Gamma (n+2)\nn\\
&-2 (m+1) (m^2 n+m (2 n^2-1)-n (n+1)) \Gamma (m+n))((-1)^{-m-n} \Gamma (m+n) \nn\\
&(m (m^2-1)  n (-1)^{m} (c (n^2-1)+24 \Delta )+24 \Delta  (-1)^m (n+1)  (m (2m n-m+n^2-1)-n))\nn\\
&+12 \Delta  (-1)^{-n} \Gamma (m+2) \Gamma (n+2) ( (n+1) (\Delta +m (\Delta +n))+ (2-m n (n+1))))\nn\\
&+\frac{1}{12} \Delta  (m+1) (\Delta +m) (-1)^{-2 m-n} (\frac{1}{n (n+1) \Gamma (m+2) \Gamma (n-1)}\Gamma (m+n) \nn\\
&(m (m^2-1) (-1)^n n (-1)^{m+n} (c (n^2-1)+24 \Delta )+24 \Delta  (-1)^m (n+1) (m (m (2 n-1)+n^2-1)-n))\nn\\
&+12 \Delta   (n-1) ((-1)^m (n+1) (\Delta +m (\Delta +n))+(-1)^{m+1} (m n (n+1)-2)))\nn\\
&\frac{1}{12} \Delta  (2 \Delta +m)  (-1)^{-2 m-n} (\frac{1}{m \Gamma (m) \Gamma (n+2)}\Gamma (m+n) \nn\\
&(m (m^2-1) (-1)^n n (-1)^{m+n} (c (n^2-1)+24 \Delta )+24 \Delta  (-1)^m (n+1)  (m (m (2 n-1)+n^2-1)-n))\nn\\
&+12 \Delta  (m+1)  ((-1)^m (n+1) (\Delta +m (\Delta +n))+(-1)^{m+1}  (m n (n+1)-2)))\nn\\
&+\frac{1}{144 (m+1) \Gamma (m-1) \Gamma (m) \Gamma (m+2)^2 \Gamma (n-1) \Gamma (n+2)}\nn\\
&((-1)^{m+1} m  (-1)^{-m-n} (c (m^2-1)+24 \Delta ) \Gamma (m+n) ((m+1) (-1)^{n+1} n (-1)^{m+n} \Gamma (m) \Gamma (m+2) \nn\\
&(c (n^2-1)+24 \Delta ) \Gamma (m+n)+12 \Delta   \Gamma (m-1) ((m+1) \Gamma (m) ((-1)^m (-n-1) \Gamma (m+2) \Gamma (n+2) \nn\\
&(\Delta +m (\Delta +n))+(-1)^m  ((m n (n+1)-2) \Gamma (m+2) \Gamma (n+2)+2 n (n+1) \Gamma (m+n)))\nn\\
&-2 (-1)^m  (n+1) (2 m n-m+n^2-1) \Gamma (m+2) \Gamma (m+n))))
\Big)+\dots\label{equ45}
\]}

\nocite{*}
\bibliography{DescendantRef}

\begin{thebibliography}{93}%
\makeatletter
\providecommand \@ifxundefined [1]{%
 \@ifx{#1\undefined}
}%
\providecommand \@ifnum [1]{%
 \ifnum #1\expandafter \@firstoftwo
 \else \expandafter \@secondoftwo
 \fi
}%
\providecommand \@ifx [1]{%
 \ifx #1\expandafter \@firstoftwo
 \else \expandafter \@secondoftwo
 \fi
}%
\providecommand \natexlab [1]{#1}%
\providecommand \enquote  [1]{``#1''}%
\providecommand \bibnamefont  [1]{#1}%
\providecommand \bibfnamefont [1]{#1}%
\providecommand \citenamefont [1]{#1}%
\providecommand \href@noop [0]{\@secondoftwo}%
\providecommand \href [0]{\begingroup \@sanitize@url \@href}%
\providecommand \@href[1]{\@@startlink{#1}\@@href}%
\providecommand \@@href[1]{\endgroup#1\@@endlink}%
\providecommand \@sanitize@url [0]{\catcode `\\12\catcode `\$12\catcode
  `\&12\catcode `\#12\catcode `\^12\catcode `\_12\catcode `\%12\relax}%
\providecommand \@@startlink[1]{}%
\providecommand \@@endlink[0]{}%
\providecommand \url  [0]{\begingroup\@sanitize@url \@url }%
\providecommand \@url [1]{\endgroup\@href {#1}{\urlprefix }}%
\providecommand \urlprefix  [0]{URL }%
\providecommand \Eprint [0]{\href }%
\providecommand \doibase [0]{https://doi.org/}%
\providecommand \selectlanguage [0]{\@gobble}%
\providecommand \bibinfo  [0]{\@secondoftwo}%
\providecommand \bibfield  [0]{\@secondoftwo}%
\providecommand \translation [1]{[#1]}%
\providecommand \BibitemOpen [0]{}%
\providecommand \bibitemStop [0]{}%
\providecommand \bibitemNoStop [0]{.\EOS\space}%
\providecommand \EOS [0]{\spacefactor3000\relax}%
\providecommand \BibitemShut  [1]{\csname bibitem#1\endcsname}%
\let\auto@bib@innerbib\@empty
\bibitem [{\citenamefont {Maldacena}(1998)}]{Maldacena:1997re}%
  \BibitemOpen
  \bibfield  {author} {\bibinfo {author} {\bibfnamefont {J.~M.}\ \bibnamefont
  {Maldacena}},\ }\bibfield  {title} {\enquote {\bibinfo {title} {{The Large N
  limit of superconformal field theories and supergravity}},}\ }\href
  {https://doi.org/10.1023/A:1026654312961} {\bibfield  {journal} {\bibinfo
  {journal} {Adv. Theor. Math. Phys.}\ }\textbf {\bibinfo {volume} {2}},\
  \bibinfo {pages} {231--252} (\bibinfo {year} {1998})},\ \Eprint
  {https://arxiv.org/abs/hep-th/9711200} {arXiv:hep-th/9711200} \BibitemShut
  {NoStop}%
\bibitem [{\citenamefont {Gubser}, \citenamefont {Klebanov},\ and\
  \citenamefont {Polyakov}(1998)}]{Gubser:1998bc}%
  \BibitemOpen
  \bibfield  {author} {\bibinfo {author} {\bibfnamefont {S.~S.}\ \bibnamefont
  {Gubser}}, \bibinfo {author} {\bibfnamefont {I.~R.}\ \bibnamefont
  {Klebanov}},\ and\ \bibinfo {author} {\bibfnamefont {A.~M.}\ \bibnamefont
  {Polyakov}},\ }\bibfield  {title} {\enquote {\bibinfo {title} {{Gauge theory
  correlators from noncritical string theory}},}\ }\href
  {https://doi.org/10.1016/S0370-2693(98)00377-3} {\bibfield  {journal}
  {\bibinfo  {journal} {Phys. Lett. B}\ }\textbf {\bibinfo {volume} {428}},\
  \bibinfo {pages} {105--114} (\bibinfo {year} {1998})},\ \Eprint
  {https://arxiv.org/abs/hep-th/9802109} {arXiv:hep-th/9802109} \BibitemShut
  {NoStop}%
\bibitem [{\citenamefont {Witten}(1998)}]{Witten:1998qj}%
  \BibitemOpen
  \bibfield  {author} {\bibinfo {author} {\bibfnamefont {E.}~\bibnamefont
  {Witten}},\ }\bibfield  {title} {\enquote {\bibinfo {title} {{Anti-de Sitter
  space and holography}},}\ }\href {https://doi.org/10.4310/ATMP.1998.v2.n2.a2}
  {\bibfield  {journal} {\bibinfo  {journal} {Adv. Theor. Math. Phys.}\
  }\textbf {\bibinfo {volume} {2}},\ \bibinfo {pages} {253--291} (\bibinfo
  {year} {1998})},\ \Eprint {https://arxiv.org/abs/hep-th/9802150}
  {arXiv:hep-th/9802150} \BibitemShut {NoStop}%
\bibitem [{\citenamefont {Casini}\ and\ \citenamefont
  {Huerta}(2004)}]{Casini:2004bw}%
  \BibitemOpen
  \bibfield  {author} {\bibinfo {author} {\bibfnamefont {H.}~\bibnamefont
  {Casini}}\ and\ \bibinfo {author} {\bibfnamefont {M.}~\bibnamefont
  {Huerta}},\ }\bibfield  {title} {\enquote {\bibinfo {title} {{A Finite
  entanglement entropy and the c-theorem}},}\ }\href
  {https://doi.org/10.1016/j.physletb.2004.08.072} {\bibfield  {journal}
  {\bibinfo  {journal} {Phys. Lett. B}\ }\textbf {\bibinfo {volume} {600}},\
  \bibinfo {pages} {142--150} (\bibinfo {year} {2004})},\ \Eprint
  {https://arxiv.org/abs/hep-th/0405111} {arXiv:hep-th/0405111} \BibitemShut
  {NoStop}%
\bibitem [{\citenamefont {Calabrese}\ and\ \citenamefont
  {Cardy}(2004)}]{Calabrese:2004eu}%
  \BibitemOpen
  \bibfield  {author} {\bibinfo {author} {\bibfnamefont {P.}~\bibnamefont
  {Calabrese}}\ and\ \bibinfo {author} {\bibfnamefont {J.~L.}\ \bibnamefont
  {Cardy}},\ }\bibfield  {title} {\enquote {\bibinfo {title} {{Entanglement
  entropy and quantum field theory}},}\ }\href
  {https://doi.org/10.1088/1742-5468/2004/06/P06002} {\bibfield  {journal}
  {\bibinfo  {journal} {J. Stat. Mech.}\ }\textbf {\bibinfo {volume} {0406}},\
  \bibinfo {pages} {P06002} (\bibinfo {year} {2004})},\ \Eprint
  {https://arxiv.org/abs/hep-th/0405152} {arXiv:hep-th/0405152} \BibitemShut
  {NoStop}%
\bibitem [{\citenamefont {Kitaev}\ and\ \citenamefont
  {Preskill}(2006)}]{Kitaev:2005dm}%
  \BibitemOpen
  \bibfield  {author} {\bibinfo {author} {\bibfnamefont {A.}~\bibnamefont
  {Kitaev}}\ and\ \bibinfo {author} {\bibfnamefont {J.}~\bibnamefont
  {Preskill}},\ }\bibfield  {title} {\enquote {\bibinfo {title} {{Topological
  entanglement entropy}},}\ }\href
  {https://doi.org/10.1103/PhysRevLett.96.110404} {\bibfield  {journal}
  {\bibinfo  {journal} {Phys. Rev. Lett.}\ }\textbf {\bibinfo {volume} {96}},\
  \bibinfo {pages} {110404} (\bibinfo {year} {2006})},\ \Eprint
  {https://arxiv.org/abs/hep-th/0510092} {arXiv:hep-th/0510092} \BibitemShut
  {NoStop}%
\bibitem [{\citenamefont {Casini}, \citenamefont {Salazar~Landea},\ and\
  \citenamefont {Torroba}(2016)}]{Casini:2016fgb}%
  \BibitemOpen
  \bibfield  {author} {\bibinfo {author} {\bibfnamefont {H.}~\bibnamefont
  {Casini}}, \bibinfo {author} {\bibfnamefont {I.}~\bibnamefont
  {Salazar~Landea}},\ and\ \bibinfo {author} {\bibfnamefont {G.}~\bibnamefont
  {Torroba}},\ }\bibfield  {title} {\enquote {\bibinfo {title} {{The g-theorem
  and quantum information theory}},}\ }\href
  {https://doi.org/10.1007/JHEP10(2016)140} {\bibfield  {journal} {\bibinfo
  {journal} {JHEP}\ }\textbf {\bibinfo {volume} {10}},\ \bibinfo {pages} {140}
  (\bibinfo {year} {2016})},\ \Eprint {https://arxiv.org/abs/1607.00390}
  {arXiv:1607.00390 [hep-th]} \BibitemShut {NoStop}%
\bibitem [{\citenamefont {Nishioka}(2018)}]{Nishioka:2018khk}%
  \BibitemOpen
  \bibfield  {author} {\bibinfo {author} {\bibfnamefont {T.}~\bibnamefont
  {Nishioka}},\ }\bibfield  {title} {\enquote {\bibinfo {title} {{Entanglement
  entropy: holography and renormalization group}},}\ }\href
  {https://doi.org/10.1103/RevModPhys.90.035007} {\bibfield  {journal}
  {\bibinfo  {journal} {Rev. Mod. Phys.}\ }\textbf {\bibinfo {volume} {90}},\
  \bibinfo {pages} {035007} (\bibinfo {year} {2018})},\ \Eprint
  {https://arxiv.org/abs/1801.10352} {arXiv:1801.10352 [hep-th]} \BibitemShut
  {NoStop}%
\bibitem [{\citenamefont {Witten}(2018)}]{Witten:2018zxz}%
  \BibitemOpen
  \bibfield  {author} {\bibinfo {author} {\bibfnamefont {E.}~\bibnamefont
  {Witten}},\ }\bibfield  {title} {\enquote {\bibinfo {title} {{APS Medal for
  Exceptional Achievement in Research: Invited article on entanglement
  properties of quantum field theory}},}\ }\href
  {https://doi.org/10.1103/RevModPhys.90.045003} {\bibfield  {journal}
  {\bibinfo  {journal} {Rev. Mod. Phys.}\ }\textbf {\bibinfo {volume} {90}},\
  \bibinfo {pages} {045003} (\bibinfo {year} {2018})},\ \Eprint
  {https://arxiv.org/abs/1803.04993} {arXiv:1803.04993 [hep-th]} \BibitemShut
  {NoStop}%
\bibitem [{\citenamefont {Casini}\ and\ \citenamefont
  {Huerta}(2022)}]{Casini:2022rlv}%
  \BibitemOpen
  \bibfield  {author} {\bibinfo {author} {\bibfnamefont {H.}~\bibnamefont
  {Casini}}\ and\ \bibinfo {author} {\bibfnamefont {M.}~\bibnamefont
  {Huerta}},\ }\bibfield  {title} {\enquote {\bibinfo {title} {{Lectures on
  entanglement in quantum field theory}},}\ }\href@noop {} {\  (\bibinfo {year}
  {2022})},\ \Eprint {https://arxiv.org/abs/2201.13310} {arXiv:2201.13310
  [hep-th]} \BibitemShut {NoStop}%
\bibitem [{\citenamefont {Van~Raamsdonk}(2010)}]{VanRaamsdonk:2010pw}%
  \BibitemOpen
  \bibfield  {author} {\bibinfo {author} {\bibfnamefont {M.}~\bibnamefont
  {Van~Raamsdonk}},\ }\bibfield  {title} {\enquote {\bibinfo {title} {{Building
  up spacetime with quantum entanglement}},}\ }\href
  {https://doi.org/10.1142/S0218271810018529} {\bibfield  {journal} {\bibinfo
  {journal} {Gen. Rel. Grav.}\ }\textbf {\bibinfo {volume} {42}},\ \bibinfo
  {pages} {2323--2329} (\bibinfo {year} {2010})},\ \Eprint
  {https://arxiv.org/abs/1005.3035} {arXiv:1005.3035 [hep-th]} \BibitemShut
  {NoStop}%
\bibitem [{\citenamefont {Maldacena}\ and\ \citenamefont
  {Susskind}(2013)}]{Maldacena:2013xja}%
  \BibitemOpen
  \bibfield  {author} {\bibinfo {author} {\bibfnamefont {J.}~\bibnamefont
  {Maldacena}}\ and\ \bibinfo {author} {\bibfnamefont {L.}~\bibnamefont
  {Susskind}},\ }\bibfield  {title} {\enquote {\bibinfo {title} {{Cool horizons
  for entangled black holes}},}\ }\href
  {https://doi.org/10.1002/prop.201300020} {\bibfield  {journal} {\bibinfo
  {journal} {Fortsch. Phys.}\ }\textbf {\bibinfo {volume} {61}},\ \bibinfo
  {pages} {781--811} (\bibinfo {year} {2013})},\ \Eprint
  {https://arxiv.org/abs/1306.0533} {arXiv:1306.0533 [hep-th]} \BibitemShut
  {NoStop}%
\bibitem [{\citenamefont {Rangamani}\ and\ \citenamefont
  {Takayanagi}(2017)}]{Rangamani:2016dms}%
  \BibitemOpen
  \bibfield  {author} {\bibinfo {author} {\bibfnamefont {M.}~\bibnamefont
  {Rangamani}}\ and\ \bibinfo {author} {\bibfnamefont {T.}~\bibnamefont
  {Takayanagi}},\ }\href {https://doi.org/10.1007/978-3-319-52573-0} {\emph
  {\bibinfo {title} {{Holographic Entanglement Entropy}}}},\ Vol.\ \bibinfo
  {volume} {931}\ (\bibinfo  {publisher} {Springer},\ \bibinfo {year} {2017})\
  \Eprint {https://arxiv.org/abs/1609.01287} {arXiv:1609.01287 [hep-th]}
  \BibitemShut {NoStop}%
\bibitem [{\citenamefont {Hawking}(1976)}]{Hawking:1976ra}%
  \BibitemOpen
  \bibfield  {author} {\bibinfo {author} {\bibfnamefont {S.~W.}\ \bibnamefont
  {Hawking}},\ }\bibfield  {title} {\enquote {\bibinfo {title} {{Breakdown of
  Predictability in Gravitational Collapse}},}\ }\href
  {https://doi.org/10.1103/PhysRevD.14.2460} {\bibfield  {journal} {\bibinfo
  {journal} {Phys. Rev. D}\ }\textbf {\bibinfo {volume} {14}},\ \bibinfo
  {pages} {2460--2473} (\bibinfo {year} {1976})}\BibitemShut {NoStop}%
\bibitem [{\citenamefont {Mathur}(2009)}]{Mathur:2009hf}%
  \BibitemOpen
  \bibfield  {author} {\bibinfo {author} {\bibfnamefont {S.~D.}\ \bibnamefont
  {Mathur}},\ }\bibfield  {title} {\enquote {\bibinfo {title} {{The Information
  paradox: A Pedagogical introduction}},}\ }\href
  {https://doi.org/10.1088/0264-9381/26/22/224001} {\bibfield  {journal}
  {\bibinfo  {journal} {Class. Quant. Grav.}\ }\textbf {\bibinfo {volume}
  {26}},\ \bibinfo {pages} {224001} (\bibinfo {year} {2009})},\ \Eprint
  {https://arxiv.org/abs/0909.1038} {arXiv:0909.1038 [hep-th]} \BibitemShut
  {NoStop}%
\bibitem [{\citenamefont {Almheiri}\ \emph {et~al.}(2013)\citenamefont
  {Almheiri}, \citenamefont {Marolf}, \citenamefont {Polchinski},\ and\
  \citenamefont {Sully}}]{Almheiri:2012rt}%
  \BibitemOpen
  \bibfield  {author} {\bibinfo {author} {\bibfnamefont {A.}~\bibnamefont
  {Almheiri}}, \bibinfo {author} {\bibfnamefont {D.}~\bibnamefont {Marolf}},
  \bibinfo {author} {\bibfnamefont {J.}~\bibnamefont {Polchinski}},\ and\
  \bibinfo {author} {\bibfnamefont {J.}~\bibnamefont {Sully}},\ }\bibfield
  {title} {\enquote {\bibinfo {title} {{Black Holes: Complementarity or
  Firewalls?}}}\ }\href {https://doi.org/10.1007/JHEP02(2013)062} {\bibfield
  {journal} {\bibinfo  {journal} {JHEP}\ }\textbf {\bibinfo {volume} {02}},\
  \bibinfo {pages} {062} (\bibinfo {year} {2013})},\ \Eprint
  {https://arxiv.org/abs/1207.3123} {arXiv:1207.3123 [hep-th]} \BibitemShut
  {NoStop}%
\bibitem [{\citenamefont {Penington}(2020)}]{Penington:2019npb}%
  \BibitemOpen
  \bibfield  {author} {\bibinfo {author} {\bibfnamefont {G.}~\bibnamefont
  {Penington}},\ }\bibfield  {title} {\enquote {\bibinfo {title} {{Entanglement
  Wedge Reconstruction and the Information Paradox}},}\ }\href
  {https://doi.org/10.1007/JHEP09(2020)002} {\bibfield  {journal} {\bibinfo
  {journal} {JHEP}\ }\textbf {\bibinfo {volume} {09}},\ \bibinfo {pages} {002}
  (\bibinfo {year} {2020})},\ \Eprint {https://arxiv.org/abs/1905.08255}
  {arXiv:1905.08255 [hep-th]} \BibitemShut {NoStop}%
\bibitem [{\citenamefont {Almheiri}\ \emph {et~al.}(2019)\citenamefont
  {Almheiri}, \citenamefont {Engelhardt}, \citenamefont {Marolf},\ and\
  \citenamefont {Maxfield}}]{Almheiri:2019psf}%
  \BibitemOpen
  \bibfield  {author} {\bibinfo {author} {\bibfnamefont {A.}~\bibnamefont
  {Almheiri}}, \bibinfo {author} {\bibfnamefont {N.}~\bibnamefont
  {Engelhardt}}, \bibinfo {author} {\bibfnamefont {D.}~\bibnamefont {Marolf}},\
  and\ \bibinfo {author} {\bibfnamefont {H.}~\bibnamefont {Maxfield}},\
  }\bibfield  {title} {\enquote {\bibinfo {title} {{The entropy of bulk quantum
  fields and the entanglement wedge of an evaporating black hole}},}\ }\href
  {https://doi.org/10.1007/JHEP12(2019)063} {\bibfield  {journal} {\bibinfo
  {journal} {JHEP}\ }\textbf {\bibinfo {volume} {12}},\ \bibinfo {pages} {063}
  (\bibinfo {year} {2019})},\ \Eprint {https://arxiv.org/abs/1905.08762}
  {arXiv:1905.08762 [hep-th]} \BibitemShut {NoStop}%
\bibitem [{\citenamefont {Nakata}\ \emph {et~al.}(2021)\citenamefont {Nakata},
  \citenamefont {Takayanagi}, \citenamefont {Taki}, \citenamefont {Tamaoka},\
  and\ \citenamefont {Wei}}]{Nakata:2020luh}%
  \BibitemOpen
  \bibfield  {author} {\bibinfo {author} {\bibfnamefont {Y.}~\bibnamefont
  {Nakata}}, \bibinfo {author} {\bibfnamefont {T.}~\bibnamefont {Takayanagi}},
  \bibinfo {author} {\bibfnamefont {Y.}~\bibnamefont {Taki}}, \bibinfo {author}
  {\bibfnamefont {K.}~\bibnamefont {Tamaoka}},\ and\ \bibinfo {author}
  {\bibfnamefont {Z.}~\bibnamefont {Wei}},\ }\bibfield  {title} {\enquote
  {\bibinfo {title} {{New holographic generalization of entanglement
  entropy}},}\ }\href {https://doi.org/10.1103/PhysRevD.103.026005} {\bibfield
  {journal} {\bibinfo  {journal} {Phys. Rev. D}\ }\textbf {\bibinfo {volume}
  {103}},\ \bibinfo {pages} {026005} (\bibinfo {year} {2021})},\ \Eprint
  {https://arxiv.org/abs/2005.13801} {arXiv:2005.13801 [hep-th]} \BibitemShut
  {NoStop}%
\bibitem [{\citenamefont {Guo}, \citenamefont {He},\ and\ \citenamefont
  {Zhang}(2022{\natexlab{a}})}]{Guo:2022jzs}%
  \BibitemOpen
  \bibfield  {author} {\bibinfo {author} {\bibfnamefont {W.-z.}\ \bibnamefont
  {Guo}}, \bibinfo {author} {\bibfnamefont {S.}~\bibnamefont {He}},\ and\
  \bibinfo {author} {\bibfnamefont {Y.-X.}\ \bibnamefont {Zhang}},\ }\bibfield
  {title} {\enquote {\bibinfo {title} {{Constructible reality condition of
  pseudo entropy via pseudo-Hermiticity}},}\ }\href@noop {} {\  (\bibinfo
  {year} {2022}{\natexlab{a}})},\ \Eprint {https://arxiv.org/abs/2209.07308}
  {arXiv:2209.07308 [hep-th]} \BibitemShut {NoStop}%
\bibitem [{Note1()}]{Note1}%
  \BibitemOpen
  \bibinfo {note} {We thank the anonymous referee for bringing this to our
  attention.}\BibitemShut {Stop}%
\bibitem [{\citenamefont {Akal}\ \emph {et~al.}(2022)\citenamefont {Akal},
  \citenamefont {Kawamoto}, \citenamefont {Ruan}, \citenamefont {Takayanagi},\
  and\ \citenamefont {Wei}}]{Akal:2021dqt}%
  \BibitemOpen
  \bibfield  {author} {\bibinfo {author} {\bibfnamefont {I.}~\bibnamefont
  {Akal}}, \bibinfo {author} {\bibfnamefont {T.}~\bibnamefont {Kawamoto}},
  \bibinfo {author} {\bibfnamefont {S.-M.}\ \bibnamefont {Ruan}}, \bibinfo
  {author} {\bibfnamefont {T.}~\bibnamefont {Takayanagi}},\ and\ \bibinfo
  {author} {\bibfnamefont {Z.}~\bibnamefont {Wei}},\ }\bibfield  {title}
  {\enquote {\bibinfo {title} {{Page curve under final state projection}},}\
  }\href {https://doi.org/10.1103/PhysRevD.105.126026} {\bibfield  {journal}
  {\bibinfo  {journal} {Phys. Rev. D}\ }\textbf {\bibinfo {volume} {105}},\
  \bibinfo {pages} {126026} (\bibinfo {year} {2022})},\ \Eprint
  {https://arxiv.org/abs/2112.08433} {arXiv:2112.08433 [hep-th]} \BibitemShut
  {NoStop}%
\bibitem [{\citenamefont {Aharonov}\ and\ \citenamefont
  {Vaidman}(2008)}]{aharonov2008two}%
  \BibitemOpen
  \bibfield  {author} {\bibinfo {author} {\bibfnamefont {Y.}~\bibnamefont
  {Aharonov}}\ and\ \bibinfo {author} {\bibfnamefont {L.}~\bibnamefont
  {Vaidman}},\ }\bibfield  {title} {\enquote {\bibinfo {title} {The two-state
  vector formalism: an updated review},}\ }\href@noop {} {\bibfield  {journal}
  {\bibinfo  {journal} {Time in quantum mechanics}\ ,\ \bibinfo {pages}
  {399--447}} (\bibinfo {year} {2008})}\BibitemShut {NoStop}%
\bibitem [{\citenamefont {Mollabashi}\ \emph
  {et~al.}(2021{\natexlab{a}})\citenamefont {Mollabashi}, \citenamefont
  {Shiba}, \citenamefont {Takayanagi}, \citenamefont {Tamaoka},\ and\
  \citenamefont {Wei}}]{Mollabashi:2020yie}%
  \BibitemOpen
  \bibfield  {author} {\bibinfo {author} {\bibfnamefont {A.}~\bibnamefont
  {Mollabashi}}, \bibinfo {author} {\bibfnamefont {N.}~\bibnamefont {Shiba}},
  \bibinfo {author} {\bibfnamefont {T.}~\bibnamefont {Takayanagi}}, \bibinfo
  {author} {\bibfnamefont {K.}~\bibnamefont {Tamaoka}},\ and\ \bibinfo {author}
  {\bibfnamefont {Z.}~\bibnamefont {Wei}},\ }\bibfield  {title} {\enquote
  {\bibinfo {title} {{Pseudo Entropy in Free Quantum Field Theories}},}\ }\href
  {https://doi.org/10.1103/PhysRevLett.126.081601} {\bibfield  {journal}
  {\bibinfo  {journal} {Phys. Rev. Lett.}\ }\textbf {\bibinfo {volume} {126}},\
  \bibinfo {pages} {081601} (\bibinfo {year} {2021}{\natexlab{a}})},\ \Eprint
  {https://arxiv.org/abs/2011.09648} {arXiv:2011.09648 [hep-th]} \BibitemShut
  {NoStop}%
\bibitem [{\citenamefont {Camilo}\ and\ \citenamefont
  {Prudenziati}(2021)}]{Camilo:2021dtt}%
  \BibitemOpen
  \bibfield  {author} {\bibinfo {author} {\bibfnamefont {G.}~\bibnamefont
  {Camilo}}\ and\ \bibinfo {author} {\bibfnamefont {A.}~\bibnamefont
  {Prudenziati}},\ }\bibfield  {title} {\enquote {\bibinfo {title} {{Twist
  operators and pseudo entropies in two-dimensional momentum space}},}\
  }\href@noop {} {\  (\bibinfo {year} {2021})},\ \Eprint
  {https://arxiv.org/abs/2101.02093} {arXiv:2101.02093 [hep-th]} \BibitemShut
  {NoStop}%
\bibitem [{\citenamefont {Mollabashi}\ \emph
  {et~al.}(2021{\natexlab{b}})\citenamefont {Mollabashi}, \citenamefont
  {Shiba}, \citenamefont {Takayanagi}, \citenamefont {Tamaoka},\ and\
  \citenamefont {Wei}}]{Mollabashi:2021xsd}%
  \BibitemOpen
  \bibfield  {author} {\bibinfo {author} {\bibfnamefont {A.}~\bibnamefont
  {Mollabashi}}, \bibinfo {author} {\bibfnamefont {N.}~\bibnamefont {Shiba}},
  \bibinfo {author} {\bibfnamefont {T.}~\bibnamefont {Takayanagi}}, \bibinfo
  {author} {\bibfnamefont {K.}~\bibnamefont {Tamaoka}},\ and\ \bibinfo {author}
  {\bibfnamefont {Z.}~\bibnamefont {Wei}},\ }\bibfield  {title} {\enquote
  {\bibinfo {title} {{Aspects of pseudoentropy in field theories}},}\ }\href
  {https://doi.org/10.1103/PhysRevResearch.3.033254} {\bibfield  {journal}
  {\bibinfo  {journal} {Phys. Rev. Res.}\ }\textbf {\bibinfo {volume} {3}},\
  \bibinfo {pages} {033254} (\bibinfo {year} {2021}{\natexlab{b}})},\ \Eprint
  {https://arxiv.org/abs/2106.03118} {arXiv:2106.03118 [hep-th]} \BibitemShut
  {NoStop}%
\bibitem [{\citenamefont {Nishioka}, \citenamefont {Takayanagi},\ and\
  \citenamefont {Taki}(2021)}]{Nishioka:2021cxe}%
  \BibitemOpen
  \bibfield  {author} {\bibinfo {author} {\bibfnamefont {T.}~\bibnamefont
  {Nishioka}}, \bibinfo {author} {\bibfnamefont {T.}~\bibnamefont
  {Takayanagi}},\ and\ \bibinfo {author} {\bibfnamefont {Y.}~\bibnamefont
  {Taki}},\ }\bibfield  {title} {\enquote {\bibinfo {title} {{Topological
  pseudo entropy}},}\ }\href {https://doi.org/10.1007/JHEP09(2021)015}
  {\bibfield  {journal} {\bibinfo  {journal} {JHEP}\ }\textbf {\bibinfo
  {volume} {09}},\ \bibinfo {pages} {015} (\bibinfo {year} {2021})},\ \Eprint
  {https://arxiv.org/abs/2107.01797} {arXiv:2107.01797 [hep-th]} \BibitemShut
  {NoStop}%
\bibitem [{\citenamefont {Goto}, \citenamefont {Nozaki},\ and\ \citenamefont
  {Tamaoka}(2021)}]{Goto:2021kln}%
  \BibitemOpen
  \bibfield  {author} {\bibinfo {author} {\bibfnamefont {K.}~\bibnamefont
  {Goto}}, \bibinfo {author} {\bibfnamefont {M.}~\bibnamefont {Nozaki}},\ and\
  \bibinfo {author} {\bibfnamefont {K.}~\bibnamefont {Tamaoka}},\ }\bibfield
  {title} {\enquote {\bibinfo {title} {{Subregion spectrum form factor via
  pseudoentropy}},}\ }\href {https://doi.org/10.1103/PhysRevD.104.L121902}
  {\bibfield  {journal} {\bibinfo  {journal} {Phys. Rev. D}\ }\textbf {\bibinfo
  {volume} {104}},\ \bibinfo {pages} {L121902} (\bibinfo {year} {2021})},\
  \Eprint {https://arxiv.org/abs/2109.00372} {arXiv:2109.00372 [hep-th]}
  \BibitemShut {NoStop}%
\bibitem [{\citenamefont {Mukherjee}(2022)}]{Mukherjee:2022jac}%
  \BibitemOpen
  \bibfield  {author} {\bibinfo {author} {\bibfnamefont {J.}~\bibnamefont
  {Mukherjee}},\ }\bibfield  {title} {\enquote {\bibinfo {title} {{Pseudo
  Entropy in U(1) gauge theory}},}\ }\href
  {https://doi.org/10.1007/JHEP10(2022)016} {\bibfield  {journal} {\bibinfo
  {journal} {JHEP}\ }\textbf {\bibinfo {volume} {10}},\ \bibinfo {pages} {016}
  (\bibinfo {year} {2022})},\ \Eprint {https://arxiv.org/abs/2205.08179}
  {arXiv:2205.08179 [hep-th]} \BibitemShut {NoStop}%
\bibitem [{\citenamefont {Guo}, \citenamefont {He},\ and\ \citenamefont
  {Zhang}(2022{\natexlab{b}})}]{Guo:2022sfl}%
  \BibitemOpen
  \bibfield  {author} {\bibinfo {author} {\bibfnamefont {W.-z.}\ \bibnamefont
  {Guo}}, \bibinfo {author} {\bibfnamefont {S.}~\bibnamefont {He}},\ and\
  \bibinfo {author} {\bibfnamefont {Y.-X.}\ \bibnamefont {Zhang}},\ }\bibfield
  {title} {\enquote {\bibinfo {title} {{On the real-time evolution of
  pseudo-entropy in 2d CFTs}},}\ }\href
  {https://doi.org/10.1007/JHEP09(2022)094} {\bibfield  {journal} {\bibinfo
  {journal} {JHEP}\ }\textbf {\bibinfo {volume} {09}},\ \bibinfo {pages} {094}
  (\bibinfo {year} {2022}{\natexlab{b}})},\ \Eprint
  {https://arxiv.org/abs/2206.11818} {arXiv:2206.11818 [hep-th]} \BibitemShut
  {NoStop}%
\bibitem [{\citenamefont {Miyaji}(2021)}]{Miyaji:2021lcq}%
  \BibitemOpen
  \bibfield  {author} {\bibinfo {author} {\bibfnamefont {M.}~\bibnamefont
  {Miyaji}},\ }\bibfield  {title} {\enquote {\bibinfo {title} {{Island for
  gravitationally prepared state and pseudo entanglement wedge}},}\ }\href
  {https://doi.org/10.1007/JHEP12(2021)013} {\bibfield  {journal} {\bibinfo
  {journal} {JHEP}\ }\textbf {\bibinfo {volume} {12}},\ \bibinfo {pages} {013}
  (\bibinfo {year} {2021})},\ \Eprint {https://arxiv.org/abs/2109.03830}
  {arXiv:2109.03830 [hep-th]} \BibitemShut {NoStop}%
\bibitem [{\citenamefont {Ishiyama}\ \emph {et~al.}(2022)\citenamefont
  {Ishiyama}, \citenamefont {Kojima}, \citenamefont {Matsui},\ and\
  \citenamefont {Tamaoka}}]{Ishiyama:2022odv}%
  \BibitemOpen
  \bibfield  {author} {\bibinfo {author} {\bibfnamefont {Y.}~\bibnamefont
  {Ishiyama}}, \bibinfo {author} {\bibfnamefont {R.}~\bibnamefont {Kojima}},
  \bibinfo {author} {\bibfnamefont {S.}~\bibnamefont {Matsui}},\ and\ \bibinfo
  {author} {\bibfnamefont {K.}~\bibnamefont {Tamaoka}},\ }\bibfield  {title}
  {\enquote {\bibinfo {title} {{Notes on pseudo entropy amplification}},}\
  }\href {https://doi.org/10.1093/ptep/ptac112} {\bibfield  {journal} {\bibinfo
   {journal} {PTEP}\ }\textbf {\bibinfo {volume} {2022}},\ \bibinfo {pages}
  {093B10} (\bibinfo {year} {2022})},\ \Eprint
  {https://arxiv.org/abs/2206.14551} {arXiv:2206.14551 [hep-th]} \BibitemShut
  {NoStop}%
\bibitem [{\citenamefont {Bhattacharya}, \citenamefont {Bhattacharyya},\ and\
  \citenamefont {Maulik}(2022)}]{Bhattacharya:2022wlp}%
  \BibitemOpen
  \bibfield  {author} {\bibinfo {author} {\bibfnamefont {A.}~\bibnamefont
  {Bhattacharya}}, \bibinfo {author} {\bibfnamefont {A.}~\bibnamefont
  {Bhattacharyya}},\ and\ \bibinfo {author} {\bibfnamefont {S.}~\bibnamefont
  {Maulik}},\ }\bibfield  {title} {\enquote {\bibinfo {title}
  {{Pseudocomplexity of purification for free scalar field theories}},}\ }\href
  {https://doi.org/10.1103/PhysRevD.106.086010} {\bibfield  {journal} {\bibinfo
   {journal} {Phys. Rev. D}\ }\textbf {\bibinfo {volume} {106}},\ \bibinfo
  {pages} {086010} (\bibinfo {year} {2022})},\ \Eprint
  {https://arxiv.org/abs/2209.00049} {arXiv:2209.00049 [hep-th]} \BibitemShut
  {NoStop}%
\bibitem [{\citenamefont {Doi}\ \emph {et~al.}(2022)\citenamefont {Doi},
  \citenamefont {Harper}, \citenamefont {Mollabashi}, \citenamefont
  {Takayanagi},\ and\ \citenamefont {Taki}}]{Doi:2022iyj}%
  \BibitemOpen
  \bibfield  {author} {\bibinfo {author} {\bibfnamefont {K.}~\bibnamefont
  {Doi}}, \bibinfo {author} {\bibfnamefont {J.}~\bibnamefont {Harper}},
  \bibinfo {author} {\bibfnamefont {A.}~\bibnamefont {Mollabashi}}, \bibinfo
  {author} {\bibfnamefont {T.}~\bibnamefont {Takayanagi}},\ and\ \bibinfo
  {author} {\bibfnamefont {Y.}~\bibnamefont {Taki}},\ }\bibfield  {title}
  {\enquote {\bibinfo {title} {{Pseudo Entropy in dS/CFT and Time-like
  Entanglement Entropy}},}\ }\href@noop {} {\  (\bibinfo {year} {2022})},\
  \Eprint {https://arxiv.org/abs/2210.09457} {arXiv:2210.09457 [hep-th]}
  \BibitemShut {NoStop}%
\bibitem [{\citenamefont {Li}, \citenamefont {Xiao},\ and\ \citenamefont
  {Yang}(2022)}]{Li:2022tsv}%
  \BibitemOpen
  \bibfield  {author} {\bibinfo {author} {\bibfnamefont {Z.}~\bibnamefont
  {Li}}, \bibinfo {author} {\bibfnamefont {Z.-Q.}\ \bibnamefont {Xiao}},\ and\
  \bibinfo {author} {\bibfnamefont {R.-Q.}\ \bibnamefont {Yang}},\ }\bibfield
  {title} {\enquote {\bibinfo {title} {{On holographic time-like entanglement
  entropy}},}\ }\href@noop {} {\  (\bibinfo {year} {2022})},\ \Eprint
  {https://arxiv.org/abs/2211.14883} {arXiv:2211.14883 [hep-th]} \BibitemShut
  {NoStop}%
\bibitem [{\citenamefont {Jiang}\ \emph
  {et~al.}(2023{\natexlab{a}})\citenamefont {Jiang}, \citenamefont {Wang},
  \citenamefont {Wu},\ and\ \citenamefont {Yang}}]{Jiang:2023loq}%
  \BibitemOpen
  \bibfield  {author} {\bibinfo {author} {\bibfnamefont {X.}~\bibnamefont
  {Jiang}}, \bibinfo {author} {\bibfnamefont {P.}~\bibnamefont {Wang}},
  \bibinfo {author} {\bibfnamefont {H.}~\bibnamefont {Wu}},\ and\ \bibinfo
  {author} {\bibfnamefont {H.}~\bibnamefont {Yang}},\ }\bibfield  {title}
  {\enquote {\bibinfo {title} {{Timelike entanglement entropy in
  $\text{dS}_3/\text{CFT}_2$}},}\ }\href@noop {} {\  (\bibinfo {year}
  {2023}{\natexlab{a}})},\ \Eprint {https://arxiv.org/abs/2304.10376}
  {arXiv:2304.10376 [hep-th]} \BibitemShut {NoStop}%
\bibitem [{\citenamefont {Jiang}\ \emph
  {et~al.}(2023{\natexlab{b}})\citenamefont {Jiang}, \citenamefont {Wang},
  \citenamefont {Wu},\ and\ \citenamefont {Yang}}]{Jiang:2023ffu}%
  \BibitemOpen
  \bibfield  {author} {\bibinfo {author} {\bibfnamefont {X.}~\bibnamefont
  {Jiang}}, \bibinfo {author} {\bibfnamefont {P.}~\bibnamefont {Wang}},
  \bibinfo {author} {\bibfnamefont {H.}~\bibnamefont {Wu}},\ and\ \bibinfo
  {author} {\bibfnamefont {H.}~\bibnamefont {Yang}},\ }\bibfield  {title}
  {\enquote {\bibinfo {title} {{Timelike entanglement entropy and $T\bar{T}$
  deformation}},}\ }\href@noop {} {\  (\bibinfo {year} {2023}{\natexlab{b}})},\
  \Eprint {https://arxiv.org/abs/2302.13872} {arXiv:2302.13872 [hep-th]}
  \BibitemShut {NoStop}%
\bibitem [{\citenamefont {Wang}, \citenamefont {Wu},\ and\ \citenamefont
  {Yang}(2020)}]{Wang:2018jva}%
  \BibitemOpen
  \bibfield  {author} {\bibinfo {author} {\bibfnamefont {P.}~\bibnamefont
  {Wang}}, \bibinfo {author} {\bibfnamefont {H.}~\bibnamefont {Wu}},\ and\
  \bibinfo {author} {\bibfnamefont {H.}~\bibnamefont {Yang}},\ }\bibfield
  {title} {\enquote {\bibinfo {title} {{Fix the dual geometries of $T\bar{T}$
  deformed CFT$_2$ and highly excited states of CFT$_2$}},}\ }\href
  {https://doi.org/10.1140/epjc/s10052-020-08680-7} {\bibfield  {journal}
  {\bibinfo  {journal} {Eur. Phys. J. C}\ }\textbf {\bibinfo {volume} {80}},\
  \bibinfo {pages} {1117} (\bibinfo {year} {2020})},\ \Eprint
  {https://arxiv.org/abs/1811.07758} {arXiv:1811.07758 [hep-th]} \BibitemShut
  {NoStop}%
\bibitem [{\citenamefont {Kamenev}(2005)}]{kamenev2005manybody}%
  \BibitemOpen
  \bibfield  {author} {\bibinfo {author} {\bibfnamefont {A.}~\bibnamefont
  {Kamenev}},\ }\href@noop {} {\enquote {\bibinfo {title} {Many-body theory of
  non-equilibrium systems},}\ } (\bibinfo {year} {2005}),\ \Eprint
  {https://arxiv.org/abs/cond-mat/0412296} {arXiv:cond-mat/0412296
  [cond-mat.dis-nn]} \BibitemShut {NoStop}%
\bibitem [{\citenamefont {Stefanucci}\ and\ \citenamefont
  {Van~Leeuwen}(2013)}]{stefanucci2013nonequilibrium}%
  \BibitemOpen
  \bibfield  {author} {\bibinfo {author} {\bibfnamefont {G.}~\bibnamefont
  {Stefanucci}}\ and\ \bibinfo {author} {\bibfnamefont {R.}~\bibnamefont
  {Van~Leeuwen}},\ }\href@noop {} {\emph {\bibinfo {title} {Nonequilibrium
  many-body theory of quantum systems: a modern introduction}}}\ (\bibinfo
  {publisher} {Cambridge University Press},\ \bibinfo {year}
  {2013})\BibitemShut {NoStop}%
\bibitem [{\citenamefont {Deutsch}(1991)}]{deutsch1991quantum}%
  \BibitemOpen
  \bibfield  {author} {\bibinfo {author} {\bibfnamefont {J.~M.}\ \bibnamefont
  {Deutsch}},\ }\bibfield  {title} {\enquote {\bibinfo {title} {Quantum
  statistical mechanics in a closed system},}\ }\href@noop {} {\bibfield
  {journal} {\bibinfo  {journal} {Physical review a}\ }\textbf {\bibinfo
  {volume} {43}},\ \bibinfo {pages} {2046} (\bibinfo {year}
  {1991})}\BibitemShut {NoStop}%
\bibitem [{\citenamefont {Srednicki}(1994)}]{srednicki1994chaos}%
  \BibitemOpen
  \bibfield  {author} {\bibinfo {author} {\bibfnamefont {M.}~\bibnamefont
  {Srednicki}},\ }\bibfield  {title} {\enquote {\bibinfo {title} {Chaos and
  quantum thermalization},}\ }\href@noop {} {\bibfield  {journal} {\bibinfo
  {journal} {Physical review e}\ }\textbf {\bibinfo {volume} {50}},\ \bibinfo
  {pages} {888} (\bibinfo {year} {1994})}\BibitemShut {NoStop}%
\bibitem [{\citenamefont {Rigol}, \citenamefont {Dunjko},\ and\ \citenamefont
  {Olshanii}(2008)}]{rigol2008thermalization}%
  \BibitemOpen
  \bibfield  {author} {\bibinfo {author} {\bibfnamefont {M.}~\bibnamefont
  {Rigol}}, \bibinfo {author} {\bibfnamefont {V.}~\bibnamefont {Dunjko}},\ and\
  \bibinfo {author} {\bibfnamefont {M.}~\bibnamefont {Olshanii}},\ }\bibfield
  {title} {\enquote {\bibinfo {title} {Thermalization and its mechanism for
  generic isolated quantum systems},}\ }\href@noop {} {\bibfield  {journal}
  {\bibinfo  {journal} {Nature}\ }\textbf {\bibinfo {volume} {452}},\ \bibinfo
  {pages} {854--858} (\bibinfo {year} {2008})}\BibitemShut {NoStop}%
\bibitem [{\citenamefont {Calabrese}\ and\ \citenamefont
  {Cardy}(2005)}]{Calabrese:2005in}%
  \BibitemOpen
  \bibfield  {author} {\bibinfo {author} {\bibfnamefont {P.}~\bibnamefont
  {Calabrese}}\ and\ \bibinfo {author} {\bibfnamefont {J.~L.}\ \bibnamefont
  {Cardy}},\ }\bibfield  {title} {\enquote {\bibinfo {title} {{Evolution of
  entanglement entropy in one-dimensional systems}},}\ }\href
  {https://doi.org/10.1088/1742-5468/2005/04/P04010} {\bibfield  {journal}
  {\bibinfo  {journal} {J. Stat. Mech.}\ }\textbf {\bibinfo {volume} {0504}},\
  \bibinfo {pages} {P04010} (\bibinfo {year} {2005})},\ \Eprint
  {https://arxiv.org/abs/cond-mat/0503393} {arXiv:cond-mat/0503393}
  \BibitemShut {NoStop}%
\bibitem [{\citenamefont {Calabrese}\ and\ \citenamefont
  {Cardy}(2007{\natexlab{a}})}]{calabrese2007quantum}%
  \BibitemOpen
  \bibfield  {author} {\bibinfo {author} {\bibfnamefont {P.}~\bibnamefont
  {Calabrese}}\ and\ \bibinfo {author} {\bibfnamefont {J.}~\bibnamefont
  {Cardy}},\ }\bibfield  {title} {\enquote {\bibinfo {title} {Quantum quenches
  in extended systems},}\ }\href@noop {} {\bibfield  {journal} {\bibinfo
  {journal} {Journal of Statistical Mechanics: Theory and Experiment}\ }\textbf
  {\bibinfo {volume} {2007}},\ \bibinfo {pages} {P06008} (\bibinfo {year}
  {2007}{\natexlab{a}})}\BibitemShut {NoStop}%
\bibitem [{\citenamefont {Calabrese}\ and\ \citenamefont
  {Cardy}(2007{\natexlab{b}})}]{Calabrese:2007mtj}%
  \BibitemOpen
  \bibfield  {author} {\bibinfo {author} {\bibfnamefont {P.}~\bibnamefont
  {Calabrese}}\ and\ \bibinfo {author} {\bibfnamefont {J.}~\bibnamefont
  {Cardy}},\ }\bibfield  {title} {\enquote {\bibinfo {title} {{Entanglement and
  correlation functions following a local quench: a conformal field theory
  approach}},}\ }\href {https://doi.org/10.1088/1742-5468/2007/10/P10004}
  {\bibfield  {journal} {\bibinfo  {journal} {J. Stat. Mech.}\ }\textbf
  {\bibinfo {volume} {0710}},\ \bibinfo {pages} {P10004} (\bibinfo {year}
  {2007}{\natexlab{b}})},\ \Eprint {https://arxiv.org/abs/0708.3750}
  {arXiv:0708.3750 [cond-mat.stat-mech]} \BibitemShut {NoStop}%
\bibitem [{\citenamefont {Eisler}\ and\ \citenamefont
  {Peschel}(2007)}]{eisler2007evolution}%
  \BibitemOpen
  \bibfield  {author} {\bibinfo {author} {\bibfnamefont {V.}~\bibnamefont
  {Eisler}}\ and\ \bibinfo {author} {\bibfnamefont {I.}~\bibnamefont
  {Peschel}},\ }\bibfield  {title} {\enquote {\bibinfo {title} {Evolution of
  entanglement after a local quench},}\ }\href@noop {} {\bibfield  {journal}
  {\bibinfo  {journal} {Journal of Statistical Mechanics: Theory and
  Experiment}\ }\textbf {\bibinfo {volume} {2007}},\ \bibinfo {pages} {P06005}
  (\bibinfo {year} {2007})}\BibitemShut {NoStop}%
\bibitem [{\citenamefont {Alcaraz}, \citenamefont {Berganza},\ and\
  \citenamefont {Sierra}(2011{\natexlab{a}})}]{alcaraz2011entanglement}%
  \BibitemOpen
  \bibfield  {author} {\bibinfo {author} {\bibfnamefont {F.~C.}\ \bibnamefont
  {Alcaraz}}, \bibinfo {author} {\bibfnamefont {M.~I.}\ \bibnamefont
  {Berganza}},\ and\ \bibinfo {author} {\bibfnamefont {G.}~\bibnamefont
  {Sierra}},\ }\bibfield  {title} {\enquote {\bibinfo {title} {Entanglement of
  low-energy excitations in conformal field theory},}\ }\href@noop {}
  {\bibfield  {journal} {\bibinfo  {journal} {Physical Review Letters}\
  }\textbf {\bibinfo {volume} {106}},\ \bibinfo {pages} {201601} (\bibinfo
  {year} {2011}{\natexlab{a}})}\BibitemShut {NoStop}%
\bibitem [{\citenamefont {Nozaki}, \citenamefont {Numasawa},\ and\
  \citenamefont {Takayanagi}(2014)}]{Nozaki:2014hna}%
  \BibitemOpen
  \bibfield  {author} {\bibinfo {author} {\bibfnamefont {M.}~\bibnamefont
  {Nozaki}}, \bibinfo {author} {\bibfnamefont {T.}~\bibnamefont {Numasawa}},\
  and\ \bibinfo {author} {\bibfnamefont {T.}~\bibnamefont {Takayanagi}},\
  }\bibfield  {title} {\enquote {\bibinfo {title} {{Quantum Entanglement of
  Local Operators in Conformal Field Theories}},}\ }\href
  {https://doi.org/10.1103/PhysRevLett.112.111602} {\bibfield  {journal}
  {\bibinfo  {journal} {Phys. Rev. Lett.}\ }\textbf {\bibinfo {volume} {112}},\
  \bibinfo {pages} {111602} (\bibinfo {year} {2014})},\ \Eprint
  {https://arxiv.org/abs/1401.0539} {arXiv:1401.0539 [hep-th]} \BibitemShut
  {NoStop}%
\bibitem [{\citenamefont {Alcaraz}, \citenamefont {Berganza},\ and\
  \citenamefont {Sierra}(2011{\natexlab{b}})}]{Alcaraz:2011tn}%
  \BibitemOpen
  \bibfield  {author} {\bibinfo {author} {\bibfnamefont {F.~C.}\ \bibnamefont
  {Alcaraz}}, \bibinfo {author} {\bibfnamefont {M.~I.}\ \bibnamefont
  {Berganza}},\ and\ \bibinfo {author} {\bibfnamefont {G.}~\bibnamefont
  {Sierra}},\ }\bibfield  {title} {\enquote {\bibinfo {title} {{Entanglement of
  low-energy excitations in Conformal Field Theory}},}\ }\href
  {https://doi.org/10.1103/PhysRevLett.106.201601} {\bibfield  {journal}
  {\bibinfo  {journal} {Phys. Rev. Lett.}\ }\textbf {\bibinfo {volume} {106}},\
  \bibinfo {pages} {201601} (\bibinfo {year} {2011}{\natexlab{b}})},\ \Eprint
  {https://arxiv.org/abs/1101.2881} {arXiv:1101.2881 [cond-mat.stat-mech]}
  \BibitemShut {NoStop}%
\bibitem [{\citenamefont {He}\ \emph {et~al.}(2014)\citenamefont {He},
  \citenamefont {Numasawa}, \citenamefont {Takayanagi},\ and\ \citenamefont
  {Watanabe}}]{He:2014mwa}%
  \BibitemOpen
  \bibfield  {author} {\bibinfo {author} {\bibfnamefont {S.}~\bibnamefont
  {He}}, \bibinfo {author} {\bibfnamefont {T.}~\bibnamefont {Numasawa}},
  \bibinfo {author} {\bibfnamefont {T.}~\bibnamefont {Takayanagi}},\ and\
  \bibinfo {author} {\bibfnamefont {K.}~\bibnamefont {Watanabe}},\ }\bibfield
  {title} {\enquote {\bibinfo {title} {{Quantum dimension as entanglement
  entropy in two dimensional conformal field theories}},}\ }\href
  {https://doi.org/10.1103/PhysRevD.90.041701} {\bibfield  {journal} {\bibinfo
  {journal} {Phys. Rev. D}\ }\textbf {\bibinfo {volume} {90}},\ \bibinfo
  {pages} {041701} (\bibinfo {year} {2014})},\ \Eprint
  {https://arxiv.org/abs/1403.0702} {arXiv:1403.0702 [hep-th]} \BibitemShut
  {NoStop}%
\bibitem [{\citenamefont {Nozaki}(2014)}]{Nozaki:2014uaa}%
  \BibitemOpen
  \bibfield  {author} {\bibinfo {author} {\bibfnamefont {M.}~\bibnamefont
  {Nozaki}},\ }\bibfield  {title} {\enquote {\bibinfo {title} {{Notes on
  Quantum Entanglement of Local Operators}},}\ }\href
  {https://doi.org/10.1007/JHEP10(2014)147} {\bibfield  {journal} {\bibinfo
  {journal} {JHEP}\ }\textbf {\bibinfo {volume} {10}},\ \bibinfo {pages} {147}
  (\bibinfo {year} {2014})},\ \Eprint {https://arxiv.org/abs/1405.5875}
  {arXiv:1405.5875 [hep-th]} \BibitemShut {NoStop}%
\bibitem [{\citenamefont {Caputa}, \citenamefont {Nozaki},\ and\ \citenamefont
  {Takayanagi}(2014)}]{Caputa:2014vaa}%
  \BibitemOpen
  \bibfield  {author} {\bibinfo {author} {\bibfnamefont {P.}~\bibnamefont
  {Caputa}}, \bibinfo {author} {\bibfnamefont {M.}~\bibnamefont {Nozaki}},\
  and\ \bibinfo {author} {\bibfnamefont {T.}~\bibnamefont {Takayanagi}},\
  }\bibfield  {title} {\enquote {\bibinfo {title} {{Entanglement of local
  operators in large-N conformal field theories}},}\ }\href
  {https://doi.org/10.1093/ptep/ptu122} {\bibfield  {journal} {\bibinfo
  {journal} {PTEP}\ }\textbf {\bibinfo {volume} {2014}},\ \bibinfo {pages}
  {093B06} (\bibinfo {year} {2014})},\ \Eprint
  {https://arxiv.org/abs/1405.5946} {arXiv:1405.5946 [hep-th]} \BibitemShut
  {NoStop}%
\bibitem [{\citenamefont {Caputa}\ \emph {et~al.}(2015)\citenamefont {Caputa},
  \citenamefont {Sim\'on}, \citenamefont {\v{S}tikonas},\ and\ \citenamefont
  {Takayanagi}}]{Caputa:2014eta}%
  \BibitemOpen
  \bibfield  {author} {\bibinfo {author} {\bibfnamefont {P.}~\bibnamefont
  {Caputa}}, \bibinfo {author} {\bibfnamefont {J.}~\bibnamefont {Sim\'on}},
  \bibinfo {author} {\bibfnamefont {A.}~\bibnamefont {\v{S}tikonas}},\ and\
  \bibinfo {author} {\bibfnamefont {T.}~\bibnamefont {Takayanagi}},\ }\bibfield
   {title} {\enquote {\bibinfo {title} {{Quantum Entanglement of Localized
  Excited States at Finite Temperature}},}\ }\href
  {https://doi.org/10.1007/JHEP01(2015)102} {\bibfield  {journal} {\bibinfo
  {journal} {JHEP}\ }\textbf {\bibinfo {volume} {01}},\ \bibinfo {pages} {102}
  (\bibinfo {year} {2015})},\ \Eprint {https://arxiv.org/abs/1410.2287}
  {arXiv:1410.2287 [hep-th]} \BibitemShut {NoStop}%
\bibitem [{\citenamefont {Guo}\ and\ \citenamefont {He}(2015)}]{Guo:2015uwa}%
  \BibitemOpen
  \bibfield  {author} {\bibinfo {author} {\bibfnamefont {W.-Z.}\ \bibnamefont
  {Guo}}\ and\ \bibinfo {author} {\bibfnamefont {S.}~\bibnamefont {He}},\
  }\bibfield  {title} {\enquote {\bibinfo {title} {{R\'enyi entropy of locally
  excited states with thermal and boundary effect in 2D CFTs}},}\ }\href
  {https://doi.org/10.1007/JHEP04(2015)099} {\bibfield  {journal} {\bibinfo
  {journal} {JHEP}\ }\textbf {\bibinfo {volume} {04}},\ \bibinfo {pages} {099}
  (\bibinfo {year} {2015})},\ \Eprint {https://arxiv.org/abs/1501.00757}
  {arXiv:1501.00757 [hep-th]} \BibitemShut {NoStop}%
\bibitem [{\citenamefont {Caputa}\ and\ \citenamefont
  {Veliz-Osorio}(2015)}]{Caputa:2015tua}%
  \BibitemOpen
  \bibfield  {author} {\bibinfo {author} {\bibfnamefont {P.}~\bibnamefont
  {Caputa}}\ and\ \bibinfo {author} {\bibfnamefont {A.}~\bibnamefont
  {Veliz-Osorio}},\ }\bibfield  {title} {\enquote {\bibinfo {title}
  {{Entanglement constant for conformal families}},}\ }\href
  {https://doi.org/10.1103/PhysRevD.92.065010} {\bibfield  {journal} {\bibinfo
  {journal} {Phys. Rev. D}\ }\textbf {\bibinfo {volume} {92}},\ \bibinfo
  {pages} {065010} (\bibinfo {year} {2015})},\ \Eprint
  {https://arxiv.org/abs/1507.00582} {arXiv:1507.00582 [hep-th]} \BibitemShut
  {NoStop}%
\bibitem [{\citenamefont {Chen}\ \emph {et~al.}(2015)\citenamefont {Chen},
  \citenamefont {Guo}, \citenamefont {He},\ and\ \citenamefont
  {Wu}}]{Chen:2015usa}%
  \BibitemOpen
  \bibfield  {author} {\bibinfo {author} {\bibfnamefont {B.}~\bibnamefont
  {Chen}}, \bibinfo {author} {\bibfnamefont {W.-Z.}\ \bibnamefont {Guo}},
  \bibinfo {author} {\bibfnamefont {S.}~\bibnamefont {He}},\ and\ \bibinfo
  {author} {\bibfnamefont {J.-q.}\ \bibnamefont {Wu}},\ }\bibfield  {title}
  {\enquote {\bibinfo {title} {{Entanglement Entropy for Descendent Local
  Operators in 2D CFTs}},}\ }\href {https://doi.org/10.1007/JHEP10(2015)173}
  {\bibfield  {journal} {\bibinfo  {journal} {JHEP}\ }\textbf {\bibinfo
  {volume} {10}},\ \bibinfo {pages} {173} (\bibinfo {year} {2015})},\ \Eprint
  {https://arxiv.org/abs/1507.01157} {arXiv:1507.01157 [hep-th]} \BibitemShut
  {NoStop}%
\bibitem [{\citenamefont {Caputa}, \citenamefont {Numasawa},\ and\
  \citenamefont {Veliz-Osorio}(2016)}]{Caputa:2016tgt}%
  \BibitemOpen
  \bibfield  {author} {\bibinfo {author} {\bibfnamefont {P.}~\bibnamefont
  {Caputa}}, \bibinfo {author} {\bibfnamefont {T.}~\bibnamefont {Numasawa}},\
  and\ \bibinfo {author} {\bibfnamefont {A.}~\bibnamefont {Veliz-Osorio}},\
  }\bibfield  {title} {\enquote {\bibinfo {title} {{Out-of-time-ordered
  correlators and purity in rational conformal field theories}},}\ }\href
  {https://doi.org/10.1093/ptep/ptw157} {\bibfield  {journal} {\bibinfo
  {journal} {PTEP}\ }\textbf {\bibinfo {volume} {2016}},\ \bibinfo {pages}
  {113B06} (\bibinfo {year} {2016})},\ \Eprint
  {https://arxiv.org/abs/1602.06542} {arXiv:1602.06542 [hep-th]} \BibitemShut
  {NoStop}%
\bibitem [{\citenamefont {Numasawa}(2016)}]{Numasawa:2016kmo}%
  \BibitemOpen
  \bibfield  {author} {\bibinfo {author} {\bibfnamefont {T.}~\bibnamefont
  {Numasawa}},\ }\bibfield  {title} {\enquote {\bibinfo {title} {{Scattering
  effect on entanglement propagation in RCFTs}},}\ }\href
  {https://doi.org/10.1007/JHEP12(2016)061} {\bibfield  {journal} {\bibinfo
  {journal} {JHEP}\ }\textbf {\bibinfo {volume} {12}},\ \bibinfo {pages} {061}
  (\bibinfo {year} {2016})},\ \Eprint {https://arxiv.org/abs/1610.06181}
  {arXiv:1610.06181 [hep-th]} \BibitemShut {NoStop}%
\bibitem [{\citenamefont {He}(2019)}]{He:2017lrg}%
  \BibitemOpen
  \bibfield  {author} {\bibinfo {author} {\bibfnamefont {S.}~\bibnamefont
  {He}},\ }\bibfield  {title} {\enquote {\bibinfo {title} {{Conformal bootstrap
  to R\'enyi entropy in 2D Liouville and super-Liouville CFTs}},}\ }\href
  {https://doi.org/10.1103/PhysRevD.99.026005} {\bibfield  {journal} {\bibinfo
  {journal} {Phys. Rev. D}\ }\textbf {\bibinfo {volume} {99}},\ \bibinfo
  {pages} {026005} (\bibinfo {year} {2019})},\ \Eprint
  {https://arxiv.org/abs/1711.00624} {arXiv:1711.00624 [hep-th]} \BibitemShut
  {NoStop}%
\bibitem [{\citenamefont {Guo}, \citenamefont {He},\ and\ \citenamefont
  {Luo}(2018)}]{Guo:2018lqq}%
  \BibitemOpen
  \bibfield  {author} {\bibinfo {author} {\bibfnamefont {W.-Z.}\ \bibnamefont
  {Guo}}, \bibinfo {author} {\bibfnamefont {S.}~\bibnamefont {He}},\ and\
  \bibinfo {author} {\bibfnamefont {Z.-X.}\ \bibnamefont {Luo}},\ }\bibfield
  {title} {\enquote {\bibinfo {title} {{Entanglement entropy in (1+1)D CFTs
  with multiple local excitations}},}\ }\href
  {https://doi.org/10.1007/JHEP05(2018)154} {\bibfield  {journal} {\bibinfo
  {journal} {JHEP}\ }\textbf {\bibinfo {volume} {05}},\ \bibinfo {pages} {154}
  (\bibinfo {year} {2018})},\ \Eprint {https://arxiv.org/abs/1802.08815}
  {arXiv:1802.08815 [hep-th]} \BibitemShut {NoStop}%
\bibitem [{\citenamefont {Apolo}\ \emph {et~al.}(2019)\citenamefont {Apolo},
  \citenamefont {He}, \citenamefont {Song}, \citenamefont {Xu},\ and\
  \citenamefont {Zheng}}]{Apolo:2018oqv}%
  \BibitemOpen
  \bibfield  {author} {\bibinfo {author} {\bibfnamefont {L.}~\bibnamefont
  {Apolo}}, \bibinfo {author} {\bibfnamefont {S.}~\bibnamefont {He}}, \bibinfo
  {author} {\bibfnamefont {W.}~\bibnamefont {Song}}, \bibinfo {author}
  {\bibfnamefont {J.}~\bibnamefont {Xu}},\ and\ \bibinfo {author}
  {\bibfnamefont {J.}~\bibnamefont {Zheng}},\ }\bibfield  {title} {\enquote
  {\bibinfo {title} {{Entanglement and chaos in warped conformal field
  theories}},}\ }\href {https://doi.org/10.1007/JHEP04(2019)009} {\bibfield
  {journal} {\bibinfo  {journal} {JHEP}\ }\textbf {\bibinfo {volume} {04}},\
  \bibinfo {pages} {009} (\bibinfo {year} {2019})},\ \Eprint
  {https://arxiv.org/abs/1812.10456} {arXiv:1812.10456 [hep-th]} \BibitemShut
  {NoStop}%
\bibitem [{\citenamefont {Caputa}\ \emph {et~al.}(2019)\citenamefont {Caputa},
  \citenamefont {Numasawa}, \citenamefont {Shimaji}, \citenamefont
  {Takayanagi},\ and\ \citenamefont {Wei}}]{Caputa:2019avh}%
  \BibitemOpen
  \bibfield  {author} {\bibinfo {author} {\bibfnamefont {P.}~\bibnamefont
  {Caputa}}, \bibinfo {author} {\bibfnamefont {T.}~\bibnamefont {Numasawa}},
  \bibinfo {author} {\bibfnamefont {T.}~\bibnamefont {Shimaji}}, \bibinfo
  {author} {\bibfnamefont {T.}~\bibnamefont {Takayanagi}},\ and\ \bibinfo
  {author} {\bibfnamefont {Z.}~\bibnamefont {Wei}},\ }\bibfield  {title}
  {\enquote {\bibinfo {title} {{Double Local Quenches in 2D CFTs and
  Gravitational Force}},}\ }\href {https://doi.org/10.1007/JHEP09(2019)018}
  {\bibfield  {journal} {\bibinfo  {journal} {JHEP}\ }\textbf {\bibinfo
  {volume} {09}},\ \bibinfo {pages} {018} (\bibinfo {year} {2019})},\ \Eprint
  {https://arxiv.org/abs/1905.08265} {arXiv:1905.08265 [hep-th]} \BibitemShut
  {NoStop}%
\bibitem [{\citenamefont {Bianchi}, \citenamefont {De~Angelis},\ and\
  \citenamefont {Meineri}(2022)}]{Bianchi:2022ulu}%
  \BibitemOpen
  \bibfield  {author} {\bibinfo {author} {\bibfnamefont {L.}~\bibnamefont
  {Bianchi}}, \bibinfo {author} {\bibfnamefont {S.}~\bibnamefont
  {De~Angelis}},\ and\ \bibinfo {author} {\bibfnamefont {M.}~\bibnamefont
  {Meineri}},\ }\bibfield  {title} {\enquote {\bibinfo {title} {{Radiation,
  entanglement and islands from a boundary local quench}},}\ }\href@noop {} {\
  (\bibinfo {year} {2022})},\ \Eprint {https://arxiv.org/abs/2203.10103}
  {arXiv:2203.10103 [hep-th]} \BibitemShut {NoStop}%
\bibitem [{Note2()}]{Note2}%
  \BibitemOpen
  \bibinfo {note} {See \cite {Miyaji:2015woj,
  Miyaji:2016fse,Zhang:2019kwu,Wen:2015qwa,Kudler-Flam:2018qjo,Kudler-Flam:2019oru,Kudler-Flam:2020url,Kudler-Flam:2020yml,Kudler-Flam:2020xqu}
  for studies on other information quantities (such as information metric,
  negativity, reflected entropy, etc) in local or global quantum quenches in
  CFTs.}\BibitemShut {Stop}%
\bibitem [{\citenamefont {Asplund}\ \emph
  {et~al.}(2015{\natexlab{a}})\citenamefont {Asplund}, \citenamefont
  {Bernamonti}, \citenamefont {Galli},\ and\ \citenamefont
  {Hartman}}]{Asplund:2014coa}%
  \BibitemOpen
  \bibfield  {author} {\bibinfo {author} {\bibfnamefont {C.~T.}\ \bibnamefont
  {Asplund}}, \bibinfo {author} {\bibfnamefont {A.}~\bibnamefont {Bernamonti}},
  \bibinfo {author} {\bibfnamefont {F.}~\bibnamefont {Galli}},\ and\ \bibinfo
  {author} {\bibfnamefont {T.}~\bibnamefont {Hartman}},\ }\bibfield  {title}
  {\enquote {\bibinfo {title} {{Holographic Entanglement Entropy from 2d CFT:
  Heavy States and Local Quenches}},}\ }\href
  {https://doi.org/10.1007/JHEP02(2015)171} {\bibfield  {journal} {\bibinfo
  {journal} {JHEP}\ }\textbf {\bibinfo {volume} {02}},\ \bibinfo {pages} {171}
  (\bibinfo {year} {2015}{\natexlab{a}})},\ \Eprint
  {https://arxiv.org/abs/1410.1392} {arXiv:1410.1392 [hep-th]} \BibitemShut
  {NoStop}%
\bibitem [{\citenamefont {Asplund}\ \emph
  {et~al.}(2015{\natexlab{b}})\citenamefont {Asplund}, \citenamefont
  {Bernamonti}, \citenamefont {Galli},\ and\ \citenamefont
  {Hartman}}]{Asplund:2015eha}%
  \BibitemOpen
  \bibfield  {author} {\bibinfo {author} {\bibfnamefont {C.~T.}\ \bibnamefont
  {Asplund}}, \bibinfo {author} {\bibfnamefont {A.}~\bibnamefont {Bernamonti}},
  \bibinfo {author} {\bibfnamefont {F.}~\bibnamefont {Galli}},\ and\ \bibinfo
  {author} {\bibfnamefont {T.}~\bibnamefont {Hartman}},\ }\bibfield  {title}
  {\enquote {\bibinfo {title} {{Entanglement Scrambling in 2d Conformal Field
  Theory}},}\ }\href {https://doi.org/10.1007/JHEP09(2015)110} {\bibfield
  {journal} {\bibinfo  {journal} {JHEP}\ }\textbf {\bibinfo {volume} {09}},\
  \bibinfo {pages} {110} (\bibinfo {year} {2015}{\natexlab{b}})},\ \Eprint
  {https://arxiv.org/abs/1506.03772} {arXiv:1506.03772 [hep-th]} \BibitemShut
  {NoStop}%
\bibitem [{\citenamefont {Suzuki}, \citenamefont {Takayanagi},\ and\
  \citenamefont {Umemoto}(2019)}]{Suzuki:2019xdq}%
  \BibitemOpen
  \bibfield  {author} {\bibinfo {author} {\bibfnamefont {Y.}~\bibnamefont
  {Suzuki}}, \bibinfo {author} {\bibfnamefont {T.}~\bibnamefont {Takayanagi}},\
  and\ \bibinfo {author} {\bibfnamefont {K.}~\bibnamefont {Umemoto}},\
  }\bibfield  {title} {\enquote {\bibinfo {title} {{Entanglement Wedges from
  the Information Metric in Conformal Field Theories}},}\ }\href
  {https://doi.org/10.1103/PhysRevLett.123.221601} {\bibfield  {journal}
  {\bibinfo  {journal} {Phys. Rev. Lett.}\ }\textbf {\bibinfo {volume} {123}},\
  \bibinfo {pages} {221601} (\bibinfo {year} {2019})},\ \Eprint
  {https://arxiv.org/abs/1908.09939} {arXiv:1908.09939 [hep-th]} \BibitemShut
  {NoStop}%
\bibitem [{\citenamefont {Nozaki}, \citenamefont {Numasawa},\ and\
  \citenamefont {Takayanagi}(2013)}]{Nozaki:2013wia}%
  \BibitemOpen
  \bibfield  {author} {\bibinfo {author} {\bibfnamefont {M.}~\bibnamefont
  {Nozaki}}, \bibinfo {author} {\bibfnamefont {T.}~\bibnamefont {Numasawa}},\
  and\ \bibinfo {author} {\bibfnamefont {T.}~\bibnamefont {Takayanagi}},\
  }\bibfield  {title} {\enquote {\bibinfo {title} {{Holographic Local Quenches
  and Entanglement Density}},}\ }\href
  {https://doi.org/10.1007/JHEP05(2013)080} {\bibfield  {journal} {\bibinfo
  {journal} {JHEP}\ }\textbf {\bibinfo {volume} {05}},\ \bibinfo {pages} {080}
  (\bibinfo {year} {2013})},\ \Eprint {https://arxiv.org/abs/1302.5703}
  {arXiv:1302.5703 [hep-th]} \BibitemShut {NoStop}%
\bibitem [{\citenamefont {Kusuki}\ and\ \citenamefont
  {Miyaji}(2020)}]{Kusuki:2019avm}%
  \BibitemOpen
  \bibfield  {author} {\bibinfo {author} {\bibfnamefont {Y.}~\bibnamefont
  {Kusuki}}\ and\ \bibinfo {author} {\bibfnamefont {M.}~\bibnamefont
  {Miyaji}},\ }\bibfield  {title} {\enquote {\bibinfo {title} {{Entanglement
  Entropy after Double Excitation as an Interaction Measure}},}\ }\href
  {https://doi.org/10.1103/PhysRevLett.124.061601} {\bibfield  {journal}
  {\bibinfo  {journal} {Phys. Rev. Lett.}\ }\textbf {\bibinfo {volume} {124}},\
  \bibinfo {pages} {061601} (\bibinfo {year} {2020})},\ \Eprint
  {https://arxiv.org/abs/1908.03351} {arXiv:1908.03351 [hep-th]} \BibitemShut
  {NoStop}%
\bibitem [{\citenamefont {Kawamoto}\ \emph {et~al.}(2022)\citenamefont
  {Kawamoto}, \citenamefont {Mori}, \citenamefont {Suzuki}, \citenamefont
  {Takayanagi},\ and\ \citenamefont {Ugajin}}]{Kawamoto:2022etl}%
  \BibitemOpen
  \bibfield  {author} {\bibinfo {author} {\bibfnamefont {T.}~\bibnamefont
  {Kawamoto}}, \bibinfo {author} {\bibfnamefont {T.}~\bibnamefont {Mori}},
  \bibinfo {author} {\bibfnamefont {Y.-k.}\ \bibnamefont {Suzuki}}, \bibinfo
  {author} {\bibfnamefont {T.}~\bibnamefont {Takayanagi}},\ and\ \bibinfo
  {author} {\bibfnamefont {T.}~\bibnamefont {Ugajin}},\ }\bibfield  {title}
  {\enquote {\bibinfo {title} {{Holographic local operator quenches in
  BCFTs}},}\ }\href {https://doi.org/10.1007/JHEP05(2022)060} {\bibfield
  {journal} {\bibinfo  {journal} {JHEP}\ }\textbf {\bibinfo {volume} {05}},\
  \bibinfo {pages} {060} (\bibinfo {year} {2022})},\ \Eprint
  {https://arxiv.org/abs/2203.03851} {arXiv:2203.03851 [hep-th]} \BibitemShut
  {NoStop}%
\bibitem [{\citenamefont {Moore}\ and\ \citenamefont
  {Seiberg}(1989)}]{Moore:1988ss}%
  \BibitemOpen
  \bibfield  {author} {\bibinfo {author} {\bibfnamefont {G.~W.}\ \bibnamefont
  {Moore}}\ and\ \bibinfo {author} {\bibfnamefont {N.}~\bibnamefont
  {Seiberg}},\ }\bibfield  {title} {\enquote {\bibinfo {title} {{Naturality in
  Conformal Field Theory}},}\ }\href
  {https://doi.org/10.1016/0550-3213(89)90511-7} {\bibfield  {journal}
  {\bibinfo  {journal} {Nucl. Phys. B}\ }\textbf {\bibinfo {volume} {313}},\
  \bibinfo {pages} {16--40} (\bibinfo {year} {1989})}\BibitemShut {NoStop}%
\bibitem [{\citenamefont {He}\ \emph {et~al.}(2023)\citenamefont {He},
  \citenamefont {Zhang}, \citenamefont {Zhao},\ and\ \citenamefont
  {Zhao}}]{He:2023syy}%
  \BibitemOpen
  \bibfield  {author} {\bibinfo {author} {\bibfnamefont {S.}~\bibnamefont
  {He}}, \bibinfo {author} {\bibfnamefont {Y.-X.}\ \bibnamefont {Zhang}},
  \bibinfo {author} {\bibfnamefont {L.}~\bibnamefont {Zhao}},\ and\ \bibinfo
  {author} {\bibfnamefont {Z.-X.}\ \bibnamefont {Zhao}},\ }\bibfield  {title}
  {\enquote {\bibinfo {title} {{Entanglement and Pseudo Entanglement Dynamics
  versus Fusion in CFT}},}\ }\href@noop {} {\  (\bibinfo {year} {2023})},\
  \Eprint {https://arxiv.org/abs/2312.02679} {arXiv:2312.02679 [hep-th]}
  \BibitemShut {NoStop}%
\bibitem [{\citenamefont {Shi}, \citenamefont {Kato},\ and\ \citenamefont
  {Kim}(2020)}]{Shi:2019mlt}%
  \BibitemOpen
  \bibfield  {author} {\bibinfo {author} {\bibfnamefont {B.}~\bibnamefont
  {Shi}}, \bibinfo {author} {\bibfnamefont {K.}~\bibnamefont {Kato}},\ and\
  \bibinfo {author} {\bibfnamefont {I.~H.}\ \bibnamefont {Kim}},\ }\bibfield
  {title} {\enquote {\bibinfo {title} {{Fusion rules from entanglement}},}\
  }\href {https://doi.org/10.1016/j.aop.2020.168164} {\bibfield  {journal}
  {\bibinfo  {journal} {Annals Phys.}\ }\textbf {\bibinfo {volume} {418}},\
  \bibinfo {pages} {168164} (\bibinfo {year} {2020})},\ \Eprint
  {https://arxiv.org/abs/1906.09376} {arXiv:1906.09376 [cond-mat.str-el]}
  \BibitemShut {NoStop}%
\bibitem [{\citenamefont {Verlinde}(1988)}]{Verlinde:1988sn}%
  \BibitemOpen
  \bibfield  {author} {\bibinfo {author} {\bibfnamefont {E.~P.}\ \bibnamefont
  {Verlinde}},\ }\bibfield  {title} {\enquote {\bibinfo {title} {{Fusion Rules
  and Modular Transformations in 2D Conformal Field Theory}},}\ }\href
  {https://doi.org/10.1016/0550-3213(88)90603-7} {\bibfield  {journal}
  {\bibinfo  {journal} {Nucl. Phys. B}\ }\textbf {\bibinfo {volume} {300}},\
  \bibinfo {pages} {360--376} (\bibinfo {year} {1988})}\BibitemShut {NoStop}%
\bibitem [{Note3()}]{Note3}%
  \BibitemOpen
  \bibinfo {note} {Due to the orthogonality of primary operators with different
  conformal dimensions in the sense of two-point function, we usually have to
  choose the same primary operators for constructing the transition matrix in
  specific models \cite {Guo:2022sfl}.}\BibitemShut {Stop}%
\bibitem [{\citenamefont {Fuchs}(1991)}]{fuchs1991quantum}%
  \BibitemOpen
  \bibfield  {author} {\bibinfo {author} {\bibfnamefont {J.}~\bibnamefont
  {Fuchs}},\ }\bibfield  {title} {\enquote {\bibinfo {title} {Quantum
  dimensions},}\ }\href@noop {} {\bibfield  {journal} {\bibinfo  {journal}
  {Commun. Theor. Phys.}\ }\textbf {\bibinfo {volume} {1}},\ \bibinfo {pages}
  {59--109} (\bibinfo {year} {1991})}\BibitemShut {NoStop}%
\bibitem [{\citenamefont {Di~Francesco}, \citenamefont {Mathieu},\ and\
  \citenamefont {Senechal}(1997)}]{DiFrancesco:1997nk}%
  \BibitemOpen
  \bibfield  {author} {\bibinfo {author} {\bibfnamefont {P.}~\bibnamefont
  {Di~Francesco}}, \bibinfo {author} {\bibfnamefont {P.}~\bibnamefont
  {Mathieu}},\ and\ \bibinfo {author} {\bibfnamefont {D.}~\bibnamefont
  {Senechal}},\ }\href {https://doi.org/10.1007/978-1-4612-2256-9} {\emph
  {\bibinfo {title} {{Conformal Field Theory}}}},\ Graduate Texts in
  Contemporary Physics\ (\bibinfo  {publisher} {Springer-Verlag},\ \bibinfo
  {address} {New York},\ \bibinfo {year} {1997})\BibitemShut {NoStop}%
\bibitem [{Note4()}]{Note4}%
  \BibitemOpen
  \bibinfo {note} {Here the following equation to simplify the result has been
  used \par $\DOTSB \sum@ \slimits@ \limits _{k=1}^{m-1} \protect \frac {(k+m)
  (-k+m+1) (k+n-1)!}{(k+1)! (n-2)!}=\protect \frac {2 \left (m^2 n+m \left (2
  n^2-1\right )-n (n+1)\right ) \Gamma (m+n)}{\Gamma (m+1) \Gamma (n+2)}-m
  (m+1) n+2$}\BibitemShut {NoStop}%
\bibitem [{\citenamefont {Kudler-Flam}, \citenamefont {MacCormack},\ and\
  \citenamefont {Ryu}(2019)}]{Kudler-Flam:2019oru}%
  \BibitemOpen
  \bibfield  {author} {\bibinfo {author} {\bibfnamefont {J.}~\bibnamefont
  {Kudler-Flam}}, \bibinfo {author} {\bibfnamefont {I.}~\bibnamefont
  {MacCormack}},\ and\ \bibinfo {author} {\bibfnamefont {S.}~\bibnamefont
  {Ryu}},\ }\bibfield  {title} {\enquote {\bibinfo {title} {{Holographic
  entanglement contour, bit threads, and the entanglement tsunami}},}\ }\href
  {https://doi.org/10.1088/1751-8121/ab2dae} {\bibfield  {journal} {\bibinfo
  {journal} {J. Phys. A}\ }\textbf {\bibinfo {volume} {52}},\ \bibinfo {pages}
  {325401} (\bibinfo {year} {2019})},\ \Eprint
  {https://arxiv.org/abs/1902.04654} {arXiv:1902.04654 [hep-th]} \BibitemShut
  {NoStop}%
\bibitem [{\citenamefont {Wen}, \citenamefont {Chang},\ and\ \citenamefont
  {Ryu}(2015)}]{Wen:2015qwa}%
  \BibitemOpen
  \bibfield  {author} {\bibinfo {author} {\bibfnamefont {X.}~\bibnamefont
  {Wen}}, \bibinfo {author} {\bibfnamefont {P.-Y.}\ \bibnamefont {Chang}},\
  and\ \bibinfo {author} {\bibfnamefont {S.}~\bibnamefont {Ryu}},\ }\bibfield
  {title} {\enquote {\bibinfo {title} {{Entanglement negativity after a local
  quantum quench in conformal field theories}},}\ }\href
  {https://doi.org/10.1103/PhysRevB.92.075109} {\bibfield  {journal} {\bibinfo
  {journal} {Phys. Rev. B}\ }\textbf {\bibinfo {volume} {92}},\ \bibinfo
  {pages} {075109} (\bibinfo {year} {2015})},\ \Eprint
  {https://arxiv.org/abs/1501.00568} {arXiv:1501.00568 [cond-mat.stat-mech]}
  \BibitemShut {NoStop}%
\bibitem [{\citenamefont {Kudler-Flam}\ and\ \citenamefont
  {Ryu}(2019)}]{Kudler-Flam:2018qjo}%
  \BibitemOpen
  \bibfield  {author} {\bibinfo {author} {\bibfnamefont {J.}~\bibnamefont
  {Kudler-Flam}}\ and\ \bibinfo {author} {\bibfnamefont {S.}~\bibnamefont
  {Ryu}},\ }\bibfield  {title} {\enquote {\bibinfo {title} {{Entanglement
  negativity and minimal entanglement wedge cross sections in holographic
  theories}},}\ }\href {https://doi.org/10.1103/PhysRevD.99.106014} {\bibfield
  {journal} {\bibinfo  {journal} {Phys. Rev. D}\ }\textbf {\bibinfo {volume}
  {99}},\ \bibinfo {pages} {106014} (\bibinfo {year} {2019})},\ \Eprint
  {https://arxiv.org/abs/1808.00446} {arXiv:1808.00446 [hep-th]} \BibitemShut
  {NoStop}%
\bibitem [{\citenamefont {Miyaji}\ \emph {et~al.}(2015)\citenamefont {Miyaji},
  \citenamefont {Numasawa}, \citenamefont {Shiba}, \citenamefont {Takayanagi},\
  and\ \citenamefont {Watanabe}}]{Miyaji:2015woj}%
  \BibitemOpen
  \bibfield  {author} {\bibinfo {author} {\bibfnamefont {M.}~\bibnamefont
  {Miyaji}}, \bibinfo {author} {\bibfnamefont {T.}~\bibnamefont {Numasawa}},
  \bibinfo {author} {\bibfnamefont {N.}~\bibnamefont {Shiba}}, \bibinfo
  {author} {\bibfnamefont {T.}~\bibnamefont {Takayanagi}},\ and\ \bibinfo
  {author} {\bibfnamefont {K.}~\bibnamefont {Watanabe}},\ }\bibfield  {title}
  {\enquote {\bibinfo {title} {{Distance between Quantum States and
  Gauge-Gravity Duality}},}\ }\href
  {https://doi.org/10.1103/PhysRevLett.115.261602} {\bibfield  {journal}
  {\bibinfo  {journal} {Phys. Rev. Lett.}\ }\textbf {\bibinfo {volume} {115}},\
  \bibinfo {pages} {261602} (\bibinfo {year} {2015})},\ \Eprint
  {https://arxiv.org/abs/1507.07555} {arXiv:1507.07555 [hep-th]} \BibitemShut
  {NoStop}%
\bibitem [{\citenamefont {Miyaji}(2016)}]{Miyaji:2016fse}%
  \BibitemOpen
  \bibfield  {author} {\bibinfo {author} {\bibfnamefont {M.}~\bibnamefont
  {Miyaji}},\ }\bibfield  {title} {\enquote {\bibinfo {title} {{Butterflies
  from Information Metric}},}\ }\href {https://doi.org/10.1007/JHEP09(2016)002}
  {\bibfield  {journal} {\bibinfo  {journal} {JHEP}\ }\textbf {\bibinfo
  {volume} {09}},\ \bibinfo {pages} {002} (\bibinfo {year} {2016})},\ \Eprint
  {https://arxiv.org/abs/1607.01467} {arXiv:1607.01467 [hep-th]} \BibitemShut
  {NoStop}%
\bibitem [{\citenamefont {Zhang}\ and\ \citenamefont
  {Calabrese}(2020)}]{Zhang:2019kwu}%
  \BibitemOpen
  \bibfield  {author} {\bibinfo {author} {\bibfnamefont {J.}~\bibnamefont
  {Zhang}}\ and\ \bibinfo {author} {\bibfnamefont {P.}~\bibnamefont
  {Calabrese}},\ }\bibfield  {title} {\enquote {\bibinfo {title} {{Subsystem
  distance after a local operator quench}},}\ }\href
  {https://doi.org/10.1007/JHEP02(2020)056} {\bibfield  {journal} {\bibinfo
  {journal} {JHEP}\ }\textbf {\bibinfo {volume} {02}},\ \bibinfo {pages} {056}
  (\bibinfo {year} {2020})},\ \Eprint {https://arxiv.org/abs/1911.04797}
  {arXiv:1911.04797 [hep-th]} \BibitemShut {NoStop}%
\bibitem [{\citenamefont {Kudler-Flam}, \citenamefont {Kusuki},\ and\
  \citenamefont {Ryu}(2020)}]{Kudler-Flam:2020url}%
  \BibitemOpen
  \bibfield  {author} {\bibinfo {author} {\bibfnamefont {J.}~\bibnamefont
  {Kudler-Flam}}, \bibinfo {author} {\bibfnamefont {Y.}~\bibnamefont
  {Kusuki}},\ and\ \bibinfo {author} {\bibfnamefont {S.}~\bibnamefont {Ryu}},\
  }\bibfield  {title} {\enquote {\bibinfo {title} {{Correlation measures and
  the entanglement wedge cross-section after quantum quenches in
  two-dimensional conformal field theories}},}\ }\href
  {https://doi.org/10.1007/JHEP04(2020)074} {\bibfield  {journal} {\bibinfo
  {journal} {JHEP}\ }\textbf {\bibinfo {volume} {04}},\ \bibinfo {pages} {074}
  (\bibinfo {year} {2020})},\ \Eprint {https://arxiv.org/abs/2001.05501}
  {arXiv:2001.05501 [hep-th]} \BibitemShut {NoStop}%
\bibitem [{\citenamefont {Kudler-Flam}, \citenamefont {Kusuki},\ and\
  \citenamefont {Ryu}(2021)}]{Kudler-Flam:2020xqu}%
  \BibitemOpen
  \bibfield  {author} {\bibinfo {author} {\bibfnamefont {J.}~\bibnamefont
  {Kudler-Flam}}, \bibinfo {author} {\bibfnamefont {Y.}~\bibnamefont
  {Kusuki}},\ and\ \bibinfo {author} {\bibfnamefont {S.}~\bibnamefont {Ryu}},\
  }\bibfield  {title} {\enquote {\bibinfo {title} {{The quasi-particle picture
  and its breakdown after local quenches: mutual information, negativity, and
  reflected entropy}},}\ }\href {https://doi.org/10.1007/JHEP03(2021)146}
  {\bibfield  {journal} {\bibinfo  {journal} {JHEP}\ }\textbf {\bibinfo
  {volume} {03}},\ \bibinfo {pages} {146} (\bibinfo {year} {2021})},\ \Eprint
  {https://arxiv.org/abs/2008.11266} {arXiv:2008.11266 [hep-th]} \BibitemShut
  {NoStop}%
\bibitem [{\citenamefont {Kudler-Flam}\ \emph {et~al.}(2021)\citenamefont
  {Kudler-Flam}, \citenamefont {Nozaki}, \citenamefont {Ryu},\ and\
  \citenamefont {Tan}}]{Kudler-Flam:2020yml}%
  \BibitemOpen
  \bibfield  {author} {\bibinfo {author} {\bibfnamefont {J.}~\bibnamefont
  {Kudler-Flam}}, \bibinfo {author} {\bibfnamefont {M.}~\bibnamefont {Nozaki}},
  \bibinfo {author} {\bibfnamefont {S.}~\bibnamefont {Ryu}},\ and\ \bibinfo
  {author} {\bibfnamefont {M.~T.}\ \bibnamefont {Tan}},\ }\bibfield  {title}
  {\enquote {\bibinfo {title} {{Entanglement of local operators and the
  butterfly effect}},}\ }\href
  {https://doi.org/10.1103/PhysRevResearch.3.033182} {\bibfield  {journal}
  {\bibinfo  {journal} {Phys. Rev. Res.}\ }\textbf {\bibinfo {volume} {3}},\
  \bibinfo {pages} {033182} (\bibinfo {year} {2021})},\ \Eprint
  {https://arxiv.org/abs/2005.14243} {arXiv:2005.14243 [hep-th]} \BibitemShut
  {NoStop}%
\bibitem [{\citenamefont {Unruh}(1976)}]{Unruh:1976db}%
  \BibitemOpen
  \bibfield  {author} {\bibinfo {author} {\bibfnamefont {W.~G.}\ \bibnamefont
  {Unruh}},\ }\bibfield  {title} {\enquote {\bibinfo {title} {{Notes on black
  hole evaporation}},}\ }\href {https://doi.org/10.1103/PhysRevD.14.870}
  {\bibfield  {journal} {\bibinfo  {journal} {Phys. Rev. D}\ }\textbf {\bibinfo
  {volume} {14}},\ \bibinfo {pages} {870} (\bibinfo {year} {1976})}\BibitemShut
  {NoStop}%
\bibitem [{\citenamefont {Srednicki}(1993)}]{Srednicki:1993im}%
  \BibitemOpen
  \bibfield  {author} {\bibinfo {author} {\bibfnamefont {M.}~\bibnamefont
  {Srednicki}},\ }\bibfield  {title} {\enquote {\bibinfo {title} {{Entropy and
  area}},}\ }\href {https://doi.org/10.1103/PhysRevLett.71.666} {\bibfield
  {journal} {\bibinfo  {journal} {Phys. Rev. Lett.}\ }\textbf {\bibinfo
  {volume} {71}},\ \bibinfo {pages} {666--669} (\bibinfo {year} {1993})},\
  \Eprint {https://arxiv.org/abs/hep-th/9303048} {arXiv:hep-th/9303048}
  \BibitemShut {NoStop}%
\bibitem [{\citenamefont {Aharonov}, \citenamefont {Bergmann},\ and\
  \citenamefont {Lebowitz}(1964)}]{PhysRev.134.B1410}%
  \BibitemOpen
  \bibfield  {author} {\bibinfo {author} {\bibfnamefont {Y.}~\bibnamefont
  {Aharonov}}, \bibinfo {author} {\bibfnamefont {P.~G.}\ \bibnamefont
  {Bergmann}},\ and\ \bibinfo {author} {\bibfnamefont {J.~L.}\ \bibnamefont
  {Lebowitz}},\ }\bibfield  {title} {\enquote {\bibinfo {title} {Time symmetry
  in the quantum process of measurement},}\ }\href
  {https://doi.org/10.1103/PhysRev.134.B1410} {\bibfield  {journal} {\bibinfo
  {journal} {Phys. Rev.}\ }\textbf {\bibinfo {volume} {134}},\ \bibinfo {pages}
  {B1410--B1416} (\bibinfo {year} {1964})}\BibitemShut {NoStop}%
\bibitem [{\citenamefont {Dressel}\ \emph {et~al.}(2014)\citenamefont
  {Dressel}, \citenamefont {Malik}, \citenamefont {Miatto}, \citenamefont
  {Jordan},\ and\ \citenamefont {Boyd}}]{RevModPhys.86.307}%
  \BibitemOpen
  \bibfield  {author} {\bibinfo {author} {\bibfnamefont {J.}~\bibnamefont
  {Dressel}}, \bibinfo {author} {\bibfnamefont {M.}~\bibnamefont {Malik}},
  \bibinfo {author} {\bibfnamefont {F.~M.}\ \bibnamefont {Miatto}}, \bibinfo
  {author} {\bibfnamefont {A.~N.}\ \bibnamefont {Jordan}},\ and\ \bibinfo
  {author} {\bibfnamefont {R.~W.}\ \bibnamefont {Boyd}},\ }\bibfield  {title}
  {\enquote {\bibinfo {title} {Colloquium: Understanding quantum weak values:
  Basics and applications},}\ }\href
  {https://doi.org/10.1103/RevModPhys.86.307} {\bibfield  {journal} {\bibinfo
  {journal} {Rev. Mod. Phys.}\ }\textbf {\bibinfo {volume} {86}},\ \bibinfo
  {pages} {307--316} (\bibinfo {year} {2014})}\BibitemShut {NoStop}%
\bibitem [{\citenamefont {Gaberdiel}(1994)}]{Gaberdiel:1994fs}%
  \BibitemOpen
  \bibfield  {author} {\bibinfo {author} {\bibfnamefont {M.}~\bibnamefont
  {Gaberdiel}},\ }\bibfield  {title} {\enquote {\bibinfo {title} {{A General
  transformation formula for conformal fields}},}\ }\href
  {https://doi.org/10.1016/0370-2693(94)90026-4} {\bibfield  {journal}
  {\bibinfo  {journal} {Phys. Lett. B}\ }\textbf {\bibinfo {volume} {325}},\
  \bibinfo {pages} {366--370} (\bibinfo {year} {1994})},\ \Eprint
  {https://arxiv.org/abs/hep-th/9401166} {arXiv:hep-th/9401166} \BibitemShut
  {NoStop}%
\end{thebibliography}%

\end{document}